\documentclass[fleqn,usenatbib]{mnras}

\usepackage[T1]{fontenc}
\usepackage{ae,aecompl}

\usepackage{graphicx}	
\usepackage{amsmath}	
\usepackage{amssymb}

\usepackage{txfonts}	

\newcommand{\ctxt}[1]{{#1}}
\newcommand{\btxt}[1]{{#1}}	

\title[RV survey of Carina's O stars]{A radial velocity survey of
  the Carina Nebula's O-type stars}

\author[M. M. Kiminki and N. Smith]{Megan
  M. Kiminki$^{1}$\thanks{E-mail:
    \href{mailto:mbagley@email.arizona.edu}{mbagley@email.arizona.edu}}
  and Nathan Smith$^{1}$ \\ $^{1}$Steward Observatory, University of
  Arizona, 933 N.  Cherry Avenue, Tucson, AZ 85721, USA}

\date{Accepted 2018 March 14. Received 2018 March 11; in original form 2017 June 17.}

\pubyear{2018}

\begin{document}
\label{firstpage}
\pagerange{\pageref{firstpage}--\pageref{lastpage}} \maketitle

%%%%%%%%%%%%%%%%%%%%%%%%%%%%%%%%%%%%%%%%%%%%%%%%%%%%%%%%%%%%%%%%%%%%%%%
\begin{abstract}
We have obtained multi-epoch observations of 31 O-type stars in the Carina Nebula using the CHIRON spectrograph on the CTIO/SMARTS 1.5-m telescope.  We measure their radial velocities to 1--2 km s$^{-1}$ precision and present new or updated orbital solutions for the binary systems HD 92607, HD 93576, HDE 303312, and HDE 305536.  We also compile radial velocities from the literature for 32 additional O-type and evolved massive stars in the region.  The combined data set shows a mean heliocentric radial velocity of \btxt{0.6 km s$^{-1}$.  We calculate a velocity dispersion of $\le9.1$ km s$^{-1}$,} consistent with an unbound, substructured OB association.  The Tr 14 cluster shows a marginally significant 5 km s$^{-1}$ radial velocity offset from its neighbor Tr 16, but there are otherwise no correlations between stellar position and velocity.  The O-type stars in Cr 228 and the South Pillars region have a lower velocity dispersion than the region as a whole, supporting a model of distributed massive-star formation rather than migration from the central clusters.  We compare our stellar velocities to the Carina Nebula's molecular gas and find that Tr 14 shows a close kinematic association with the Northern Cloud.  In contrast, Tr 16 has accelerated the Southern Cloud by 10--15 km s$^{-1}$, possibly triggering further massive-star formation.   The expansion of the surrounding \ion{H}{ii} region is not symmetric about the O-type stars in radial velocity space, indicating that the ionized gas is constrained by denser material on the far side.  
\end{abstract}

\begin{keywords}
binaries: spectroscopic -- ISM: evolution -- open clusters and associations: individual: Carina Nebula -- stars: early-type -- stars: kinematics and dynamics -- stars: massive 
\end{keywords}

%%%%%%%%%%%%%%%%%%%%%%%%%%%%%%%%%%%%%%%%%%%%%%%%%%%%%%%%%%%%%%%%%%%%%%%
\section{Introduction}
\label{sec:ostars-intro}

\begin{figure*}
  \centering \includegraphics[width=0.85\textwidth, trim=0 12mm 0 8mm,
    clip]{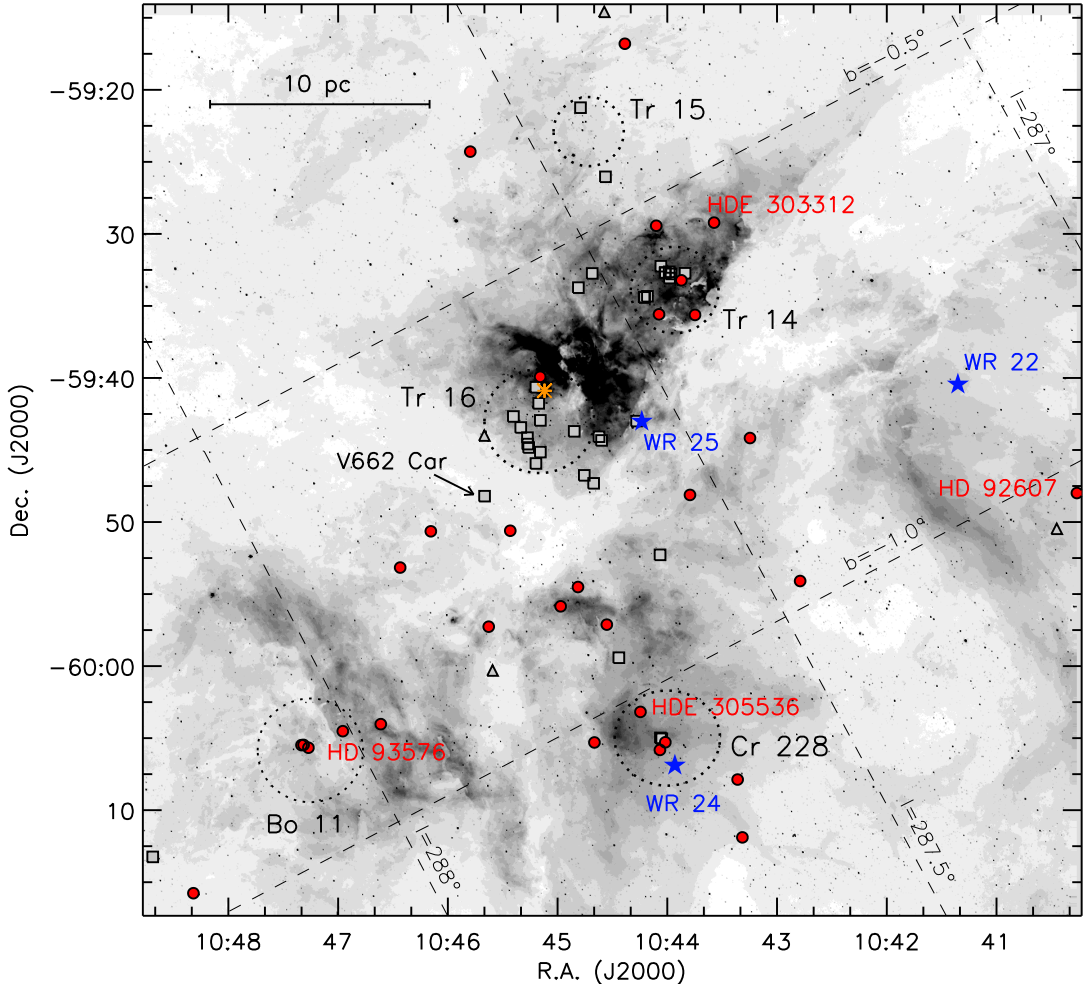}
  \caption{The O-type and evolved massive stars in the Carina Nebula,
    shown over a ground-based H$\alpha$ image from \citet{smith2010a}
    \btxt{with Galactic coordinates overlaid for reference}.  Red
    circles are O-type stars with new CHIRON observations; we identify
    by name the four spectroscopic binaries with new orbital solutions
    (see Section \ref{sec:ostars-binaries}).  Gray squares are O-type
    stars in version 3.1 of the Galactic O-Star Catalog
    \citep{maizapellaniz2013,sota2014} that were not targeted with
    CHIRON.  The V662 Car system (see Section
    \ref{subsubsec:ostars-sc}) is highlighted.  Gray triangles are
    additional O-type stars confirmed by \citet{alexander2016}.  The
    three WNH stars are marked with blue stars, and $\eta$ Car is
    indicated by an orange asterisk.  The large dashed circles mark
    the approximate locations of the major clusters in the region.}
  \label{fig:ostars-carina}
\end{figure*}

The Carina Nebula is one of the most dramatic star-forming regions in
the nearby Galaxy.  It is home to more than 70 O-type stars
\citep{smith2006a,gagne2011,alexander2016}, including the defining
star of the O2 spectral type \citep{walborn2002a}, as well as three
late-type hydrogen-rich Wolf-Rayet stars \citep[WNH
  stars;][]{smithconti2008} and the remarkable luminous blue variable
$\eta$ Carinae \citep{davidsonhumphreys1997}.  The distribution of
these O-type and evolved massive stars is shown in Figure
\ref{fig:ostars-carina}.  Roughly half belong to two central clusters,
Trumpler (Tr) 14 and Tr 16, while the rest are spread across an area
more than 30 pc in diameter.  Most of the more distributed stellar
population is found in the South Pillars, a region of active star
formation to the south of Tr 16 \citep{smith2000,smith2010b}.  The
combined ionizing radiation and stellar winds from the clustered and
distributed massive stars \citep{smithbrooks2007} has had a major
impact on their surroundings, shaping spectacular dust pillars
\citep{smith2000,smith2010b}, powering a developing superbubble
\citep{smith2000}, and potentially triggering the observed ongoing
star formation
\citep{megeath1996,smith2000,rathborne2004,smith2005c,smith2010b}.
The feedback-dominated environment of the Carina Nebula is the closest
analogue to giant starburst regions like 30 Doradus
\citep[e.g.,][]{doran2013}.

%----------------------------------------------------------------------
%----------------------------------------------------------------------

Due to the physical extent and complex structure of the Carina Nebula,
the relationships between its various clusters and subclusters have
been the subject of considerable debate.  Photometric and
spectroscopic surveys of the stellar populations of Tr 14 and Tr 16
have usually placed the two clusters at a common distance in the range
2.0 to 3.5 kpc
\citep{feinstein1973,walborn1982b,walborn1982a,the1980d,the1980b,turnermoffat1980,tapia1988,cudworth1993,masseyjohnson1993,tovmassian1994,tapia2003,hur2012}.
Other studies concluded that Tr 14 lies 1--2 kpc behind Tr 16
\citep{walborn1973,morrell1988} or vice versa \citep{carraro2004}.
The smaller Tr 15 cluster likewise alternated between being considered
a foreground \citep{thevleeming1971} or background \citep{walborn1973}
cluster, until recent X-ray data revealed a stellar bridge between Tr
15 and Tr 14 \citep{feigelson2011}.  The open cluster Collinder (Cr)
228 and the distributed massive stars of the South Pillars region have
been variously treated as an extension of Tr 16
\citep{walborn1995,smithbrooks2008}, a distinct but same-distance
cluster \citep[][and see also \citealt{smith2010b} and
  \citealt{feigelson2011}]{herbst1976,turnermoffat1980,tapia1988,tovmassian1994,massey2001},
or a foreground cluster
\citep{feinstein1976,forte1978,carraropatat2001}.  The optical studies
were complicated by the variable extinction across the region and the
unusually high ratio of total-to-selective extinction $R_V$
\citep{herbst1976,forte1978,the1980d,the1980b,smith1987,tovmassian1994,carraro2004,mohrsmith2017}.

Independent of the many spectrophotometric distance studies of the
region, measurements of the expansion of $\eta$ Car's Homunculus
Nebula have placed that star at a firm distance of 2.3 kpc
\citep{allenhillier1993,davidson2001,smith2006a}.  As there is strong
evidence that $\eta$ Car is a member of Tr 16
\citep{walbornliller1977,allen1979} and gas pillars across the Carina
Nebula complex show the influence of Tr 16's feedback
\citep{smith2000,smith2010b}, a general consensus has arisen that Tr
14, Tr 16, Cr 228, and the rest of the stars in the Carina Nebula
belong to a single massive association at 2.3 kpc (e.g.,
\citealt{smithbrooks2008}, although see \citealt{hur2012}).
Preliminary results from \emph{Gaia} Data Release 1
\citep[DR1;][]{gaia2016a,gaiabrown2016,lindegren2016}\btxt{, which
  includes 43 of the Carina Nebula's O-type stars,} support this
interpretation.  \btxt{The inferred parallax-based distances, as computed by
  \citet{astraatmadjabailerjones2016} using a prior based on the
  distribution of stars in the Milky Way, are distributed unimodally
  around $\sim2$~kpc \citep{smithstassun2017}.}

Even with the spatial link between the Carina Nebula's components now
relatively secure, the formation of its distributed massive population
remains somewhat uncertain.  The O-type stars currently seen among the
South Pillars may have been born there, a possible example of an OB
association forming through distributed, hierarchical star formation
\citep[e.g.,][]{efremovelmegreen1998,clark2005}.  But no massive
protostars have been detected among the forming stellar population in
the South Pillars \citep{gaczkowski2013}, and \citet{preibisch2011d}
argue that the dense gas clouds in the region are not massive enough
to support further massive-star formation.  If the distributed O-type
stars did not form in situ, they may have migrated out from Tr 16
either through the classic expansion of a cluster after gas dispersal
\citep{tutukov1978,hills1980,ladalada1991,ladalada2003} or as the
result of cluster-cluster interaction \citep{gieles2013} between Tr 14
and Tr 16.  \citet{kiminki2017} did not see any such outward migration
in the local proper motions of bow-shock-associated massive stars in
the South Pillars, but their sample size was limited.

In this paper, we explore the relationships between the components of
the Carina Nebula and the origins of its massive-star populations
through a survey of the radial velocities (RVs) of its O-type and
evolved massive stars.  These stellar RVs can be compared to the
kinematics of the Carina Nebula's \ion{H}{ii} regions
\citep[e.g.,][]{damiani2016} and molecular gas
\citep[e.g.,][]{rebolledo2016}, allowing a direct assessment of the
impact of massive-star feedback on the \ctxt{interstellar} medium.  The RVs of
massive stars can be strongly affected by stellar binarity
\citep[e.g.,][]{gieles2010}, but the effects can be constrained with
multi-epoch observations.

The paper is organized as follows: In Section \ref{sec:ostars-obs}, we
describe our multi-epoch spectroscopic observing campaign and our RV
measurements, and summarize the additional data compiled from the
literature.  In Section \ref{sec:ostars-binaries}, we discuss the
O-type stars with variable RVs and present orbital solutions for four
spectroscopic binary systems.  Section \ref{sec:ostars-res} discusses
the RV distributions of the various O-star populations in the Carina
Nebula and compares the observed stellar kinematics to the motions of
the region's molecular and ionized gas.  Our conclusions are
summarized in Section \ref{sec:ostars-conc}.  \btxt{Projected
  distances between sources are given using the $\eta$ Car distance of
  2.3 kpc.}

%%%%%%%%%%%%%%%%%%%%%%%%%%%%%%%%%%%%%%%%%%%%%%%%%%%%%%%%%%%%%%%%%%%%%%%
\section{Observations and Data Analysis}
\label{sec:ostars-obs}

\subsection{Target selection}
\label{subsec:ostars-targets}

Our list of O-type stars in the Carina Nebula is drawn from version
3.1 of the Galactic O-Star Catalog
\citep[GOSC;][]{maizapellaniz2013,sota2014}.  The GOSC v3.1 lists 68
objects, including the Of/WNH system WR 25, with right ascensions
between 10:40:00 and 10:49:00 and declinations between $-60$:20:00 and
$-59$:10:00.  An additional four O-type systems in this coordinate
range, originally suspected on the basis of their X-ray emission
\citep{povich2011}, were recently spectroscopically confirmed by
\citet{alexander2016}.  Adding $\eta$ Car and the remaining two WNH
stars brings the total number of known systems with O-type and evolved
massive primaries to 75.  These 75 systems are shown in Figure
\ref{fig:ostars-carina}.

We selected 31 of these systems for new spectroscopic observations.
In choosing targets, we prioritized stars that had zero or few prior
RV measurements, or whose only existing RV data had high
uncertainties.  Most of our target stars are thus outside of Tr 14 and
Tr 16, as those clusters have been the targets of multiple
spectroscopic campaigns and binary fraction analyses
\citep{penny1993,garcia1998,albacetecolombo2001,albacetecolombo2002,morrell2001,rauw2001,rauw2009,naze2005,niemela2006}.
We also prioritized sources brighter than $V=11$ mag.

\subsection{Spectroscopy}
\label{subsec:ostars-spec}

We obtained high-resolution spectra of our 31 target stars with the
CHIRON echelle spectrograph \citep{tokovinin2013} on the CTIO 1.5-m
telescope operated by the SMARTS Consortium.  Observations were taken
in queue operation using the fiber mode configuration, which provides
a resolution of R $\sim$25,000 over a wavelength range of 4100--8900
\AA.  All 31 stars were observed 2--4 times each in Nov--Dec 2014.
Observations of a given star were spaced 7--14 days apart to minimize
the chances of catching a short-period binary at the same phase, as
there is a relative lack of massive binaries in this period range
\citep{kiminkikobulnicky2012,kobulnicky2014}.  Follow-up observations
of eight stars showing possible RV variations were obtained in Oct
2015--Jan 2016.  Throughout our observing campaign, exposure times
were 60--1800 s, designed to achieve signal-to-noise (S/N) ratios of
50--70 at 5500 \AA.  The resulting S/N ratios ranged from 30 to 150
with a median of 77.  ThAr calibration spectra were taken before
moving the telescope after observing each star; bias and flat-field
observations were taken at the beginning and end of each night as part
of CHIRON's standard queue observing protocol.

\begin{figure*}
  \centering \includegraphics[width=\textwidth, trim=2mm 3mm 2mm 2mm,
    clip]{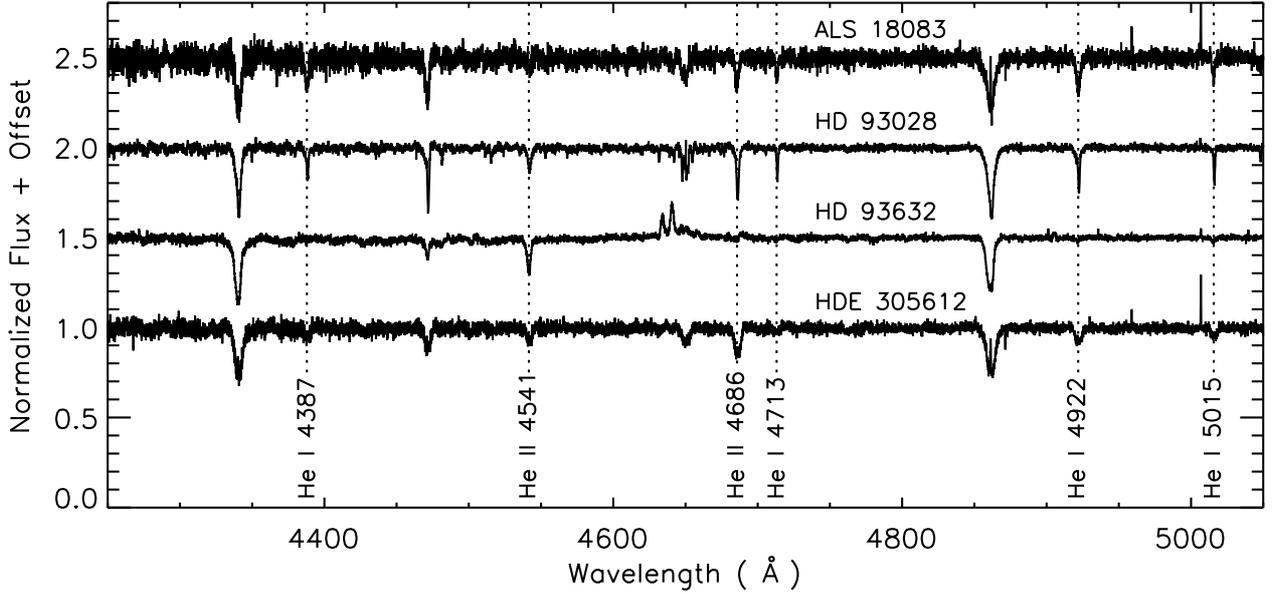}
  \caption{Continuum-normalized CHIRON spectra of four O-type stars in
    our observed sample, illustrating the range of S/N achieved in the
    wavelength range 4250--5050 \AA. Stellar radial velocities were
    measured using fits to the marked \ion{He}{i} and \ion{He}{ii}
    absorption lines \btxt{and} \ion{He}{i} $\lambda\lambda$5876,
    7065, where present \btxt{without emission components}.}
  \label{fig:ostars-spectra}
\end{figure*}

Data were reduced in {\tt IRAF}\footnote{{\tt IRAF} is distributed by the National
  Optical Astronomy Observatory, which is operated by the Association
  of Universities for Research in Astronomy (AURA) under a cooperative
  agreement with the National Science Foundation.} using standard
procedures, including bias subtraction, flat-fielding, cosmic ray
correction, and extraction of 74 echelle orders.  Wavelength
calibration was performed using the corresponding ThAr spectrum for
each star, and wavelengths were corrected to a heliocentric frame.
The wavelength scale of each order is good to an rms of 0.01
\AA~($\sim$0.6 km s$^{-1}$ at 5000 \AA).  Echelle orders were
continuum-normalized before being combined into a single spectrum for
each exposure.  A representative sample of reduced spectra is shown in
Figure \ref{fig:ostars-spectra}, focusing on a wavelength region that
covers most of the lines used for RV measurement (see Section
\ref{subsec:ostars-rv}).

%----------------------------------------------------------------------
%----------------------------------------------------------------------

\subsection{New radial velocities}
\label{subsec:ostars-rv} 

We adapt the method of \citet{sana2013} for measuring stellar RVs,
performing Gaussian fits to a set of He lines and fitting all lines
and epochs for a given star simultaneously.  The final fit forces all
lines at a given epoch to have the same RV, and assumes that the width
and amplitude of a given spectral line are constant across epochs.
The uncertainties on initial, single-line fits are used to weight each
line in the final overall fit.  We use the IDL curve-fitting package
{\tt MPFIT} \citep{markwardt2009}.

Like \citet{sana2013}, \btxt{we fit to the He absorption lines for two
  reasons.  First, these lines} are present across all O subtypes
(unlike most metal lines), allowing us to apply a consistent approach
to our full sample of observed stars.  \btxt{Second, the He lines are
  relatively less sensitive to wind effects compared to the hydrogen
  Balmer series (\citealt{bohannangarmany1978}; see also discussion in
  \citealt{sana2013}).}  Initially, we fit \ion{He}{i}
$\lambda\lambda$4387, 4471, 4713, 4922, 5015, 5876, 6678, 7065 and
\ion{He}{ii} $\lambda\lambda$4541, 4686.  The rest wavelengths adopted
for these lines \btxt{are given in Table \ref{tab:ostars-rest}}.
After the first round of fits, we compared the RVs from individual
line fits to the RVs from fitting all lines at once.  The triplet
blend \ion{He}{i} $\lambda$4471 was systematically blueshifted by
$\sim$10 km s$^{-1}$ relative to the overall results, and the singlet
line \ion{He}{i} $\lambda$6678 was systematically redshifted by a
similar amount.  We therefore removed those two lines and refit all
epochs of the observed stars.  \btxt{We did not see a systematic
  offset in the RVs of \ion{He}{ii}~$\lambda4686$, likely because the
  majority of our targets are mid- to late-type main-sequence O stars,
  in which this line is not strongly affected by winds
  \citep{sana2013}.  Wind emission in \ion{He}{ii}~$\lambda4686$ is
  apparent in the spectra of the supergiant HD 93632 (see Figure
  \ref{fig:ostars-spectra}), and we excluded this line in that star's
  final fit.  On a star-by-star basis, we also excluded other lines
  that were affected by emission from stellar winds or from the
  surrounding \ion{H}{ii} region, as well as lines that were not
  present above the noise level of the continuum.}  With these
exclusions, an average of seven absorption lines were used to fit the
RVs of each star. The measured RVs for each star at each epoch are
given in Table \ref{tab:ostars-rvdata}.  The median uncertainty in RV
for the single-lined spectra is 1.2 km s$^{-1}$.

%----------------------------------------------------------------------
\begin{table}
  \caption{\btxt{Rest wavelengths used to compute radial velocities.}}
  \label{tab:ostars-rest}
  \begin{tabular}{lll}
    \hline
    Line & Wavelength (\AA) & Reference$^{\textrm{a}}$ \\
    \hline
    \ion{He}{i}  $\lambda4387$ & 4387.9296 & NIST \\
    \ion{He}{i}  $\lambda4471^{\textrm{b}}$  & 4471.4802 & NIST \\
    \ion{He}{ii} $\lambda4541$ & 4541.591  & PvH  \\
    \ion{He}{ii} $\lambda4686$ & 4685.71   & PvH  \\
    \ion{He}{i}  $\lambda4713$ & 4713.1457 & NIST \\
    \ion{He}{i}  $\lambda4922$ & 4921.9313 & NIST \\
    \ion{He}{i}  $\lambda5015$ & 5015.6783 & NIST \\
    \ion{He}{i}  $\lambda5876$ & 5875.621  & NIST \\
    \ion{He}{i}  $\lambda6678^{\textrm{b}}$ & 6678.151  & NIST \\
    \ion{He}{i}  $\lambda7065$ & 7065.190  & NIST \\
    \hline
  \end{tabular}
  \begin{tabular}{l}
    $^{\textrm{a}}$NIST = NIST Atomic Spectra Database \citep{kramida2016}; \\
    ~PvH = Peter van Hoof's Atomic Line List, v2.04 \\
    ~(\url{http://www.pa.uky.edu/~peter/atomic/}). \\
    $^{\textrm{b}}$Excluded from final fits. \\
  \end{tabular}
\end{table}
%----------------------------------------------------------------------

%----------------------------------------------------------------------
\begin{table}
  \caption{Heliocentric radial velocities measured from CHIRON
    observations of O-type stars in the Carina Nebula.  \btxt{Dates are
    given for the midpoints of the exposures.}  This table is available
    in its entirety as Supplementary Material to the online version of
    this article.  A portion is shown here for guidance regarding its
    form and content.}
  \label{tab:ostars-rvdata}
  \begin{tabular}{l@{~~~~~~}r@{~~~~~}r@{~~~~}r@{~~~~}r@{~~~~}r}
    \hline 
    Name & HJD & RV1 & $\sigma(\textrm{RV1})$ & RV2 & $\sigma(\textrm{RV2})$ \\
     & $-2400000$ & (km s$^{-1}$) & (km s$^{-1}$) & (km s$^{-1}$) & (km s$^{-1}$) \\
    \hline 
    ALS 15204     & 56987.814 &  -60.0 &   4.7 & ... & ... \\
    ALS 15204     & 56994.850 &    1.0 &   2.3 & ... & ... \\
    \vspace{3pt}
    ALS 15204     & 57003.836 &   43.1 &   2.0 & ... & ... \\
    ALS 15206     & 56992.858 &   24.5 &   0.4 & ... & ... \\
    ALS 15206     & 57000.851 &   29.8 &   0.5 & ... & ... \\
    ALS 15206     & 57008.788 &   34.1 &   0.3 & ... & ... \\
    \hline
  \end{tabular}
\end{table}
%----------------------------------------------------------------------

One of our observed sources, HD 92607, is a double-lined spectroscopic
binary (SB2), first identified by \citet{sexton2015}.  We adapted our
RV measurement procedure to fit a double Gaussian at all epochs.  This
process required several steps: First, we fit \ion{He}{i}
$\lambda$5876 for the six epochs in which the components were clearly
separated.  Then, we fixed the width and amplitude of the \ion{He}{i}
$\lambda$5876 components and fit to the four epochs with blended
components.  These fits gave us initial estimates for the RVs at all
epochs, which we used as the starting point for fits to \ion{He}{i}
$\lambda\lambda$4922, 5015 and \ion{He}{ii} $\lambda$4686.  Again, we
fit these lines in the well-separated epochs first, then fixed widths and
amplitudes for fits to the blended epochs.  The final fit was to all
four lines at all epochs.  Because the two components of HD 92607 are
of similar spectral type, several iterations of line fitting and
orbital analysis (see Section \ref{subsec:ostars-92607}) were required to
confidently identify which line component was associated with which
star at a given epoch.  The median uncertainty on the RVs of the
components of HD 92607 is 7.1 km s$^{-1}$, and ranges as high as
30--40 km s$^{-1}$ in epochs where the lines are highly blended.

As a check on our wavelength calibration, we obtained CHIRON spectra
of the RV standard stars HIP 51722 (spectral type F7--8 V) and HIP
53719 (K1--2 IV), both from the catalog of \citet{soubiran2013}, in
Nov 2014.  As the spectra of these standard stars do not have
\ion{He}{i} lines, their RVs could not be measured by the same method
used for our science spectra.  Instead, we measured their RVs by
cross-correlation with high-resolution stellar spectra from
\citet{coelho2014}, using the {\tt IRAF} task {\tt fxcor}.  The measured RVs
of both standard stars agree with their expected values within 0.4 km
s$^{-1}$.

\subsection{Radial velocities from the literature}
\label{subsec:ostars-litdata}

Where available, we also compiled heliocentric RVs from the literature
for the Carina Nebula's O-type and evolved massive stars.  The adopted
RVs and associated uncertainties for stars not observed with CHIRON
are given in Table \ref{tab:ostars-lit} along with the corresponding
references.  \btxt{Nearly all of these RVs were measured from various
  subsets of He absorption lines, often in conjunction with metal
  and/or hydrogen Balmer lines.  The exceptions are: $\eta$ Car, where
  the systemic radial velocity was derived from H$_2$ emission from
  the Homunculus \citep{smith2004}; the three WNH stars, discussed
  further below; and the \citet{conti1979} RVs for the Of supergiant
  HD 93129A, where we use RVs from the ``Group 1'' narrow metal
  lines.}  Where \btxt{a} system is a spectroscopic binary with a
\btxt{published} orbital solution, we give the systemic
velocity; \btxt{for other sources with observations at multiple
  epochs,} we give the weighted mean \btxt{of the available
  velocities.}  For visual binaries that are resolvable at the
few-arcsecond level (e.g., HD 93161AB), we use only data that specify
which component was observed.  We also note whether systems were
classified as constant-RV or binary in the multiplicity survey of
\citet{chini2012}.  Additional notes on the spectroscopic multiplicity
of a system are given in the references column.  We also incorporate
literature data for the following CHIRON targets: ALS 15206
\citep{huanggies2006}, ALS 15207 \citep{garcia1998}, HD 93028
\citep{feast1957,conti1977}, HDE 303308 \citep{conti1977}, HDE 305518
\citep{huanggies2006}, and HDE 305619 \citep{humphreys1973}.  In no
case does the inclusion of these sources' literature RVs shift their
overall weighted mean RV by more than 1.5 km s$^{-1}$.

%----------------------------------------------------------------------
\begin{table*}
  \caption{Mean or systemic heliocentric radial velocities adopted from the literature.}
  \label{tab:ostars-lit}
  \begin{tabular}{lrrcl}
    \hline
    Name & $\overline{\textrm{RV}}$ or $\gamma$ & $\sigma({\overline{\textrm{RV}}})$ & Spectral flag from                & References \\
         & (km s$^{-1}$)                        & (km s$^{-1}$)                      & \citet{chini2012}$^{\mathrm{a}}$  &            \\
    \hline
    ALS 15229		  &   14.7 &   3.5 & ... & \citet{garcia1998}			   \\
    CPD-58 2611   	  &    4.7 &   5.0 &   C$^{\mathrm{b}}$ & \citet{penny1993,garcia1998}    	   \\
    CPD-58 2620	  	  &   -0.6 &   5.8 & SB1 & \citet{penny1993,garcia1998}    	   \\
    CPD-59 2591		  &   -2.9 &  10.\phantom{0} & ... & \citet{huanggies2006}		   \\
    CPD-59 2624		  &   13.5 &  10.\phantom{0} & ... & \citet{huanggies2006}; SB2 in \citet{alexander2016}	     \\
    CPD-59 2626		  &  -22.7 &  10.\phantom{0} & ... & \citet{huanggies2006}		   \\
    CPD-59 2627		  &  -15.9 &  10.\phantom{0} & ... & \citet{huanggies2006}		   \\
    CPD-59 2629		  &   -8.8 &  12.8 & ... & SB1 in \citet{williams2011}		   \\
    CPD-59 2635		  &    0.\phantom{0} &  1.\phantom{0} &  SB2 & SB2 solution from \citet{albacetecolombo2001}         \\
    CPD-59 2636		  &    4.0 &  5.7 &  SB2 & Quadruple system; \citet{albacetecolombo2002}	 \\
    CPD-59 2641		  &   -4.5 &  1.8 &  SB2 & SB2 solution from \citet{rauw2009}  			   \\
    $\eta$ Carinae		  &   -8.1 &  1.\phantom{0} & ... & \citet{smith2004}        	   \\
    HD 93128	  	  &    5.0 &   5.3 &   C$^{\mathrm{b}}$ & \citet{penny1993,garcia1998}    	   \\
    HD 93129A	  	  &  -10.1 &  18.7 &   C$^{\mathrm{b}}$ & \citet{conti1979,penny1993}		   \\
    HD 93129B	  	  &    7.1 &   2.2 & ... & \citet{penny1993}		   	   \\
    HD 93160	  	  &  -12.9 &  17.9 & SB1 & \citet{thackeray1973,conti1977}	   \\
    HD 93161A	  	  &    1.1 &   3.6 & SB2 & SB2 data from \citet{naze2005}          \\
    HD 93161B	  	  &  -25.4 &  28.4 & ... & \citet{naze2005}            		   \\
    HD 93130	  	  &   59.9 &   3.1 & SB2 & \citet{conti1977}		   	   \\
    HD 93204		  &    8.5 &   3.8 &   C & \citet{conti1977}			   \\
    HD 93205		  &   -2.9 &   0.9 & SB2 & SB2 solution from \citet{morrell2001}   \\
    HD 93250		  &   -5.2 &   7.6 &   C$^{\mathrm{b}}$ & \citet{thackeray1973,rauw2009,williams2011}           \\
    HD 93343		  &   -2.8 &  54.4 & SB2 & SB2 data from \citet{rauw2009}			   \\
    HD 93403		  &  -14.2 &   5.4 & SB2 & SB2 solution from \citet{rauw2000}	   \\
    HD 93843		  &   -9.9 &   0.4 &   C & \citet{feast1957,conti1977}		   \\
    QZ Car		  &   -8.7 &  15.5 & SB2 & Quadruple system; \citet{morrisonconti1980,mayer2001} \\
    Trumpler 14-9         &    5.6 &   3.8 & ... & \citet{penny1993,garcia1998}	   	   \\
    V572 Car              &   -3.4 &   3.7 & SB2 & Triple system; \citet{rauw2001}	   \\
    V573 Car		  &   -5.\phantom{0} &   4.\phantom{0} & ... & SB2 solution from \citet{solivellaniemela1999}	 \\
    V662 Car		  &  -15.\phantom{0} &   2.\phantom{0} & ... & SB2 solution from \citet{niemela2006}   \\
    WR 22$^{\mathrm{c}}$  &  -27.5 &  ...  & ... & SB1 solution from \citet{schweickhardt1999}	         \\
    WR 24$^{\mathrm{c}}$  &  -34.0 &  13.3 & ... & \citet{conti1979} 		           \\
    WR 25$^{\mathrm{c}}$  &  -34.6 &   0.5 & ... & SB1 solution from \citet{gamen2006}	   \\
    \hline
  \end{tabular}
  \begin{tabular}{l}
     $^{\mathrm{a}}$C = constant; SB1 = single-lined spectroscopic binary; SB2 = double-lined spectroscopic binary. \\ 
     $^{\mathrm{b}}$Although classified as constant in \citet{chini2012}, these stars meet our criteria for significant RV variability ($P(\chi^2,\nu) < 0.01$; see Section \ref{sec:ostars-binaries}) \\
     and have RV amplitudes $>20$ km s$^{-1}$. \\
     $^{\mathrm{c}}$\btxt{Reported RVs are from \ion{N}{iv} $\lambda4058$ emission; see discussion in Section \ref{subsec:ostars-litdata}.}\\
  \end{tabular}
\end{table*}

%----------------------------------------------------------------------

\btxt{As mentioned above, the three WNH stars are a special case.}
Like all Wolf-Rayet stars, these sources have strong stellar winds,
\btxt{with} their spectral features formed at various depths within
those winds \citep[e.g.,][]{crowther2007}.  \btxt{Consequently, the
  observed RVs vary greatly depending on which lines are measured.
  Relative to the stars' systemic velocities, the hydrogen Balmer and
  \ion{He}{ii} absorption lines of Wolf-Rayet spectra tend to be
  highly blueshifted
  \citep{moffat1978,moffatseggewiss1978,conti1979,masseyconti1981,rauw1996},
  while their \ion{He}{ii} emission lines tend to be highly redshifted
  \citep{moffat1978,moffatseggewiss1978,massey1980,niemelamoffat1982,schweickhardt1999,schnurr2008a}.}
\btxt{In WN and WNH spectra, certain narrow \ion{N}{iii}, \ion{N}{iv},
  and \ion{Si}{iv} emission lines are thought to be relatively better
  tracers of the true stellar RVs
  \citep[e.g.,][]{moffat1978,moffatseggewiss1978,moffatseggewiss1979,conti1979}.}
\btxt{For the Carina Nebula's WNH stars, we adopt RVs measured from
  \ion{N}{iv}~$\lambda4058$ emission.}

\btxt{However, it is clear that the adopted RVs for the WNH stars
  are still affected by winds.  Using the spectral features of the
  O-type companion to WR 22, \citet{rauw1996} calculated the true
  systemic velocity of the binary to be $21.3\pm7.0$~km~s$^{-1}$,
  while \citet{schweickhardt1999} report $\approx0\pm15$~km~s$^{-1}$
  for the same source.  The \ion{N}{iv}~$\lambda4058$ line in the
  spectrum of WR 22 is thus blueshifted by $\sim25$--$50$~km~s$^{-1}$.  In
  other WN+O binaries, this line shows relative blueshifts of
  0--80~km~s$^{-1}$
  \citep{niemela1980,niemelamoffat1982,collado2013,collado2015,munoz2017}.}
\btxt{Because the exact magnitude of the shift is uncertain and
  seems to vary by star, we do not attempt to correct the
  \ion{N}{iv}~$\lambda4058$ RVs for wind effects, and} we consider the
three WNH stars as a separate group in our analysis.  Note that
although $\eta$ Car is also an evolved massive star with a high
mass-loss rate, its RV was measured by a different method
\citep{smith2004} and it does not suffer from the same bias.

In compiling literature RVs for Table \ref{tab:ostars-lit}, we
\btxt{found} that the RVs in
\citet{levato1990,levato1991a,levato1991b} are notably discrepant with
later observations, including our own.  These three studies measured
RVs \btxt{from an unspecified subset of hydrogen Balmer and He
  absorption lines} for a total of 92 candidate members of Cr 228, Tr
16, and Tr 14, using the spectrograph on the 1-m Yale CTIO telescope
\citep[see][]{levato1986}.  \btxt{They computed average cluster RVs of
  $-23$, $-26$, and $-29$~km~s$^{-1}$ for Cr 228, Tr 16, and Tr 14,
  respectively.  By contrast, the surveys of \citet{penny1993} and
  \citet{garcia1998} found $+2.8$~km~s$^{-1}$ and $+6.0$~km~s$^{-1}$,
  respectively, for the average RV of Tr 14.  \citet{penny1993} also
  reported very different results from \citet{levato1991b}, in both RV
  values and variability, for individual sources in Tr 14.  Across the
  Carina Nebula, we find that the
  \citet{levato1990,levato1991a,levato1991b} data are typically 10--60
  km s$^{-1}$ more negative than other literature values for the same
  stars (see Table \ref{tab:ostars-lit}) and our CHIRON observations.
  We therefore exclude} all data from the Levato et al. studies,
except \btxt{for specific binary orbits as discussed} in Section
\ref{sec:ostars-binaries}.

With the inclusion of the literature data in Table \ref{tab:ostars-lit}, we
have RVs for 63 of the 75 O-type and evolved massive stars in the
Carina Nebula.  Four of the twelve stars without RV data are those
that were newly confirmed by \citet{alexander2016}.  These four are
relatively highly extincted, with a median $A_V$ of 5.7 mag compared
to the median $A_V$ of 2.2 mag for the rest of the O-type stars in the
region \citep{gagne2011,povich2011b}.  \citet{alexander2016} postulate
that one of these, OBc 3, is a coincidentally aligned background star.
The other eight stars without RV data are neither systematically more
extincted nor systematically fainter at visual wavelengths than the 63
sources with RV data.  Some of these eight have been observed as part
of the OWN RV survey (\citealt{barba2010}; see
\citealt{sota2014,maizapellaniz2016}) but currently lack published
RVs.

%%%%%%%%%%%%%%%%%%%%%%%%%%%%%%%%%%%%%%%%%%%%%%%%%%%%%%%%%%%%%%%%%%%%%%%
\section{New and Updated Binary Orbits}
\label{sec:ostars-binaries}

%----------------------------------------------------------------------
\begin{table*}
  \caption{Orbital and physical parameters of new and updated binary solutions.}
  \label{tab:ostars-orbits}
  \begin{tabular}{lcccc}
    \hline
    Element & HD 92607 & HD 93576 & HDE 303312 & HDE 305536 \\
    \hline
    $P$ (d)                  & 3.6993 (0.0001)   & 1.852102 (0.000002) & 9.4111 (fixed)            & 1.88535 (0.00002) \\
    $e$                      & 0.00 (fixed)      & 0.075 (0.009)     & 0.58 (fixed)              & 0.129 (0.008)     \\
    $\gamma$ (km s$^{-1}$)    &  8.8 (1.6)        & -8.4 (0.6)        & 2.3 (0.6)                 &  2.3 (0.2)        \\
    $\omega$ (\degr)         & 90 (fixed)        & 201 (9)          & 192 (2)                   & 56 (3)            \\
    $T_p$ (HJD$-$2400000)    & 56965.910 (0.009) & 57005.94 (0.05)   & 56980.58 (0.03)           & 56975.34 (0.02) \\
    $K_1$ (km s$^{-1}$)       & 224 (3)           & 85.0 (0.8)        & 71 (1)                    & 37.3 (0.3)        \\
    $K_2$ (km s$^{-1}$)       & 242 (3)           & ...               & ...                       & ...               \\
    $f(M_1,M_2)$ ($\textrm{M}_{\sun}$)  & ...               & 0.117 (0.004)     & 0.19 (0.01)               & 0.0099 (0.0004)   \\
    $M_1$sin$^3i$ ($\textrm{M}_{\sun}$) & 20.2 (0.6)        & ...               & ...                       & ...               \\
    $M_2$sin$^3i$ ($\textrm{M}_{\sun}$) & 18.7 (0.6)        & ...               & ...                       & ...               \\
    $a_1$sin$i$ ($\textrm{R}_{\sun}$)   & 16.4 (0.2)        & 3.10 (0.03)       & 10.7 (0.2)                & 1.38 (0.01)       \\
    $a_2$sin$i$ ($\textrm{R}_{\sun}$)   & 17.7 (0.2)        & ...               & ...                       & ...               \\
    rms$_1$ (km s$^{-1}$)     & 19.7              & 10.9               & 5.6                       & 2.8               \\
    rms$_2$ (km s$^{-1}$)     & 23.7              & ...               & ...                       & ...               \\
    \hline
  \end{tabular}
  \begin{tabular}{l}
    \btxt{One-sigma} uncertainties for each quantity are listed in parentheses. \\
  \end{tabular}
\end{table*}

%----------------------------------------------------------------------

Multiplicity is ubiquitous among O-type stars
\citep{garmany1980,mason2009,chini2012,sana2012,sana2013,sana2014,kobulnicky2014,moedistefano2016}
and must be considered when determining their systemic RVs.  For each
of our CHIRON sources, we compute $P(\chi^2,\nu)$, the probability
that the $\chi^2$ about the weighted mean would be exceeded by random
chance given $\nu = N_{\textrm{obs}}-1$ degrees of freedom.  Nine of the 31
observed stars display significant ($P(\chi^2,\nu) < 0.01$) RV
variations with amplitudes $>20$ km s$^{-1}$ in our CHIRON data.  As
discussed below, we are able to find periods and fit orbital solutions
to four of these spectroscopic binaries.  Two additional observed
sources meet the criteria for significant, high-amplitude variations
when their literature RVs are included.  One of these
sources is HD 93403, a known SB2 with a full orbital solution in
\citet{rauw2000}.  Our CHIRON observations are in good agreement with
the orbit of the primary star, and we adopt the systemic velocity for
HD 93403 from the \citet{rauw2000} solution.

An additional five CHIRON targets display significant ($P(\chi^2,\nu)
< 0.01$) but low-amplitude ($\Delta$RV $<20$ km s$^{-1}$) RV
variations.  These stars include HDE 305619, which was flagged as an
SB1 by \citet{chini2012} but lacks published RV data.  The other four
stars may have undetected long-period companions or may be showing
photospheric variability
\citep[e.g.,][]{garmany1980,giesbolton1986,fullerton1996,ritchie2009,martins2015}.

\btxt{We search for orbital solutions for the} stars with significant,
high-amplitude RV variations and $N_{\textrm{obs}} \ge 7$ \btxt{using
  the IDL package {\tt rvfit} \citep{iglesiasmarzoa2015}.}
\btxt{Spectroscopic binary orbits are defined by the following
  parameters:} $P$ (orbital period), $e$ (eccentricity), $\gamma$
(systemic velocity), $\omega$ (argument of periastron), $T_p$ (epoch
of periastron), $K_1$ (primary semi-amplitude), and, where applicable,
$K_2$ (secondary semi-amplitude).  \btxt{{\tt rvfit} uses an adaptive
  simulated annealing method to find the global $\chi^2$ minimum in
  the six- or seven-dimensional parameter space.  Unlike some orbital
  fitting codes, {\tt rvfit} does not use an initial estimate of the
  period.  }
In four cases, described in further detail below, \btxt{{\tt rvfit}
  converged on orbital solutions with well-fit phase-folded velocity
  curves.  Parameter uncertainties were computed} using {\tt rvfit}'s
Markov Chain Monte Carlo \btxt{(MCMC) algorithm
  \citep{iglesiasmarzoa2015} with $10^5$ points.  The MCMC provides
  the probability distribution of each parameter marginalized over all
  other parameters.  For our data, these marginalized histograms are
  well approximated by Gaussian distributions and we take their
  standard deviations as the one-sigma uncertainties on the orbital
  parameters.}

\btxt{During our initial analysis of the variable-RV sources, we
  computed} the frequency power spectrum of each source using
A. W. Fullerton's IDL implementation of the one-dimensional {\tt
  CLEAN} algorithm \citep{roberts1987}.  \btxt{We confirm the periods
  returned by {\tt rvfit} by checking against these periodograms,
  which are presented in Appendix \ref{sec:ostars-appendix}.}

\subsection{HD 92607}
\label{subsec:ostars-92607}

HD 92607 was first noted as an SB2 by \citet{sexton2015}, who
classified its components as O8.5 V + O9 V.  It is on the far western
side of the Carina Nebula, approximately 25 pc from Tr 14 and 16 (see
Figure \ref{fig:ostars-carina}).  The lower-mass stellar population in
this area, by the edge of the Carina Nebula's northern molecular
cloud, is relatively young \citep[$<1$ Myr;][]{kumar2014}.  HD 92607
is associated with an extended arc of 24 $\mu$m emission, suggestive
of a stellar wind bow shock, that points to the southeast
\citep{sexton2015}.

We obtained 12 CHIRON spectra of HD 92607 in 2014--2015.  As described
in Section \ref{subsec:ostars-rv}, we adapted our RV-fitting procedure
to fit two Gaussian profiles to \ion{He}{i} $\lambda\lambda$4922,
5015, 5876 and \ion{He}{ii} $\lambda$4686 across all 12 epochs, using
fits to the well-separated epochs to fix the widths and amplitudes of
the lines.  We used {\tt rvfit} to fit an orbit to the primary and
secondary RVs simultaneously.  \ctxt{We fit a common apparent systemic
  velocity to both components.}\footnote{\ctxt{We used an F-test to
    compare our single-$\gamma$ solution to the best-fitting orbit
    computed with different $\gamma$ values for each component.  The
    latter did not provide a significant improvement in fit
    ($p=0.27$).}}

\begin{figure}
  \centering
  \includegraphics[width=\columnwidth, trim=3mm 3mm 1mm 2mm, clip]{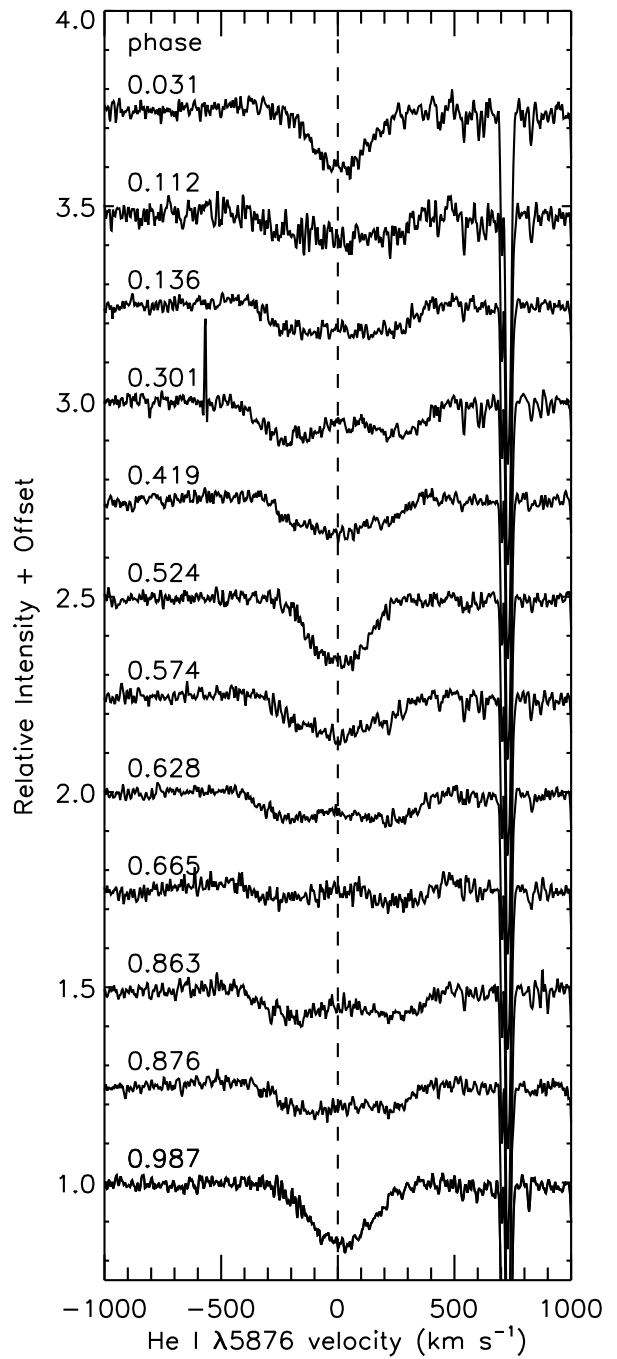}
  \caption{\ion{He}{i} $\lambda$5876 in velocity space and in order of
    phase for CHIRON observations of the SB2 HD 92607, for an orbital
    period of 3.6993 d.  The narrow absorption line on the right is
    interstellar \ion{Na}{I} $\lambda$5890.}
  \label{fig:ostars-ph92607}
\end{figure}

\begin{figure}
  \centering
  \includegraphics[width=\columnwidth]{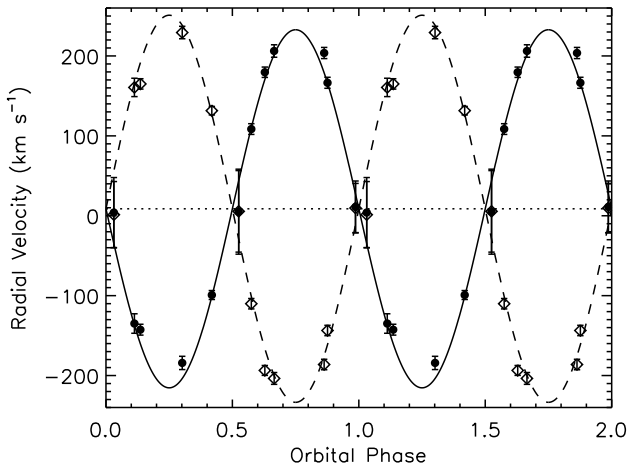}
  \caption{Radial velocity curve and orbital solution for the SB2 HD
    92607 at the best-fitting orbital period of 3.6993 d.  Filled circles
    and open diamonds correspond to the primary and secondary
    components, respectively.  The dotted line marks the systemic
    velocity.}
  \label{fig:ostars-rv92607}
\end{figure}

The best-fitting period for HD 92607 is 3.6993 $\pm$ 0.0001 d.  Figure
\ref{fig:ostars-ph92607} shows the progression of \ion{He}{i}
$\lambda$5876 over this period.  After our initial orbital fit
converged on an eccentricity of zero, we fixed $e=0$ for further
fitting, allowing {\tt rvfit} to treat $\omega$ and $T_p$ (which are
ill-defined at very low eccentricities) in a consistent manner
\citep[see][]{iglesiasmarzoa2015}.  The final best-fitting orbital
solution is shown in Figure \ref{fig:ostars-rv92607}, and the orbital
parameters are listed in Table \ref{tab:ostars-orbits}.  In the
best-fitting orbit, the minimum masses of the primary and secondary
stars are $M_1$sin$^3i = 20.2 \pm 0.6$ $\textrm{M}_{\sun}$ and $M_2$sin$^3i =
18.7 \pm 0.6$ $\textrm{M}_{\sun}$, respectively, giving a mass ratio of $q =
0.93 \pm 0.04$.  But their spectral types of O8.5 V and O9 V
\citep{sexton2015}, according to the ``observational'' scale of
\citet{martins2005}, have somewhat lower expected masses of $M_1=18.8$
$\textrm{M}_{\sun}$ and $M_2=17.1$ $\textrm{M}_{\sun}$.  The apparent discrepancy could
be the result of the normal uncertainties on spectral-type
classification; for instance, \citet{gagne2011} list HD 92607 as an
O8 V star, although they do not account for its spectroscopic
binarity.  Or perhaps the components of HD 92607 are slightly older,
and hence cooler for their masses, than average main-sequence
stars of their types.  In any case, the HD 92607 system is likely
close to edge-on ($i\sim90$\degr), as a smaller inclination would
imply implausibly large masses for its component stars.

%----------------------------------------------------------------------
%----------------------------------------------------------------------

%----------------------------------------------------------------------
%----------------------------------------------------------------------

\subsection{HD 93576}
\label{subsec:ostars-93576}

HD 93576 is one of five O-type systems in the open cluster Bochum (Bo) 11 in
Carina's South Pillars region.  Its primary is of type O9--9.5 IV--V
\citep{sota2014,sexton2015}.  It is associated with an extended arc of
8 $\mu$m emission, which points away from Bo 11 and has infrared
colors indicative of a stellar wind bow shock \citep{sexton2015}.  Its
local proper motion confirms that it is moving
away from Bo 11 at $\lesssim15$ km s$^{-1}$; it may have been recently
ejected from the cluster \citep{kiminki2017}.

\begin{figure}
  \centering
  \includegraphics[width=\columnwidth]{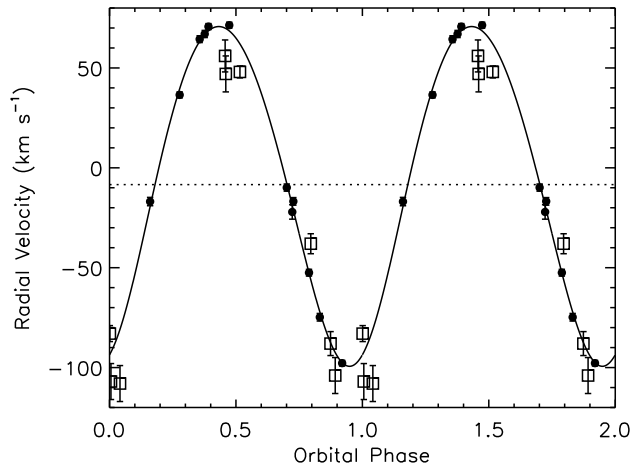}
  \caption{Radial velocity curve and orbital solution for the SB1 HD
    93576 at the best-fitting orbital period of 1.852102 d.  Solid
    circles are measurements from our CHIRON spectra \btxt{and} open
    squares are data from \citet{levato1990}. \btxt{Both data sets
      were used to compute the orbital solution.}  The dotted line
    marks the systemic velocity.}
  \label{fig:ostars-rv93576}
\end{figure}

\citet{levato1990} first noted periodic RV variations in the spectra
of HD 93576, and fit an orbital solution with a period of 2.020 d.  We
obtained 12 CHIRON observations of this system in 2014--2016.
\btxt{We used {\tt rvfit} to compute orbital solutions both with and
  without the \citet{levato1990} data.  With our CHIRON data alone, we
  find a period of 1.85201~d; including the \citet{levato1990} data
  constrains the period to $1.852102\pm0.000002$~d.  The final
  best-fitting parameters given in Table \ref{tab:ostars-orbits} and
  used to plot the orbital solution in Figure \ref{fig:ostars-rv93576}
  are from the combined data set.  As can be seen in Figure
  \ref{fig:ostars-rv93576}, many but not all of the \citet{levato1990}
  RVs for this system are 10--20 km~s$^{-1}$ blueshifted relative to
  CHIRON data at the same phase.  \citet{levato1990} reported a
  systemic velocity of $-21$~km~s$^{-1}$ for this system, while our
  fit to the combined data set finds a systemic velocity of
  $-8.4$~km~s$^{-1}$ and fitting to only our CHIRON data gives a
  systemic velocity of $-8.2$~km~s$^{-1}$.}

%----------------------------------------------------------------------
%----------------------------------------------------------------------

\btxt{The orbit of HD 93576} is minimally but significantly
non-circular, with an eccentricity of 0.075 $\pm$ 0.009.  The
resulting mass function is $f(M_1,M_2) = 0.117 \pm 0.004~\textrm{M}_{\sun}$.
Assuming the primary star has a mass of 16--18 $\textrm{M}_{\sun}$, as
appropriate for its spectral type per \citet{martins2005}, the minimum
mass of the secondary star is $\sim$3.7 $\textrm{M}_{\sun}$, roughly the mass
of a B8 V star \citep{drillinglandolt2000}.

\subsection{HDE 303312}
\label{subsec:ostars-303312}

HDE 303312 is an O9.7 IV star \citep{sota2014} located on the
outskirts of Tr 14 (see Figure \ref{fig:ostars-carina}).  There are no
RVs for this source in the literature, but \btxt{it was identified as
  an eclipsing system by \citet{otero2006} based on its $V$-band light
  curve from the All Sky Automated Survey
  \citep[ASAS;][]{pojmanski1997}.  \citet{otero2006} report a period
  of 9.4109~d and note} that it is an ``extremely eccentric'' system.
We obtained a total of 12 CHIRON observations of HDE 303312 over
2014--2015, and downloaded \btxt{the available ASAS $V$-band
  magnitudes for 2000--2009}.  In addition to using {\tt rvfit} on the
velocity data, we use the eclipsing binary modeling \btxt{programs
  {\tt PHOEBE} \citep{prsazwitter2005}} and {\tt
  NIGHTFALL}\footnote{\url{http://www.hs.uni-hamburg.de/DE/Ins/Per/Wichmann/Nightfall}}
to synthesize and compare photometric light curves.  \btxt{We find
  that the ASAS light curve folds most cleanly at a period of 9.4111~d
  rather than 9.4109~d.  We therefore adopt 9.4111~d as the
  photometric period with an estimated uncertainty of $\pm0.0002$~d.}

\begin{figure}
  \centering
  \includegraphics[width=\columnwidth]{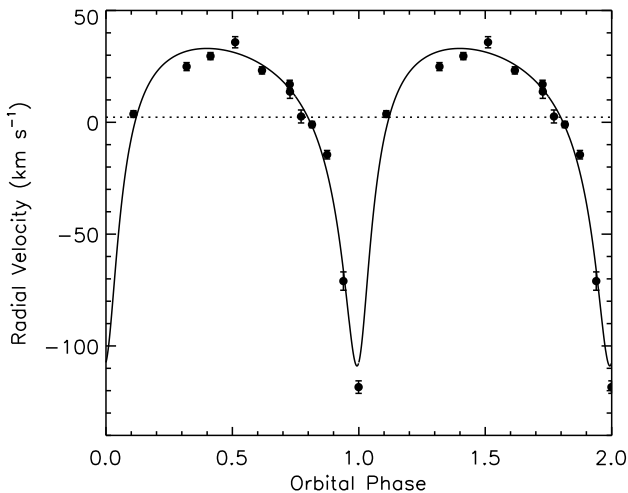}
  \caption{Radial velocity curve and orbital \btxt{solution} for the
    eclipsing SB1 HDE 303312.  The solid line shows the best-fitting
    orbital solution to the RV data with the period and eccentricity
    fixed at 9.4111 d and \btxt{0.58}, respectively\btxt{.  The dotted line
      marks the systemic velocity.}}
  \label{fig:ostars-rv303312}
\end{figure}

\btxt{Initial, unconstrained runs of {\tt rvfit} find an RV period
  between 9.41 and 9.65~d, with an eccentricity of 0.6--0.7.  These
  results confirm that the eclipsing and spectroscopic binaries are
  the same system, and we subsequently fix the period at 9.4111~d.
  With all other orbital parameters left free, the best-fitting
  solution to the RV data has $e=0.67$ and $\omega=188\degr$.}
\btxt{Matching the phase separation of the $V$-band eclipses for
  $e=0.67$ requires $\omega\approx140\degr$ or $220\degr$,
  incompatible with the shape and symmetry of the RV curve.}  \ctxt{We
  do not see any shifts in eclipse timing over the nine years of ASAS
  data that would suggest apsidal motion as an explanation for this
  discrepancy.}  \btxt{We \ctxt{therefore} iterate between the the RV
  curve and light curve to converge on values of $e$, $\omega$, and
  inclination $i$ that are consistent with the observed data.  Our
  final best fit to the RV curve, presented in Table
  \ref{tab:ostars-orbits} and plotted in Figure
  \ref{fig:ostars-rv303312}, holds the eccentricity fixed at $=0.58$
  and gives $\omega=192\pm2\degr$.  The inclination of the system is
  $80$--$85\degr$.}
  
%----------------------------------------------------------------------
%----------------------------------------------------------------------

\btxt{This best-fitting orbital solution for HDE 303312 gives a mass
  function of $f(M_1,M_2)=0.19\pm0.01~\textrm{M}_{\sun}$.}
If we take the mass of the primary star to be 15--18~M$_{\sun}$, based
on its spectral type \citep[see][]{martins2005}, the mass of the
secondary star is 4--5 $\textrm{M}_{\sun}$.  Assuming the secondary star is on
the main sequence, that mass corresponds to a spectral type of B6--7 V
\citep{drillinglandolt2000}\btxt{, which would have an effective
  temperature of $T_{\textrm{eff}}\approx13,000$--$14,000$~K
  \citep{kenyonhartmann1995}.}  \btxt{Unfortunately, the noise in the
  ASAS data prevents us from putting strong constraints on the
  relative luminosities or stellar radii of the components.  High-S/N,
  multi-wavelength photometry is needed to confirm the characteristics
  of the secondary star.}

\subsection{HDE 305536}
\label{subsec:ostars-305536}

The O9.5 V star \citep{sota2014} HDE 305536 is located close to the
optical center of the open cluster Cr 228 \citep[e.g.,][]{wu2009}, a
few pc from the WNH star WR 24.  Like HD 93576, it is associated
with an extended arc of 8 $\mu$m emission indicative of a stellar wind
bow shock, which in this case points in the general direction of Tr 16
\citep{sexton2015}.  HDE 305536 was observed multiple times by
\citet{levato1990}, who detected significant RV variations and fit a
rough orbit with a period of 2.018 d.  

\begin{figure}
  \centering
  \includegraphics[width=\columnwidth]{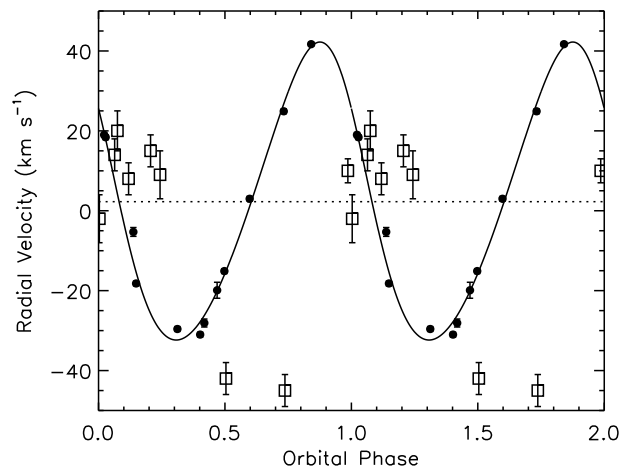}
  \caption{Radial velocity curve and orbital solution for the SB1 HDE
    305536 at the best-fitting orbital period of 1.88535 d.  Solid
    circles are measurements from our CHIRON spectra.  Open squares
    are data from \citet{levato1990}, shown for comparison \btxt{but
      not used to fit the orbital solution}.  The dotted line marks the
    systemic velocity.}
  \label{fig:ostars-rv305536}
\end{figure}

We obtained 12 CHIRON spectra of this system over 2014--2016\btxt{.
  The parameters of the best-fitting orbital solution to our data,
  which has a period of 1.88535 $\pm$ 0.00002 d, are given in Table
  \ref{tab:ostars-orbits} and the corresponding phase-folded RV curve
  is shown in Figure \ref{fig:ostars-rv305536}.  {\tt rvfit} was
  unable to converge on a solution when the RVs from
  \citet{levato1990} were included; we overplot those data, folded at
  our best-fitting CHIRON-based period, for comparison only.
  Relative to our CHIRON measurements, the \citet{levato1990} data}
appear offset by 10--30 km s$^{-1}$ in RV and/or 0.2--0.3 in phase.
As discussed in Section \ref{subsec:ostars-litdata}, many
\citet{levato1990} sources show unexplained RV offsets relative to
later observations. In this case, the apparent phase shift \btxt{may
  also suggest the possibility of orbital precession.}

%----------------------------------------------------------------------
%----------------------------------------------------------------------

\btxt{The orbit of HDE 305536} is slightly eccentric with
$e = 0.129 \pm 0.008$.  The mass function of the system is
$f(M_1,M_2)\approx0.01$ $\textrm{M}_{\sun}$.  Assuming the mass of the O9.5 V primary star
is $\sim16$ $\textrm{M}_{\sun}$ \citep{martins2005}, the minimum mass of the
secondary star is $\sim1.4$ $\textrm{M}_{\sun}$, equivalent to a mid F-type dwarf
\citep{drillinglandolt2000}.

\subsection{Additional spectroscopic binaries}
\label{subsec:ostars-nonorbits}

In addition to HD 93403 and the four spectroscopic binaries described above, six
other stars showed significant ($P(\chi^2,\nu)<0.01$) RV variations
with amplitudes $>20$ km s$^{-1}$.  These variations are larger than
typical photospheric variability \citep[e.g.,][]{martins2015}, making these
stars probable but unconfirmed spectroscopic binaries.  We
briefly discuss each of these sources below.

\emph{ALS 15204}: There are no literature RV data for this O7.5 V star
\citep{sota2014} in Tr 14.  We observed it three times with CHIRON in
2014.  In spectra taken 16 days apart, its RV changed by more than
$100$ km s$^{-1}$.  Owing to this star's relative faintness
\citep[$V=10.9$ mag;][]{hur2012}, we did not pursue follow-up
observations in 2015.

\emph{ALS 15206}: We observed ALS 15206, an O9.2 V star
\citep{sota2014} in Tr 14, a total of three times with CHIRON in Dec
2014.  Our observations show variability that is significant but
low-amplitude ($\Delta$RV$\sim10$ km s$^{-1}$).  However, its single-epoch
RV from \citet{huanggies2006} is $\sim30$ km s$^{-1}$ blueshifted
relative to our CHIRON numbers, suggesting a higher degree of
variability.  ALS 15206 is associated with an extended arc of 8 $\mu$m
emission, likely a stellar wind bow shock, that points northwest
toward the center of Tr 14 \citep{sexton2015}.  It has a local proper
motion of $\lesssim30$ km s$^{-1}$, directed to the northeast
\citep{kiminki2017}.  As with ALS 15204, we did not pursue follow-up
CHIRON observations because this star is relatively faint at visual
magnitudes \citep[$V=10.7$ mag;][]{hur2012}

\emph{ALS 15207}: This O9 V star \citep{sota2014} in Tr 14 was flagged
by \citet{levato1991b} as an SB2; however, \citet{garcia1998} observed
no line doubling nor any significant RV variations.  We obtained a
total of seven CHIRON spectra of ALS 15207 in 2014--2015; like
\citet{garcia1998}, we found that its \ion{He}{i} and \ion{He}{ii}
absorption lines were well fit with single Gaussian profiles.  We
measured RVs ranging from $-21.5$ km s$^{-1}$ to $+15.0$ km s$^{-1}$ but
were unable to constrain the period of the variability.

\emph{CPD-59 2554}: Although this O9.5 IV star \citep{sota2014} in Cr
228 was observed by \citet{levato1990} to be a constant-RV source, our
nine CHIRON spectra from 2014--2015 show significant variability, with
RVs ranging from $-22.7$ km s$^{-1}$ to $+63.4$ km s$^{-1}$.  We were
unable to constrain the period of the variability.

\emph{HD 93028}: This O9 IV star \citep{sota2014} is located about 6
pc southwest of the center of Cr 228.  \citet{levato1990} identified
it as a spectroscopic binary and fit a rough orbit with a period of
51.554 d.  We obtained four CHIRON spectra of this source in 2014,
which showed significant RV variations at an amplitude just above our
20 km s$^{-1}$ cutoff; a larger amplitude is seen when data from
\citet{feast1957} and \citet{conti1977} are included.
\citet{sota2014} report that HD 93028 is in a long-period ($\sim200$
d) spectroscopic binary, but do not provide an orbital solution.

\emph{HDE 305525}: At about 7 pc southwest of Tr 16, this O5.5 V star
\citep{sota2014} is part of the distributed population in the South
Pillars.  \citet{levato1990} detected significant RV variability in
its spectra but were unable to find an orbital solution.  Our eight
CHIRON spectra from 2014--2015 confirm RV variations with an amplitude
of more than 130 km s$^{-1}$.  Like \citet{levato1990}, we were unable
to constrain the period of the variability.

%%%%%%%%%%%%%%%%%%%%%%%%%%%%%%%%%%%%%%%%%%%%%%%%%%%%%%%%%%%%%%%%%%%%%%%
\section{Results and Discussion}
\label{sec:ostars-res}

\subsection{Distribution of stellar radial velocities}
\label{subsec:ostars-distribution}

\begin{figure}
  \centering
  \includegraphics[width=\columnwidth,trim=4mm 3mm 6mm 3mm,clip]{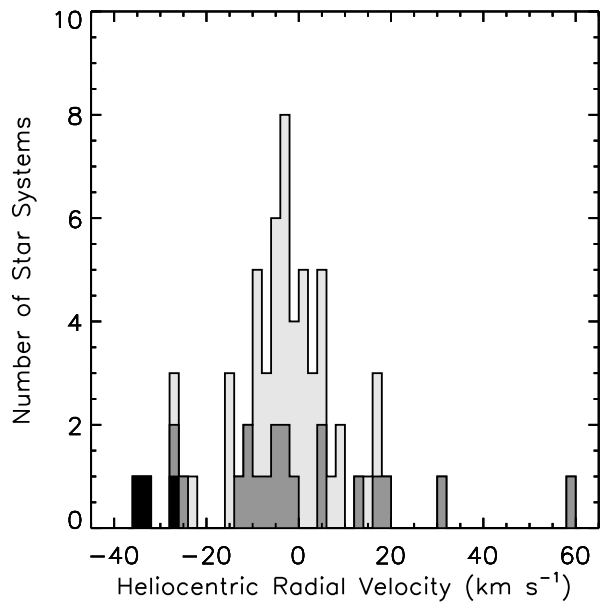}
  \caption{Histogram of the heliocentric radial velocities of the
    O-type and evolved massive stars in the Carina Nebula.
    Well-constrained sources (solved spectroscopic binaries and stars
    with little to no RV variation) are shown in light gray.  Unsolved
    binaries are shown in dark gray.  The three WNH stars, whose
    \btxt{reported} RVs are affected by strong stellar winds, are
    indicated in black.}
  \label{fig:ostars-histogram}
\end{figure}

Combining new spectroscopy with literature data, we have compiled
heliocentric RVs for 59 O-type systems in the Carina Nebula, as well
as for the LBV $\eta$ Car and the three WNH stars in the region.  We
present a histogram of these RVs in Figure \ref{fig:ostars-histogram}.  The
stars are divided into three groups, characterizing the likelihood
that the observed RVs represent their true systemic motions.  The
first group, shown in light gray in Figure \ref{fig:ostars-histogram}, are
sources with relatively reliable RVs.  These are spectroscopic
binaries with well-constrained orbital solutions, for which we plot
the computed systemic velocity, as well as stars with no known RV
variability or insignificant or low-amplitude RV variability, for
which we plot the weighted mean of the observed RVs.  In total, this
group, which we will henceforth refer to as the ``well-constrained sources,'' consists of $\eta$ Car and 40 O-type stars.  The weighted mean
RV of the well-constrained sources is 0.6 km s$^{-1}$, with a standard
deviation of 9.1 km s$^{-1}$.

%----------------------------------------------------------------------
%----------------------------------------------------------------------

The second group of sources, shown in dark gray in Figure
\ref{fig:ostars-histogram}, are those that have been identified as
spectroscopic binaries but lack orbital solutions.  These include
stars that were identified as SB1s or SB2s in the multiplicity survey
of \citet{chini2012}, stars with binarity flagged elsewhere in the
literature, and stars whose CHIRON and/or literature RVs display
significant variability with amplitudes $>20$ km s$^{-1}$.  We refer
to this group of 19 O-type systems as the ``unsolved binaries.''  For
each system, we plot the weighted mean of its measured RVs, but
caution that this number may not be representative of its true
systemic velocity, especially for systems with few published RVs.  For
instance, the most notable outlier in Figure
\ref{fig:ostars-histogram} is HD 93130, for which there is only a
single reliable published RV \citep[59.9~km~s$^{-1}$;][]{conti1977}.
However, \btxt{\citet{chini2012} classify this source as an SB2,
  suggesting} that its outlying \btxt{single-epoch} RV is the result
of its binarity rather than its relationship with the Carina Nebula.
The overall weighted mean RV of the well-constrained sources and the
unsolved binaries together remains 0.6 km s$^{-1}$, while the standard
deviation of the distribution increases to 13.5 km s$^{-1}$.

The third group of sources, shown in black in Figure
\ref{fig:ostars-histogram}, are the WNH stars.  As discussed in
Section \ref{subsec:ostars-litdata}, \btxt{the
  \ion{N}{iv}~$\lambda4058$ RVs adopted for these sources are affected
  by their strong stellar winds.  All three fall on the negative end
  of the Carina Nebula's RV distribution, offset by roughly
  $-30$~km~s$^{-1}$ from the mean velocity of the O-type systems.  We
  cannot rule out the possibility that the WNH stars are a distinct
  kinematic group or that one or more of them is moving at outlying
  speeds after having been ejected from the region's clusters.}
\btxt{However, given that the three stars show a similar RV offset,
  wind effects are most likely the primary cause.}

The overall RV distribution in Figure \ref{fig:ostars-histogram} is unimodal,
consistent with the various clusters and subclusters of the Carina
Nebula being part of a single complex at a common distance.  
We do not see any probable runaway stars.  Aside from WR 24 and
WR 25, only HD 93130 has an RV that deviates from the region's mean by
$\ge30$ km s$^{-1}$, and as described above, HD 93130 is a
poorly-studied spectroscopic binary and its given RV is unlikely to be
its true systemic velocity.  However, with RV data, we cannot rule out
the presence of runaways with high tangential velocities.  Our region
of study extends $\sim20$ pc in all directions from Tr 16, meaning a
star with a tangential velocity of 30 km s$^{-1}$ would have exited
our field in $\sim600,000$ yr.  An O-type star ejected by the recent
supernova explosion of its companion \citep[see, e.g.,][]{blaauw1961}
might still appear to be within the coordinates of the Carina Nebula,
but we see no evidence for this in the RV data.

The standard deviation of the RVs of the well-constrained sources, 9.1
km s$^{-1}$, is an upper limit on the true one-dimensional velocity
dispersion of the Carina Nebula, as this group of sources may still
contain long-period spectroscopic binaries or other undetected RV
variables.  Typical OB associations like Scorpius-Centaurus and
Perseus OB2 have one-dimensional velocity dispersions of 1--3 km
s$^{-1}$ \citep{debruijne1999,steenbrugge2003}, but these associations
are substantially smaller in mass and extent---and slightly
older---than the Carina Nebula complex
\citep[see][]{bally2008,preibischmamajek2008}.  More directly
comparable to the Carina Nebula is the Cygnus OB2 association, which
contains $>50$ O-type stars distributed across tens of parsecs
\citep{masseythompson1991,wright2015}.  In both tangential and radial
velocities, the one-dimensional velocity dispersion of Cyg OB2 is
$\sim10$ km s$^{-1}$ \citep{kiminki2007,kiminki2008,wright2016},
similar to what we have measured for the O-type stars in the Carina
Nebula.  This is about a factor of two higher than the velocity
dispersions seen in the massive, compact starburst clusters R136
\citep{henaultbrunet2012}, NGC 3603 \citep{rochau2010}, and the Arches
Cluster \citep{clarkson2012}.

\subsection{Variations between clusters}
\label{subsec:ostars-clusters}

\begin{figure}
  \centering \includegraphics[width=\columnwidth,trim=6mm 3mm 4mm
    5mm,clip]{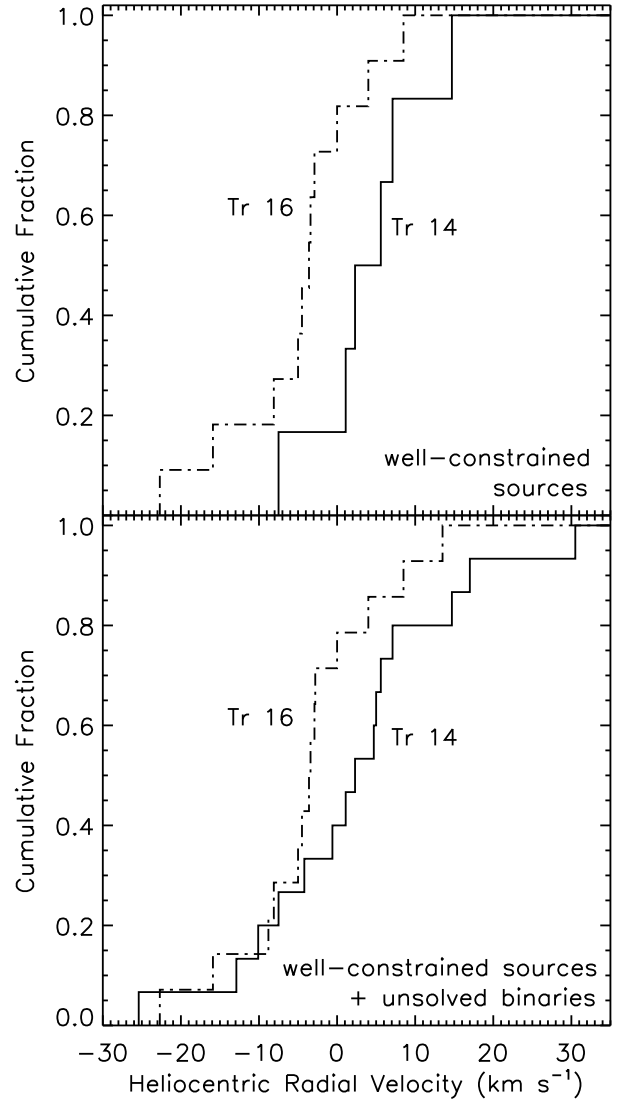}
  \caption{Cumulative histograms of the heliocentric radial velocities
    of O-type stars in Tr 14 (solid line) and O-type stars and $\eta$
    Car in Tr 16 (dot-dash line).  Both clusters are defined with a
    radius of 5\arcmin. Top: stellar systems with well-constrained RVs
    (either solved spectroscopic binaries or sources with little to no
    RV variation).  Bottom: as above, but including the mean RVs of
    unsolved spectroscopic binaries.}
  \label{fig:ostars-trumplers}
\end{figure}

In Figure \ref{fig:ostars-trumplers}, we separate out and compare the RVs of
the O-type and evolved massive stars in Tr 14 and Tr 16.  We define
membership in each cluster as being within a projected 5\arcmin~of the
cluster center.  This is the observed radius of Tr 14's
pre-main-sequence population \citep{ascenso2007} and a natural break
in its O-star distribution.  Tr 14 contains six O-type stars with
well-constrained RVs and an additional nine unsolved binaries.  Tr 16
is larger and less well-defined than Tr 14, and is elongated rather
than spherical \citep[e.g.,][]{feigelson2011}, and so is poorly defined by a single radius.  With a 5\arcmin~cutoff, Tr 16
contains 11 sources with well-constrained RVs (including $\eta$ Car),
and three unsolved binaries.  As we describe below, increasing the
radius of Tr 16 slightly increases its velocity dispersion but has
little effect on the weighted mean of its RV distribution.

%----------------------------------------------------------------------
%----------------------------------------------------------------------

The cumulative histograms in Figure \ref{fig:ostars-trumplers} suggest
that the O-type stars in Tr 14 might have slightly more positive RVs,
on average, than those in Tr 16.  The weighted mean of the
well-constrained sources in Tr 14 is $2.3 \pm 7.4$ km s$^{-1}$, in
agreement with the $2.8 \pm 4.9$ km s$^{-1}$ measured by
\citet{penny1993} and the $6.0 \pm 1.4$ km s$^{-1}$ found by
\citet{garcia1998}.  In comparison, the weighted mean of the
well-constrained sources within a 5\arcmin~radius of Tr 16 is $-3.5
\pm 8.6$ km s$^{-1}$.  Increasing the radius of Tr 16 to
10\arcmin~changes this to $-3.3 \pm 10.4$ km s$^{-1}$.  To evaluate
the significance of the apparent RV difference between clusters, we
ran a two-sided Kolmogorov-Smirnov (K-S) test, which evaluates the
probability that the two clusters were drawn from the same RV
distribution.  We estimated the uncertainty on the K-S probability by
drawing 10$^4$ combinations of each cluster's RVs, randomly permuting
each star's RV within a Gaussian distribution with a standard
deviation corresponding to that star's observed RV uncertainty.  For
the well-constrained sources (the top panel of Figure
\ref{fig:ostars-trumplers}), the probability that the two clusters
come from the same RV distribution \btxt{is $4^{+19}_{-3}$\%}, meaning the
difference in RVs is marginally significant.  Including unsolved
binaries (as in the bottom panel of Figure \ref{fig:ostars-trumplers})
brings the probability of a shared RV distribution \btxt{to $16^{+45}_{-13}$\%},
indicating that the difference is not significant.  Further study of
the kinematics of Tr 14 and Tr 16 are needed to determine if any
overall RV difference persists in their lower-mass stellar
populations.  \citet{damiani2017} found a mean RV of $-5$ km s$^{-1}$
for the FGK-type populations of Tr 14 and Tr 16 combined, with no
apparent bimodality in their RV distribution, but they did not attempt
to separate their stars by cluster.

\begin{figure*}
  \centering
  \includegraphics[width=\linewidth, trim=0 0 0 5mm, clip]{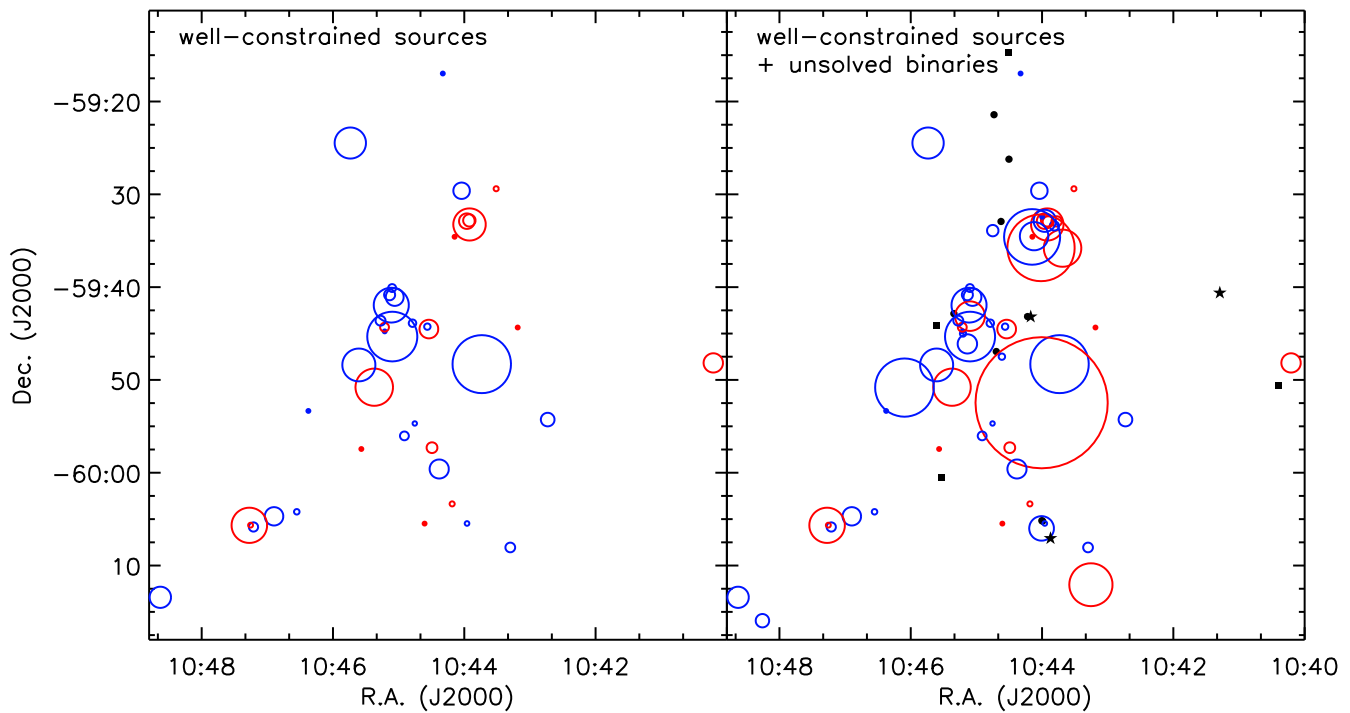}
  \caption{Maps of the heliocentric radial velocities of O-type and
    evolved massive stars in the Carina Nebula.  Left: stars
    (including $\eta$ Car) with well-constrained RVs as defined in the
    text.  Blue circles indicate negative RVs and red circles indicate
    positive RVs; circle size scales with the magnitude of the
    velocity.  Right: as left, but including known spectroscopic
    binaries that lack orbital solutions.  The filled black circles
    and squares show the positions of O-type stars from v3.1 of the
    GOSC and from \citet{alexander2016}, respectively, that lack RV
    data.  The positions of the three WNH stars are shown by black
    filled stars.}
  \label{fig:ostars-circles}
\end{figure*}

In Figure \ref{fig:ostars-circles}, we search for any additional
cluster-scale or spatial dependencies by mapping our RVs as a function
of position.  The circle representing each star is scaled by the
magnitude of its mean/systemic RV and color-coded as redshifted
(positive RV) or blueshifted (negative RV).  The left-hand plot
includes only those sources with well-constrained RVs, while the
right-hand plot shows both the well-constrained sources and the
unsolved binaries.  For clarity, we mark only the positions and not
the RVs of the three WNH stars.

%----------------------------------------------------------------------
%----------------------------------------------------------------------

The apparent velocity difference between Tr 14 and Tr 16 is visible in
Figure \ref{fig:ostars-circles}, in that the former has more red
(positive-RV) sources than the latter.  Another notable structure is
the grouping of O-type stars extending from the nominal center of Cr
228 up toward Tr 16.  These sources have low-magnitude RVs and appear
as a sequence of relatively small circles.  Their weighted mean RV
(0.3 km s$^{-1}$) is similar to that of the region as a whole, but
their velocity dispersion (4.0 km s$^{-1}$) is comparatively small.
This relatively low velocity dispersion suggests that this string of
massive stars formed in place; we would expect a population that
had migrated out from Tr 16 to have a higher velocity dispersion than
the cluster rather than vice versa.

\subsection{Comparison to molecular gas}
\label{subsec:ostars-gas}

The molecular cloud complex associated with the Carina Nebula is
composed of three main parts
\citep{degraauw1981,brooks1998,yonekura2005,rebolledo2016}.  The
Northern Cloud partially surrounds Tr 14 and extends to the northwest
\citep{degraauw1981,brooks2003}, where it connects to additional
molecular material around the Gum 31 \ion{H}{ii} region
\citep[e.g.,][]{rebolledo2016}.  The Southern Cloud coincides with the
optically dark lane between Tr 16 and the South Pillars
\citep{degraauw1981}.  The South Pillars are themselves composed of
molecular gas, shaped by feedback from Carina's OB stars
\citep{rathborne2004,yonekura2005,rebolledo2016}.  Within Tr 16 there
are only scattered molecular globules, thought to be the remnants of
the gas from which that cluster formed
\citep{coxbronfman1995,brooks2000}.

\begin{figure*}
  \centering \includegraphics[width=0.75\textwidth,trim=1mm 1mm 2mm
    2mm,clip]{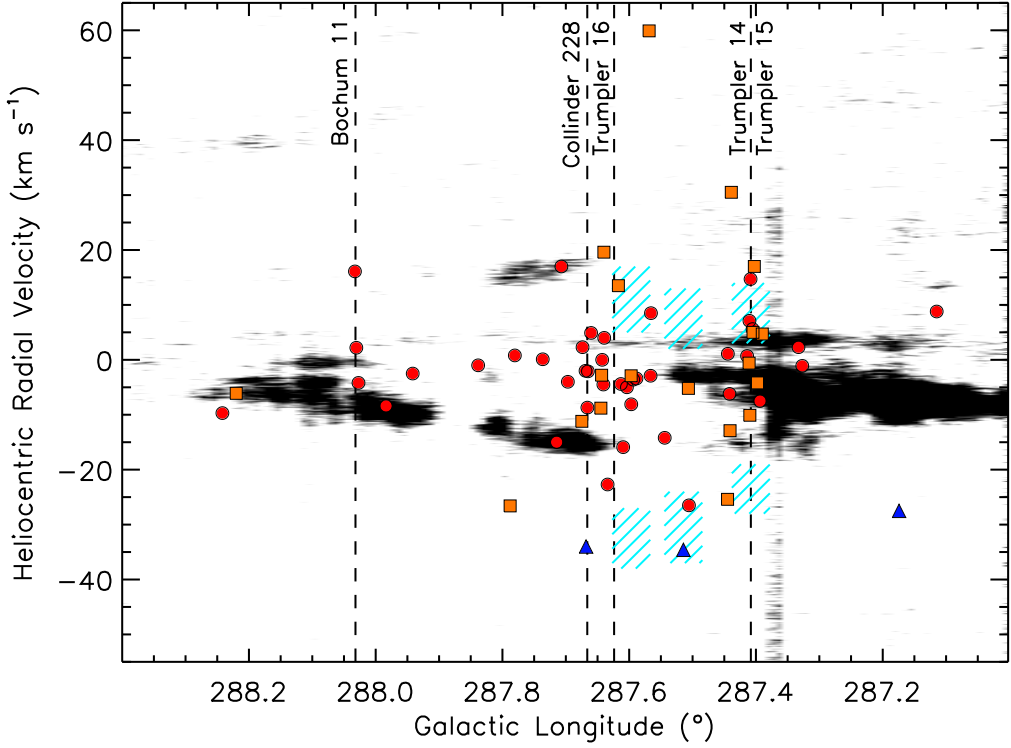}
  \caption{Position-velocity diagram of \textsuperscript{12}CO (1--0)
    emission (gray-scale) from \citet[][converted to a heliocentric
      frame]{rebolledo2016} compared to O-type and massive evolved
    stars.  Red circles are systems (including $\eta$ Car) with
    well-constrained radial velocities, orange squares are known or
    suspected spectroscopic binaries that lack orbital solutions, and
    blue triangles are WNH stars.  The hatched cyan regions show the
    RVs of the approaching and receding components of emission from
    ionized gas \citep{damiani2016} centered on the positions of
    $\eta$ Car, WR 25, and Tr 14.}
  \label{fig:ostars-coglon}
\end{figure*}

\begin{figure}
  \centering \includegraphics[width=\columnwidth,trim=2mm 3mm 2mm
    2mm,clip]{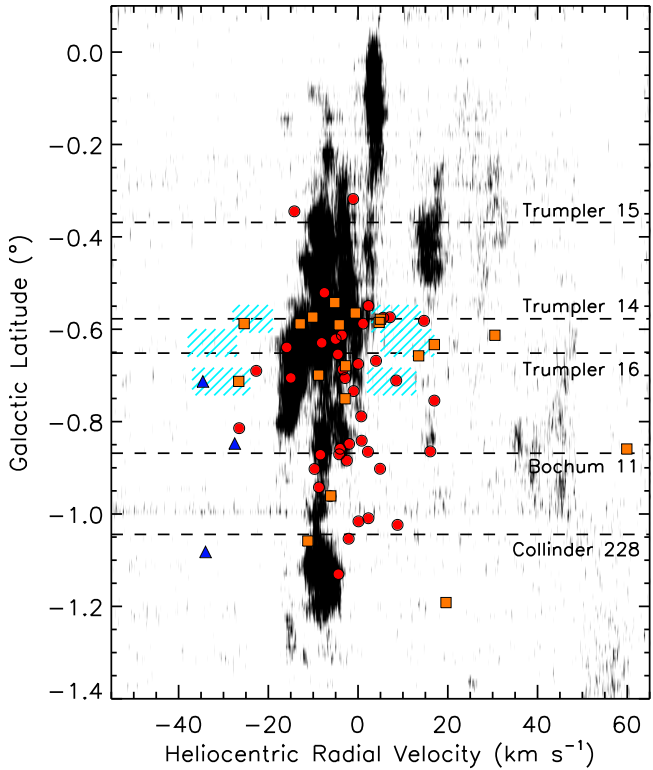}
  \caption{Similar to Figure \ref{fig:ostars-coglon}, but showing the
    heliocentric radial velocities of \textsuperscript{12}CO, ionized
    gas, and stars as a function of Galatic latitude.}
  \label{fig:ostars-coglat}
\end{figure}

\citet{rebolledo2016} observed the Carina Nebula in $^{12}$CO (1--0)
emission as part of the Mopra Southern Galactic Plane CO Survey
\citep{burton2013}.  We use their data to compare the kinematics of
Carina's molecular gas to the RVs of its O-type and evolved massive
stars.  For Figures \ref{fig:ostars-coglon} and
\ref{fig:ostars-coglat}, we collapse their three-dimensional data
cubes along Galactic latitude and Galactic longitude, respectively, to
present two-dimensional position--velocity diagrams of the dense gas
in Carina.  We convert the gas velocities from the Local Standard of
Rest to a heliocentric frame.  At the coordinates of the Carina
Nebula, $RV_{\textrm{helio}}\approx RV_{\textrm{LSR}} + 11.6$ km
s$^{-1}$.  All RVs in the following discussion are heliocentric unless
otherwise noted.

Figure \ref{fig:ostars-coglon} plots the RVs of the Carina Nebula's
O-type stars, evolved massive stars, and $^{12}$CO (1--0) emission as
a function of Galactic longitude $l$.  The $^{12}$CO fluxes from
\citet{rebolledo2016} are summed across Galactic latitudes $-1.4$ $\le
b\le$ 0.1\degr.  As in Figure \ref{fig:ostars-histogram}, the stars
are divided into three groups: stars (including $\eta$ Car) with
well-constrained systemic velocities, unsolved binaries, and WNH
stars.  We also indicate the approximate heliocentric velocities of
the redshifted and blueshifted components of ionized gas emission
lines observed by \citet{damiani2016} for shells centered on the
locations of $\eta$ Car, WR 25, and Tr 14.  In Figure
\ref{fig:ostars-coglat}, we present the same data, but plotted as a
function of Galactic latitude $b$.  The $^{12}$CO fluxes are summed
across Galactic longitudes 287.0 $\le l\le$ 288.4\degr.

%----------------------------------------------------------------------
%----------------------------------------------------------------------

%----------------------------------------------------------------------
%----------------------------------------------------------------------

The three components of Carina's molecular cloud complex separate
cleanly when plotted against Galactic latitude in Figure
\ref{fig:ostars-coglon}: the Northern Cloud at $l \approx$
287.0--287.5\degr~with heliocentric RV $\approx$ ($-10$)--0 km s$^{-1}$;
the Southern Cloud at $l \approx$ 287.6--287.8\degr~with heliocentric
RV $\approx$ ($-15$)--($-10$) km s$^{-1}$; and the South Pillars at $l
\approx$ 287.9--288.2\degr~with heliocentric RV $\approx$ ($-10$)--($-5$)
km s$^{-1}$ and a smaller part at $\sim0$ km s$^{-1}$.  The separation
is less notable in Figure \ref{fig:ostars-coglat}, as the sourthern part of
the Northern Cloud overlaps with the Southern Cloud in Galactic
longitude, but the RV offset in the Southern Cloud is still apparent.
The molecular gas at $l \approx$ 287.7--287.8\degr~, $b\approx-0.4$,
with heliocentric RV $\sim+20$ km s$^{-1}$, is thought to be
associated with the far side of the Carina arm, at a greater distance
than the Carina Nebula \citep{damiani2016}.

Figures \ref{fig:ostars-coglon} and \ref{fig:ostars-coglat} further
emphasize that the RVs of the O-type stars in the Carina Nebula are
not spatially dependent, as there is no trend in O-star RVs with
Galactic longitude or latitude.  \citet{rebolledo2016} applied the
four-arm Milky Way model of \citet{vallee2014} to the rotation curve
of \citet{mccluregriffithsdickey2007} and calculated the expected RV
for objects at various distances along the tangent of the Carina
spiral arm.  For sources on the near side of the arm, at $\sim2$ kpc,
the expected Local Standard of Rest RV is approximately $-10$ km
s$^{-1}$, which corresponds to a heliocentric RV of $\approx+2$ km
s$^{-1}$, very close to the observed mean RV of the O-type stars.
Objects at greater distances from the Sun would be expected to have
more positive RVs \citep[see][]{rebolledo2016}.  Our RV results thus
favor the OB clusters and groups in the Carina Nebula being at a
common distance of approximately 2 kpc.

\subsubsection{Tr 14, Tr 15, and the Northern Cloud}
\label{subsubsec:ostars-nc}

There is a general consensus that the relatively compact Tr 14 is the
youngest of the Carina Nebula's Trumpler clusters, at just 1--2 Myr
old
\citep{walborn1973,walborn1982b,walborn1982a,walborn1995,morrell1988,vazquez1996,smithbrooks2008,rochau2011}.
The case for its youth is strengthened by its close spatial
association with the Northern Cloud, which wraps around the west side
of the cluster \citep{degraauw1981,brooks2003,tapia2003}.  Bright
radio emission arises from multiple ionization fronts where radiation
from Tr 14 impacts dense clumps in the Northern Cloud
\citep{whiteoak1994,brooks2001}.  It is clear from Figures
\ref{fig:ostars-coglon} and \ref{fig:ostars-coglat} that the O-type
stars in Tr 14 are also kinematically associated with the Northern
Cloud, further strengthening the picture of Tr 14 as a young cluster
that has not yet dispersed its natal molecular gas.

While Tr 15 has sometimes been considered an unrelated foreground or
background cluster \citep{thevleeming1971,walborn1973}, its extended
X-ray stellar population indicates a connection to Tr 14
\citep{feigelson2011,wang2011}.  We have RV data for only two O-type
stars around Tr 15, HD 93403 and HD 93190, which appear at
$b\sim-0.35$\degr~in Figure \ref{fig:ostars-coglat}.  Their RVs are
consistent with those of the rest of the O-type stars in the Carina
Nebula, and span the RV range of the Northern Cloud.  Both are 4--5 pc
outside the core of Tr 15, and neither is associated with any group or
subcluster \citep[see][]{feigelson2011}.  Tr 15 is likely several Myr
older than Carina's other Trumpler clusters
\citep{carraro2002,tapia2003,wang2011}, and it seems likely that these
two stars drifted out from Tr 15 over the course of their lifetimes.

Off the southern edge of the Northern Cloud, on the far west side of
the Carina Nebula (see Figure \ref{fig:ostars-carina}), there are
several O-type and evolved massive stars, notably WR 22 and HD 92607.
There is a young ($<1$ Myr) pre-main-sequence population in this
region, although it does not show any clustering around the massive
stars \citep{kumar2014}.  WR 22 is a particularly puzzling system: it
is a very massive binary \citep[$\sim$55-$\textrm{M}_{\sun}$
  primary;][]{schweickhardt1999} approximately 21\arcmin~(15 pc) from
Tr 14, without any subcluster or group of its own.  Its
\btxt{\ion{N}{iv}~$\lambda4058$} RV is \btxt{somewhat} less negative
than the RVs of WR 24 and WR 25, but it is unclear whether this
represents a true difference in their systemic motions or whether it
is the result of uncharacterized \btxt{differences in their stellar
  winds}.

The SB2 HD 92607 (see Section \ref{subsec:ostars-92607}), evident on
the far right side of Figure \ref{fig:ostars-coglon}, is less extreme
in mass but still somewhat puzzling in origin.  Its systemic RV (8.8
km s$^{-1}$) is relatively positive for an O-type system in the Carina
Nebula, although still within one standard deviation of the mean.  And
despite its proximity to the Northern Cloud, it lacks a kinematic
association with the molecular gas---perhaps a hint that it did not
form in its currently observed location.  HD 92607 is associated with
a candidate bow shock in the form of a resolved 24 $\mu$m arc pointing
to the southeast \citep{sexton2015}, which might suggest an origin
outside the Carina Nebula.  However, in the active environment of this
giant \ion{H}{ii} region, bow shocks are not clear indicators of
stellar motion \citep{kiminki2017}.  Its spectroscopic parallax
\citep[using data from][]{martins2005,gagne2011,povich2011} \btxt{and
  the inferred distance from its \emph{Gaia} DR1 geometric parallax
  \citep{astraatmadjabailerjones2016}} are both consistent with HD
92607 being part of the Carina Nebula at a distance of around 2 kpc.

\subsubsection{Tr 16 and the Southern Cloud}
\label{subsubsec:ostars-sc}

Unlike the Northern Cloud, which has a close kinematic association
with the Tr 14 cluster, the Southern Cloud is offset by 10--15 km
s$^{-1}$ in RV from its neighbor Tr 16.  The Southern Cloud has the
most negative RVs of the molecular gas in Carina, being 5--10 km
s$^{-1}$ blueshifted relative to both the Northern Cloud and the
molecular gas in the South Pillars.  It coincides with a prominent,
optically dark dust lane, indicating that it lies in front of the
southeastern edge of Tr 16
\citep{dickel1974,degraauw1981,brooks1998}.  Inside Tr
16 itself, there are only small clumps of molecular gas (not visible
in the \citealt{rebolledo2016} data), and these have RVs more
consistent with those of the O-type stars in the cluster
\citep{coxbronfman1995,brooks2000}.

One of the key arguments in favor of Tr 16 being 1--2 Myr older than
Tr 14 \citep[e.g.,][]{degraauw1981,walborn1995} is that the latter (in
addition to being more compact) is still partially enclosed by its
natal molecular cloud, while the former appears to have disrupted the
gas from which it formed.  The distribution of gas RVs in Figure
\ref{fig:ostars-coglon} suggests that the \ctxt{Northern} Cloud, Southern
Cloud, and Southern Pillars were originally part of a single
continuous molecular cloud, with the massive stars in Tr 16 having
since blown out the central part of that cloud.  The Northern Cloud,
currently being eroded by Tr 14, has not been accelerated along our
line-of-sight, while the molecular gas in the South Pillars has
undergone some acceleration at its closest approach to Tr 16.

The acceleration of the Southern Cloud by the O-type stars in Tr 16 is
readily explained by the rocket effect
\citep{oortspitzer1955,ballyscoville1980,bertoldimckee1990}.  As the
neutral gas facing Tr 16 is ionized by the strong ultraviolet radition
of the cluster \citep{smith2006a}, it flows away from the surface of
the molecular cloud at roughly its sound speed.  The recoil force on
the molecular cloud causes it to accelerate away from the ionizing
source at a rate proportional to the rate at which it loses mass
through ionization.  In the simplest scenario, assuming the cloud is
initially at rest relative to the ionizing source, the cloud mass
and velocity are connected through \citep{spitzer1978}:

\begin{equation}
M_c = M_0e^{-v_c/V_{\textrm{ion}}},
\end{equation}

\noindent where $M_0$ and $M_c$ are the initial and current masses of the
molecular cloud, respectively, $v_c$ is the current velocity of the
cloud, and $V_{\textrm{ion}}$ is the velocity with which the newly ionized
material flows away from the cloud.  For the molecular cloud to have
been accelerated to $\sim$10 km s$^{-1}$, roughly the speed of sound
in ionized gas, the cloud's mass must have decreased by $\sim$60\%.
The current mass of the Southern Cloud is $\sim5\times10^4~\textrm{M}_{\sun}$
\citep{rebolledo2016}.  Accounting only for mass loss through
ionization, its initial mass would have been on the order of
$1.3\times10^5~\textrm{M}_{\sun}$, comparable to the current mass of the
Northern Cloud \citep{yonekura2005,rebolledo2016}.

Noteworthy among the O-type stars in Tr 16 is V662 Car at
$l=287.7$\degr, $b=-0.7$\degr, an eclipsing spectroscopic binary with
a systemic velocity of $-15\pm2$ km s$^{-1}$ \citep{niemela2006}.
V662 Car is the only O-type star to coincide with the Southern Cloud
in three-dimensional position--velocity space, and its relatively high
visual extinction \citep{smith1987,povich2011b} suggests that it is
behind or within the molecular gas.  The primary star's spectrum has
unusually strong \ion{He}{ii} $\lambda4686$ \citep{niemela2006},
leading it to be assigned to the luminosity class Vz \citep{sota2014},
which is associated with very young O-type stars close to the zero-age
main sequence
\citep{walborn2009,sabinsanjulian2014,walborn2014,arias2016}.  Roughly
a dozen of the O-type stars in the Carina Nebula are of class Vz
\citep{sota2014}, with the highest fraction relative to non-z O dwarfs
found in Tr 14 \citep{arias2016}.  In addition, \citet{niemela2006}
found that both components of V662 Car have smaller radii and
luminosities than expected for their spectral types, another indicator
of youth.  The combination of V662 Car's young age, its deviation from
the mean RV of the O-type stars in Tr 16, and its kinematic
association with the Southern Cloud suggest that it was formed
separately from and more recently than the body of Tr 16.  We propose
that its formation may have been triggered by the feedback-induced
acceleration of the Southern Cloud.

\subsubsection{Cr 228, Bo 11, and the molecular gas in the South Pillars}
\label{subsubsec:ostars-sp}

The relationship between the O-type stars in Bo 11 and Cr 228 and the
molecular gas in the South Pillars is difficult to interpret, because
the gas detected in $^{12}$CO is spatially separate from the O-type
stars.  Most of the $^{12}$CO (1--0) emission in the South Pillars
comes from the so-called Giant Pillar
\citep{smith2000,yonekura2005,smith2010b}, a dusty structure that
points toward Tr 16 from the southernmost part of the region, lying
roughly halfway between Bo 11 and the nominal center of Cr 228.  The
Giant Pillar is the site of current star formation
\citep{smith2010b,gaczkowski2013}, but is $\ge4$ pc from any O-type
stars.  Its gas has heliocentric RVs of ($-10$)--($-5$) km s$^{-1}$,
similar to the \ctxt{Northern} Cloud.  Another dusty structure, whose
ionization fronts also face Tr 16, is spatially coincident with Bo 11.
Some $^{12}$CO (1--0) emission, too faint to appear in Figures
\ref{fig:ostars-coglon} and \ref{fig:ostars-coglat}, is detected in this region
\citep{rebolledo2016}, but it has a heliocentric RV of $\gtrsim30$ km
s$^{-1}$ and likely belongs to the far side of the Carina spiral arm.

Most of the O-type stars in Bo 11 have RVs similar to the rest of the
O-type stars in the Carina Nebula and comparable to the gas in the
South Pillars.  The exception is HDE 305612, which, with a measured
mean RV of 16.1 km s$^{-1}$, appears to be a notable outlier.
However, HDE 305612 shows significant RV variation over three epochs
of CHIRON data---but was not flagged as an unsolved binary because the
amplitude of that variation is $<20$ km s$^{-1}$.

The O-star population of Cr 228 also has similar RVs to the rest of
the region; in Figure \ref{fig:ostars-coglon}, it is indistinguishable from
the O-star population of Tr 16.  None of the many smaller dust pillars
around Cr 228 \citep{smith2010b} are detected in $^{12}$CO (1--0)
emission \citep{rebolledo2016}, ruling out a direct comparison between
the RVs of stars and gas in this part of the Carina Nebula.  As
discussed in Section \ref{subsec:ostars-clusters}, the O-type stars in and
around Cr 228 have a lower velocity dispersion than the region as a
whole, suggesting that they were not scattered out of Tr 16.

In the far southeast corner of the Carina Nebula, approximately
12.5\arcmin~($\sim$8 pc) from Bo 11, are HD 93843 and HDE 305619.
These two stars are clearly visible on the left side of Figure
\ref{fig:ostars-coglon} as the two O-type systems with the highest
Galactic longitudes.  They are typically treated as members of the
Carina Nebula complex \citep[e.g.,][]{gagne2011}, although there are
no known gas structures or lower-mass stellar populations connecting
them to the rest of the South Pillars \citep[e.g.,][]{smith2010b}.
The spectroscopic parallax of HDE 305619 \citep[using data
  from][]{kharchenko2001,martins2005,kharchenkoroeser2009,gagne2011,sota2014}
suggests that it might be in the background at a distance of $>3$ kpc,
but \btxt{inferred distances based on \emph{Gaia} DR1
  \citep{astraatmadjabailerjones2016} place both \btxt{it and HD 93483
    at $\sim2.4$~kpc}}.  Their observed RVs are also consistent with
being part of the Carina Nebula complex, although HDE 305619
\btxt{shows significant low-amplitude variability in our CHIRON data
  and was classified as an SB1 by \citet{chini2012}.}

\subsection{Comparison to ionized gas}
\label{subsec:ostars-ion}

The ionized gas in the Carina Nebula is globally expanding at $\pm
15$--20 km s$^{-1}$, as seen in radio recombination lines
\citep{gardner1970,huchtmeierday1975,azcarate1981,brooks2001} and
optical line emission
\citep{deharvengmaucherat1975,walbornhesser1975,walsh1984,smith2004b}.
This expansion is driven by feedback from Carina's O-type and evolved
massive stars \citep{smithbrooks2007}.  It was most recently mapped by
\citet{damiani2016}, who observed more than 650 optical sightlines
across Tr 14 and Tr 16.  They identified three non-spherical expanding
shells, roughly centered on the positions of $\eta$ Car, WR 25, and Tr
14.  We represent these shells in Figures \ref{fig:ostars-coglon} and
\ref{fig:ostars-coglat} with hatched regions showing the range of observed
approaching and receding RVs for each shell.  \citet{damiani2016}
estimate that the overall expansion is centered around an RV of $-12.5$
km s$^{-1}$.  Their Tr 14 shell is centered around a slightly less
negative $-8$ km s$^{-1}$.  These values agree with prior results:
\citet{walborn1973,walborn2002b,walborn2007} place the kinematic
center of the expansion at $-14$ km s$^{-1}$, and radio data
\citep{gardner1970,huchtmeierday1975,azcarate1981} consistently
centers the expansion at a heliocentric RV of $-9$ km s$^{-1}$.

However, the RV distribution of Carina's O-type and evolved massive
stars is not aligned with the expansion of the ionized gas.  As
described in Section \ref{subsec:ostars-distribution}, the weighted
mean RV of the well-constrained O-type stars is 0.6 $\pm$ 9.1 km
s$^{-1}$, putting the kinematic center of the ionized gas expansion
roughly one standard deviation blueward of the mean O-star RV.
Figures \ref{fig:ostars-coglon} and \ref{fig:ostars-coglat} show that
the receding/redshifted components of the gas overlap in RV space with
the positive-RV tail of the O-star distribution, while the
approaching/blueshifted components of the gas have more negative RVs
than nearly all of the O-type stars.  The \btxt{\ion{N}{iv}~$\lambda4058$}
RVs of the WNH stars, including WR 25, agree with the RVs of the
approaching gas, \btxt{but as previously discussed,} this is due to
wind effects in the WNH spectra and not due to a physical association.

The most likely explanation for the kinematic asymmetry between the
gas and the O-type stars is that the receding/redshifted ionized gas
is bounded by dense, neutral material behind the Carina Nebula, while
the approaching/blueshifted ionized gas moves freely along our line of
sight.  In addition, the approaching gas around Tr 16 may be partially
composed of a photoevaporative flow off the Southern Cloud, which has
already been accelerated toward us by feedback from Tr 16 (see Section
\ref{subsubsec:ostars-sc}).  \citet{damiani2016} noted a number of
smaller-scale kinematic asymmetries in the ionized gas shells,
confirming that the expansion of the \ion{H}{ii} region is
non-spherical and is affected by density variations in the surrounding
medium.  We consequently caution against using the kinematic center of
the ionized gas to define the systemic RV of the Carina
Nebula.

%%%%%%%%%%%%%%%%%%%%%%%%%%%%%%%%%%%%%%%%%%%%%%%%%%%%%%%%%%%%%%%%%%%%%%%
\section{Conclusions}
\label{sec:ostars-conc}

We have conducted a radial velocity survey of the O-type and evolved
massive stars in the Carina Nebula.  We obtained multi-epoch echelle
spectroscopy for 31 O-type stars, and compiled published RVs for an
additional 32 systems including three WNH stars and the LBV $\eta$
Car.  With these data, we find the first spectroscopic orbital
solutions for the near-twin system HD 92607 and the eclipsing binary
HDE 303312 and provide updated orbital solutions for HD 93576 and HDE
305536.  Our further results are summarized as follows:

(1) Of the 63 O-type and evolved massive star systems with RV data,
41 have well-constrained systemic velocities.  These well-constrained
sources have a weighted mean RV of 0.6 km s$^{-1}$, comparable to
prior results for the Tr 14 cluster and to the expected radial motion
for sources at a distance of $\sim2$ kpc along the Carina \ctxt{spiral} arm.

(2) \ctxt{The standard deviation of the RVs of the well-constrained
  sources is 9.1~km~s$^{-1}$ and provides an upper limit to the
  one-dimensional velocity dispersion of the region.}  This value is
high compared to \ctxt{the velocity dispersions of} typical, less
massive OB associations and is roughly twice that of massive bound
starburst clusters.  However, it is similar to the velocity dispersion
of the large, unbound Cyg OB2 association, and is not unexpected for a
region with the content and substructure of the Carina Nebula.

(3) The overall O-star RV distribution is unimodal, favoring a common
distance to the various clusters of the Carina Nebula.

(4) There is a possible but marginally significant difference between
the RV distributions of the Tr 16 and Tr 14 clusters, with the
latter's O-star RVs $\sim5$ km s$^{-1}$ more positive, on average,
than the former's.  Kinematic study of the intermediate-mass
populations in these clusters is needed to confirm the offset.
 
(5) We do not detect any line-of-sight runaway O-type stars, nor do we
see evidence that the distributed O-type population migrated out from
Tr 14 and Tr 16.  On the contrary, the O-type stars in Cr 228 and the
South Pillars have a low velocity dispersion compared to the region as
a whole.

(6) The Tr 14 cluster is kinematically associated with the molecular
gas of the Northern Cloud, consistent with its young age.

(7) Feedback from Tr 16 has accelerated the molecular gas of the
Southern Cloud toward us, relative to the stellar population, by 10--15 km s$^{-1}$.  V662 Car, an O-type star on the
outskirts of Tr 16, may belong to a younger generation of stars
triggered by this feedback.

(8) The approaching components of the ionized gas around Tr 14 and Tr
16 show higher velocities, relative to the O-type stars, than the
receding components.  This kinematic asymmetry indicates that the
expansion of the \ion{H}{ii} region is not spherical and is likely
impacted by the distribution of dense neutral gas.
 
Our observations set the stage for the analysis of future \emph{Gaia}
data releases, which will add further kinematic dimensions to our
understanding of the Carina Nebula.  This region continues to be a
laboratory for the study of massive-star formation and the interplay
between stellar feedback and the interstellar medium.

%%%%%%%%%%%%%%%%%%%%%%%%%%%%%%%%%%%%%%%%%%%%%%%%%%%%%%%%%%%%%%%%%%%%%%%
\section*{Acknowledgements}

We would like to thank the operators of the CTIO 1.5-m telescope for
executing our CHIRON observations, and SMARTS Data/Queue Manager Emily
MacPherson for queue scheduling.  We also thank Maxwell Moe for his
guidance in using {\tt NIGHTFALL} to fit the orbit of HDE
303312\btxt{, and we thank the anonymous referee for the time spent
  reviewing this manuscript}.  This work is based on observations at
Cerro Tololo Inter-American Observatory, National Optical Astronomy
Observatory (NOAO Prop. IDs: 2014B-0235 and 2015B-0141; PI:
M. Kiminki), which is operated by the Association of Universities for
Research in Astronomy (AURA) under a cooperative agreement with the
National Science Foundation.  This paper also makes use of data
obtained using the Mopra radio telescope, a part of the Australia
Telescope National Facility which is funded by the Commonwealth of
Australia for operation as a National Facility managed by CSIRO.

\btxt{Finally, we note that while editing the final revision of this
  manuscript, we learned of the passing of Nolan R. Walborn.  His
  contributions toward our understanding of massive O-type stars in
  general have been tremendous, and particularly so for the massive
  stars in the Carina Nebula studied in this paper.}

%%%%%%%%%%%%%%%%%%%%%%%%%%%%%%%%%%%%%%%%%%%%%%%%%%%%%%%%%%%%%%%%%%%%%%%
\bibliographystyle{mnras}
\bibliography{ms} 

\begin{thebibliography}{}
\makeatletter
\relax
\def\mn@urlcharsother{\let\do\@makeother \do\$\do\&\do\#\do\^\do\_\do\%\do\~}
\def\mn@doi{\begingroup\mn@urlcharsother \@ifnextchar [ {\mn@doi@}
  {\mn@doi@[]}}
\def\mn@doi@[#1]#2{\def\@tempa{#1}\ifx\@tempa\@empty \href
  {http://dx.doi.org/#2} {doi:#2}\else \href {http://dx.doi.org/#2} {#1}\fi
  \endgroup}
\def\mn@eprint#1#2{\mn@eprint@#1:#2::\@nil}
\def\mn@eprint@arXiv#1{\href {http://arxiv.org/abs/#1} {{\tt arXiv:#1}}}
\def\mn@eprint@dblp#1{\href {http://dblp.uni-trier.de/rec/bibtex/#1.xml}
  {dblp:#1}}
\def\mn@eprint@#1:#2:#3:#4\@nil{\def\@tempa {#1}\def\@tempb {#2}\def\@tempc
  {#3}\ifx \@tempc \@empty \let \@tempc \@tempb \let \@tempb \@tempa \fi \ifx
  \@tempb \@empty \def\@tempb {arXiv}\fi \@ifundefined
  {mn@eprint@\@tempb}{\@tempb:\@tempc}{\expandafter \expandafter \csname
  mn@eprint@\@tempb\endcsname \expandafter{\@tempc}}}

\bibitem[\protect\citeauthoryear{{Albacete Colombo}, {Morrell}, {Niemela}  \&
  {Corcoran}}{{Albacete Colombo} et~al.}{2001}]{albacetecolombo2001}
{Albacete Colombo} J.~F.,  {Morrell} N.~I.,  {Niemela} V.~S.,   {Corcoran}
  M.~F.,  2001, \mn@doi [\mnras] {10.1046/j.1365-8711.2001.04497.x}, \href
  {http://adsabs.harvard.edu/abs/2001MNRAS.326...78A} {326, 78}

\bibitem[\protect\citeauthoryear{{Albacete Colombo}, {Morrell}, {Rauw},
  {Corcoran}, {Niemela}  \& {Sana}}{{Albacete Colombo}
  et~al.}{2002}]{albacetecolombo2002}
{Albacete Colombo} J.~F.,  {Morrell} N.~I.,  {Rauw} G.,  {Corcoran} M.~F.,
  {Niemela} V.~S.,   {Sana} H.,  2002, \mn@doi [\mnras]
  {10.1046/j.1365-8711.2002.05755.x}, \href
  {http://adsabs.harvard.edu/abs/2002MNRAS.336.1099A} {336, 1099}

\bibitem[\protect\citeauthoryear{{Alexander}, {Hanes}, {Povich}  \&
  {McSwain}}{{Alexander} et~al.}{2016}]{alexander2016}
{Alexander} M.~J.,  {Hanes} R.~J.,  {Povich} M.~S.,   {McSwain} M.~V.,  2016,
  \mn@doi [\aj] {10.3847/0004-6256/152/6/190}, \href
  {http://adsabs.harvard.edu/abs/2016AJ....152..190A} {152, 190}

\bibitem[\protect\citeauthoryear{{Allen}}{{Allen}}{1979}]{allen1979}
{Allen} D.~A.,  1979, \mn@doi [\mnras] {10.1093/mnras/189.1.1P}, \href
  {http://adsabs.harvard.edu/abs/1979MNRAS.189P...1A} {189, 1P}

\bibitem[\protect\citeauthoryear{{Allen} \& {Hillier}}{{Allen} \&
  {Hillier}}{1993}]{allenhillier1993}
{Allen} D.~A.,  {Hillier} D.~J.,  1993, Proceedings of the Astronomical Society
  of Australia, \href {http://adsabs.harvard.edu/abs/1993PASAu..10..338A} {10,
  338}

\bibitem[\protect\citeauthoryear{{Almeida} et~al.,}{{Almeida}
  et~al.}{2017}]{almeida2017}
{Almeida} L.~A.,  et~al., 2017, \mn@doi [\aap] {10.1051/0004-6361/201629844},
  \href {http://adsabs.harvard.edu/abs/2017A%26A...598A..84A} {598, A84}

\bibitem[\protect\citeauthoryear{{Arias} et~al.,}{{Arias}
  et~al.}{2016}]{arias2016}
{Arias} J.~I.,  et~al., 2016, \mn@doi [\aj] {10.3847/0004-6256/152/2/31}, \href
  {http://adsabs.harvard.edu/abs/2016AJ....152...31A} {152, 31}

\bibitem[\protect\citeauthoryear{{Ascenso}, {Alves}, {Vicente}  \&
  {Lago}}{{Ascenso} et~al.}{2007}]{ascenso2007}
{Ascenso} J.,  {Alves} J.,  {Vicente} S.,   {Lago} M.~T.~V.~T.,  2007, \mn@doi
  [\aap] {10.1051/0004-6361:20077210}, \href
  {http://adsabs.harvard.edu/abs/2007A%26A...476..199A} {476, 199}

\bibitem[\protect\citeauthoryear{{Astraatmadja} \&
  {Bailer-Jones}}{{Astraatmadja} \&
  {Bailer-Jones}}{2016}]{astraatmadjabailerjones2016}
{Astraatmadja} T.~L.,  {Bailer-Jones} C.~A.~L.,  2016, \mn@doi [\apj]
  {10.3847/1538-4357/833/1/119}, \href
  {http://adsabs.harvard.edu/abs/2016ApJ...833..119A} {833, 119}

\bibitem[\protect\citeauthoryear{{Azcarate}, {Cersosimo}  \&
  {Colomb}}{{Azcarate} et~al.}{1981}]{azcarate1981}
{Azcarate} I.~N.,  {Cersosimo} J.~C.,   {Colomb} F.~R.,  1981, \rmxaa, \href
  {http://adsabs.harvard.edu/abs/1981RMxAA...6..269A} {6, 269}

\bibitem[\protect\citeauthoryear{{Bally}}{{Bally}}{2008}]{bally2008}
{Bally} J.,  2008, in {Reipurth} B.,  ed., , Handbook of Star Forming Regions,
  Volume I.
ASP, San Francisco, CA, p.~459

\bibitem[\protect\citeauthoryear{{Bally} \& {Scoville}}{{Bally} \&
  {Scoville}}{1980}]{ballyscoville1980}
{Bally} J.,  {Scoville} N.~Z.,  1980, \mn@doi [\apj] {10.1086/158094}, \href
  {http://adsabs.harvard.edu/abs/1980ApJ...239..121B} {239, 121}

\bibitem[\protect\citeauthoryear{{Barb{\'a}}, {Gamen}, {Arias}, {Morrell},
  {Ma{\'{\i}}z Apell{\'a}niz}, {Alfaro}, {Walborn}  \& {Sota}}{{Barb{\'a}}
  et~al.}{2010}]{barba2010}
{Barb{\'a}} R.~H.,  {Gamen} R.,  {Arias} J.~I.,  {Morrell} N.,  {Ma{\'{\i}}z
  Apell{\'a}niz} J.,  {Alfaro} E.,  {Walborn} N.,   {Sota} A.,  2010, in
  {Rivinius} T.,  {Cur\'e} M.,  eds,  Revista Mexicana de Astronomia y
  Astrofisica Conference Series Vol. 38, The Interferometric View on Hot Stars.
  pp 30--32

\bibitem[\protect\citeauthoryear{{Bertoldi} \& {McKee}}{{Bertoldi} \&
  {McKee}}{1990}]{bertoldimckee1990}
{Bertoldi} F.,  {McKee} C.~F.,  1990, \mn@doi [\apj] {10.1086/168713}, \href
  {http://adsabs.harvard.edu/abs/1990ApJ...354..529B} {354, 529}

\bibitem[\protect\citeauthoryear{{Blaauw}}{{Blaauw}}{1961}]{blaauw1961}
{Blaauw} A.,  1961, \bain, \href
  {http://adsabs.harvard.edu/abs/1961BAN....15..265B} {15, 265}

\bibitem[\protect\citeauthoryear{{Bohannan} \& {Garmany}}{{Bohannan} \&
  {Garmany}}{1978}]{bohannangarmany1978}
{Bohannan} B.,  {Garmany} C.~D.,  1978, \mn@doi [\apj] {10.1086/156323}, \href
  {http://adsabs.harvard.edu/abs/1978ApJ...223..908B} {223, 908}

\bibitem[\protect\citeauthoryear{{Brooks}, {Whiteoak}  \& {Storey}}{{Brooks}
  et~al.}{1998}]{brooks1998}
{Brooks} K.~J.,  {Whiteoak} J.~B.,   {Storey} J.~W.~V.,  1998, \mn@doi [\pasa]
  {10.1071/AS98202}, \href {http://adsabs.harvard.edu/abs/1998PASA...15..202B}
  {15, 202}

\bibitem[\protect\citeauthoryear{{Brooks}, {Burton}, {Rathborne}, {Ashley}  \&
  {Storey}}{{Brooks} et~al.}{2000}]{brooks2000}
{Brooks} K.~J.,  {Burton} M.~G.,  {Rathborne} J.~M.,  {Ashley} M.~C.~B.,
  {Storey} J.~W.~V.,  2000, \mn@doi [\mnras]
  {10.1046/j.1365-8711.2000.03798.x}, \href
  {http://adsabs.harvard.edu/abs/2000MNRAS.319...95B} {319, 95}

\bibitem[\protect\citeauthoryear{{Brooks}, {Storey}  \& {Whiteoak}}{{Brooks}
  et~al.}{2001}]{brooks2001}
{Brooks} K.~J.,  {Storey} J.~W.~V.,   {Whiteoak} J.~B.,  2001, \mn@doi [\mnras]
  {10.1046/j.1365-8711.2001.04590.x}, \href
  {http://adsabs.harvard.edu/abs/2001MNRAS.327...46B} {327, 46}

\bibitem[\protect\citeauthoryear{{Brooks}, {Cox}, {Schneider}, {Storey},
  {Poglitsch}, {Geis}  \& {Bronfman}}{{Brooks} et~al.}{2003}]{brooks2003}
{Brooks} K.~J.,  {Cox} P.,  {Schneider} N.,  {Storey} J.~W.~V.,  {Poglitsch}
  A.,  {Geis} N.,   {Bronfman} L.,  2003, \mn@doi [\aap]
  {10.1051/0004-6361:20031406}, \href
  {http://adsabs.harvard.edu/abs/2003A%26A...412..751B} {412, 751}

\bibitem[\protect\citeauthoryear{{Burton} et~al.,}{{Burton}
  et~al.}{2013}]{burton2013}
{Burton} M.~G.,  et~al., 2013, \mn@doi [\pasa] {10.1017/pasa.2013.22}, \href
  {http://adsabs.harvard.edu/abs/2013PASA...30...44B} {30, e044}

\bibitem[\protect\citeauthoryear{{Carraro}}{{Carraro}}{2002}]{carraro2002}
{Carraro} G.,  2002, \mn@doi [\mnras] {10.1046/j.1365-8711.2002.05249.x}, \href
  {http://adsabs.harvard.edu/abs/2002MNRAS.331..785C} {331, 785}

\bibitem[\protect\citeauthoryear{{Carraro} \& {Patat}}{{Carraro} \&
  {Patat}}{2001}]{carraropatat2001}
{Carraro} G.,  {Patat} F.,  2001, \mn@doi [\aap] {10.1051/0004-6361:20011314},
  \href {http://adsabs.harvard.edu/abs/2001A%26A...379..136C} {379, 136}

\bibitem[\protect\citeauthoryear{{Carraro}, {Romaniello}, {Ventura}  \&
  {Patat}}{{Carraro} et~al.}{2004}]{carraro2004}
{Carraro} G.,  {Romaniello} M.,  {Ventura} P.,   {Patat} F.,  2004, \mn@doi
  [\aap] {10.1051/0004-6361:20034335}, \href
  {http://adsabs.harvard.edu/abs/2004A%26A...418..525C} {418, 525}

\bibitem[\protect\citeauthoryear{{Chini}, {Hoffmeister}, {Nasseri}, {Stahl}  \&
  {Zinnecker}}{{Chini} et~al.}{2012}]{chini2012}
{Chini} R.,  {Hoffmeister} V.~H.,  {Nasseri} A.,  {Stahl} O.,   {Zinnecker} H.,
   2012, \mn@doi [\mnras] {10.1111/j.1365-2966.2012.21317.x}, \href
  {http://adsabs.harvard.edu/abs/2012MNRAS.424.1925C} {424, 1925}

\bibitem[\protect\citeauthoryear{{Clark}, {Bonnell}, {Zinnecker}  \&
  {Bate}}{{Clark} et~al.}{2005}]{clark2005}
{Clark} P.~C.,  {Bonnell} I.~A.,  {Zinnecker} H.,   {Bate} M.~R.,  2005,
  \mn@doi [\mnras] {10.1111/j.1365-2966.2005.08942.x}, \href
  {http://adsabs.harvard.edu/abs/2005MNRAS.359..809C} {359, 809}

\bibitem[\protect\citeauthoryear{{Clarkson}, {Ghez}, {Morris}, {Lu}, {Stolte},
  {McCrady}, {Do}  \& {Yelda}}{{Clarkson} et~al.}{2012}]{clarkson2012}
{Clarkson} W.~I.,  {Ghez} A.~M.,  {Morris} M.~R.,  {Lu} J.~R.,  {Stolte} A.,
  {McCrady} N.,  {Do} T.,   {Yelda} S.,  2012, \mn@doi [\apj]
  {10.1088/0004-637X/751/2/132}, \href
  {http://adsabs.harvard.edu/abs/2012ApJ...751..132C} {751, 132}

\bibitem[\protect\citeauthoryear{{Coelho}}{{Coelho}}{2014}]{coelho2014}
{Coelho} P.~R.~T.,  2014, \mn@doi [\mnras] {10.1093/mnras/stu365}, \href
  {http://adsabs.harvard.edu/abs/2014MNRAS.440.1027C} {440, 1027}

\bibitem[\protect\citeauthoryear{{Collado}, {Gamen}  \& {Barb{\'a}}}{{Collado}
  et~al.}{2013}]{collado2013}
{Collado} A.,  {Gamen} R.,   {Barb{\'a}} R.~H.,  2013, \mn@doi [\aap]
  {10.1051/0004-6361/201118460}, \href
  {http://adsabs.harvard.edu/abs/2013A%26A...552A..22C} {552, A22}

\bibitem[\protect\citeauthoryear{{Collado}, {Gamen}, {Barb{\'a}}  \&
  {Morrell}}{{Collado} et~al.}{2015}]{collado2015}
{Collado} A.,  {Gamen} R.,  {Barb{\'a}} R.~H.,   {Morrell} N.,  2015, \mn@doi
  [\aap] {10.1051/0004-6361/201424863}, \href
  {http://adsabs.harvard.edu/abs/2015A%26A...581A..49C} {581, A49}

\bibitem[\protect\citeauthoryear{{Conti}, {Leep}  \& {Lorre}}{{Conti}
  et~al.}{1977}]{conti1977}
{Conti} P.~S.,  {Leep} E.~M.,   {Lorre} J.~J.,  1977, \mn@doi [\apj]
  {10.1086/155305}, \href {http://adsabs.harvard.edu/abs/1977ApJ...214..759C}
  {214, 759}

\bibitem[\protect\citeauthoryear{{Conti}, {Niemela}  \& {Walborn}}{{Conti}
  et~al.}{1979}]{conti1979}
{Conti} P.~S.,  {Niemela} V.~S.,   {Walborn} N.~R.,  1979, \mn@doi [\apj]
  {10.1086/156837}, \href {http://adsabs.harvard.edu/abs/1979ApJ...228..206C}
  {228, 206}

\bibitem[\protect\citeauthoryear{{Cox} \& {Bronfman}}{{Cox} \&
  {Bronfman}}{1995}]{coxbronfman1995}
{Cox} P.,  {Bronfman} L.,  1995, \aap, \href
  {http://adsabs.harvard.edu/abs/1995A%26A...299..583C} {299, 583}

\bibitem[\protect\citeauthoryear{{Crowther}}{{Crowther}}{2007}]{crowther2007}
{Crowther} P.~A.,  2007, \mn@doi [\araa]
  {10.1146/annurev.astro.45.051806.110615}, \href
  {http://adsabs.harvard.edu/abs/2007ARA%26A..45..177C} {45, 177}

\bibitem[\protect\citeauthoryear{{Cudworth}, {Martin}  \&
  {Degioia-Eastwood}}{{Cudworth} et~al.}{1993}]{cudworth1993}
{Cudworth} K.~M.,  {Martin} S.~C.,   {Degioia-Eastwood} K.,  1993, \mn@doi
  [\aj] {10.1086/116557}, \href
  {http://adsabs.harvard.edu/abs/1993AJ....105.1822C} {105, 1822}

\bibitem[\protect\citeauthoryear{{Damiani} et~al.,}{{Damiani}
  et~al.}{2016}]{damiani2016}
{Damiani} F.,  et~al., 2016, \mn@doi [\aap] {10.1051/0004-6361/201628169},
  \href {http://adsabs.harvard.edu/abs/2016A%26A...591A..74D} {591, A74}

\bibitem[\protect\citeauthoryear{{Damiani} et~al.,}{{Damiani}
  et~al.}{2017}]{damiani2017}
{Damiani} F.,  et~al., 2017, \mn@doi [\aap] {10.1051/0004-6361/201629020},
  \href {http://adsabs.harvard.edu/abs/2017A%26A...603A..81D} {603, A81}

\bibitem[\protect\citeauthoryear{{Davidson} \& {Humphreys}}{{Davidson} \&
  {Humphreys}}{1997}]{davidsonhumphreys1997}
{Davidson} K.,  {Humphreys} R.~M.,  1997, \mn@doi [\araa]
  {10.1146/annurev.astro.35.1.1}, \href
  {http://adsabs.harvard.edu/abs/1997ARA%26A..35....1D} {35, 1}

\bibitem[\protect\citeauthoryear{{Davidson}, {Smith}, {Gull}, {Ishibashi}  \&
  {Hillier}}{{Davidson} et~al.}{2001}]{davidson2001}
{Davidson} K.,  {Smith} N.,  {Gull} T.~R.,  {Ishibashi} K.,   {Hillier} D.~J.,
  2001, \mn@doi [\aj] {10.1086/319419}, \href
  {http://adsabs.harvard.edu/abs/2001AJ....121.1569D} {121, 1569}

\bibitem[\protect\citeauthoryear{{Deharveng} \& {Maucherat}}{{Deharveng} \&
  {Maucherat}}{1975}]{deharvengmaucherat1975}
{Deharveng} L.,  {Maucherat} M.,  1975, \aap, \href
  {http://adsabs.harvard.edu/abs/1975A%26A....41...27D} {41, 27}

\bibitem[\protect\citeauthoryear{{Dickel}}{{Dickel}}{1974}]{dickel1974}
{Dickel} H.~R.,  1974, \aap, \href
  {http://adsabs.harvard.edu/abs/1974A%26A....31...11D} {31, 11}

\bibitem[\protect\citeauthoryear{{Doran} et~al.,}{{Doran}
  et~al.}{2013}]{doran2013}
{Doran} E.~I.,  et~al., 2013, \mn@doi [\aap] {10.1051/0004-6361/201321824},
  \href {http://adsabs.harvard.edu/abs/2013A%26A...558A.134D} {558, A134}

\bibitem[\protect\citeauthoryear{{Drilling} \& {Landolt}}{{Drilling} \&
  {Landolt}}{2000}]{drillinglandolt2000}
{Drilling} J.~S.,  {Landolt} A.~U.,  2000, in {Cox} A.~N.,  ed., , Allen's
  Astrophysical Quantities.
Springer, pp 381--396

\bibitem[\protect\citeauthoryear{{Efremov} \& {Elmegreen}}{{Efremov} \&
  {Elmegreen}}{1998}]{efremovelmegreen1998}
{Efremov} Y.~N.,  {Elmegreen} B.~G.,  1998, \mn@doi [\mnras]
  {10.1046/j.1365-8711.1998.01745.x}, \href
  {http://adsabs.harvard.edu/abs/1998MNRAS.299..643E} {299, 643}

\bibitem[\protect\citeauthoryear{{Feast}, {Thackeray}  \& {Wesselink}}{{Feast}
  et~al.}{1957}]{feast1957}
{Feast} M.~W.,  {Thackeray} A.~D.,   {Wesselink} A.~J.,  1957, \memras, \href
  {http://adsabs.harvard.edu/abs/1957MmRAS..68....1F} {68, 1}

\bibitem[\protect\citeauthoryear{{Feigelson} et~al.,}{{Feigelson}
  et~al.}{2011}]{feigelson2011}
{Feigelson} E.~D.,  et~al., 2011, \mn@doi [\apjs] {10.1088/0067-0049/194/1/9},
  \href {http://adsabs.harvard.edu/abs/2011ApJS..194....9F} {194, 9}

\bibitem[\protect\citeauthoryear{{Feinstein}, {Marraco}  \&
  {Muzzio}}{{Feinstein} et~al.}{1973}]{feinstein1973}
{Feinstein} A.,  {Marraco} H.~G.,   {Muzzio} J.~C.,  1973, \aaps, \href
  {http://adsabs.harvard.edu/abs/1973A%26AS...12..331F} {12, 331}

\bibitem[\protect\citeauthoryear{{Feinstein}, {Marraco}  \&
  {Forte}}{{Feinstein} et~al.}{1976}]{feinstein1976}
{Feinstein} A.,  {Marraco} H.~G.,   {Forte} J.~C.,  1976, \aaps, \href
  {http://cdsads.u-strasbg.fr/abs/1976A%26AS...24..389F} {24, 389}

\bibitem[\protect\citeauthoryear{{Forte}}{{Forte}}{1978}]{forte1978}
{Forte} J.~C.,  1978, \mn@doi [\aj] {10.1086/112311}, \href
  {http://adsabs.harvard.edu/abs/1978AJ.....83.1199F} {83, 1199}

\bibitem[\protect\citeauthoryear{{Fullerton}, {Gies}  \& {Bolton}}{{Fullerton}
  et~al.}{1996}]{fullerton1996}
{Fullerton} A.~W.,  {Gies} D.~R.,   {Bolton} C.~T.,  1996, \mn@doi [\apjs]
  {10.1086/192285}, \href {http://adsabs.harvard.edu/abs/1996ApJS..103..475F}
  {103, 475}

\bibitem[\protect\citeauthoryear{{Fullerton}, {Massa}, {Prinja}, {Owocki}  \&
  {Cranmer}}{{Fullerton} et~al.}{1997}]{fullerton1997}
{Fullerton} A.~W.,  {Massa} D.~L.,  {Prinja} R.~K.,  {Owocki} S.~P.,
  {Cranmer} S.~R.,  1997, \aap, \href
  {http://adsabs.harvard.edu/abs/1997A%26A...327..699F} {327, 699}

\bibitem[\protect\citeauthoryear{{Gaczkowski}, {Preibisch}, {Ratzka},
  {Roccatagliata}, {Ohlendorf}  \& {Zinnecker}}{{Gaczkowski}
  et~al.}{2013}]{gaczkowski2013}
{Gaczkowski} B.,  {Preibisch} T.,  {Ratzka} T.,  {Roccatagliata} V.,
  {Ohlendorf} H.,   {Zinnecker} H.,  2013, \mn@doi [\aap]
  {10.1051/0004-6361/201219836}, \href
  {http://adsabs.harvard.edu/abs/2013A%26A...549A..67G} {549, A67}

\bibitem[\protect\citeauthoryear{{Gagn{\'e}} et~al.,}{{Gagn{\'e}}
  et~al.}{2011}]{gagne2011}
{Gagn{\'e}} M.,  et~al., 2011, \mn@doi [\apjs] {10.1088/0067-0049/194/1/5},
  \href {http://adsabs.harvard.edu/abs/2011ApJS..194....5G} {194, 5}

\bibitem[\protect\citeauthoryear{{Gaia Collaboration} et~al.,}{{Gaia
  Collaboration} et~al.}{2016a}]{gaia2016a}
{Gaia Collaboration} et~al., 2016a, \mn@doi [\aap]
  {10.1051/0004-6361/201629272}, \href
  {http://adsabs.harvard.edu/abs/2016A%26A...595A...1G} {595, A1}

\bibitem[\protect\citeauthoryear{{Gaia Collaboration} et~al.,}{{Gaia
  Collaboration} et~al.}{2016b}]{gaiabrown2016}
{Gaia Collaboration} et~al., 2016b, \mn@doi [\aap]
  {10.1051/0004-6361/201629512}, \href
  {http://adsabs.harvard.edu/abs/2016A%26A...595A...2G} {595, A2}

\bibitem[\protect\citeauthoryear{{Gamen} et~al.,}{{Gamen}
  et~al.}{2006}]{gamen2006}
{Gamen} R.,  et~al., 2006, \mn@doi [\aap] {10.1051/0004-6361:20065618}, \href
  {http://cdsads.u-strasbg.fr/abs/2006A%26A...460..777G} {460, 777}

\bibitem[\protect\citeauthoryear{{Garc{\'{\i}}a}, {Malaroda}, {Levato},
  {Morrell}  \& {Grosso}}{{Garc{\'{\i}}a} et~al.}{1998}]{garcia1998}
{Garc{\'{\i}}a} B.,  {Malaroda} S.,  {Levato} H.,  {Morrell} N.,   {Grosso} M.,
   1998, \mn@doi [\pasp] {10.1086/316117}, \href
  {http://adsabs.harvard.edu/abs/1998PASP..110...53G} {110, 53}

\bibitem[\protect\citeauthoryear{{Gardner}, {Milne}, {Mezger}  \&
  {Wilson}}{{Gardner} et~al.}{1970}]{gardner1970}
{Gardner} F.~F.,  {Milne} D.~K.,  {Mezger} P.~G.,   {Wilson} T.~L.,  1970,
  \aap, \href {http://adsabs.harvard.edu/abs/1970A%26A.....7..349G} {7, 349}

\bibitem[\protect\citeauthoryear{{Garmany}, {Conti}  \& {Massey}}{{Garmany}
  et~al.}{1980}]{garmany1980}
{Garmany} C.~D.,  {Conti} P.~S.,   {Massey} P.,  1980, \mn@doi [\apj]
  {10.1086/158537}, \href {http://adsabs.harvard.edu/abs/1980ApJ...242.1063G}
  {242, 1063}

\bibitem[\protect\citeauthoryear{{Gieles}}{{Gieles}}{2013}]{gieles2013}
{Gieles} M.,  2013, in Massive Stars: From alpha to Omega. p.~7

\bibitem[\protect\citeauthoryear{{Gieles}, {Sana}  \& {Portegies
  Zwart}}{{Gieles} et~al.}{2010}]{gieles2010}
{Gieles} M.,  {Sana} H.,   {Portegies Zwart} S.~F.,  2010, \mn@doi [\mnras]
  {10.1111/j.1365-2966.2009.15993.x}, \href
  {http://adsabs.harvard.edu/abs/2010MNRAS.402.1750G} {402, 1750}

\bibitem[\protect\citeauthoryear{{Gies} \& {Bolton}}{{Gies} \&
  {Bolton}}{1986}]{giesbolton1986}
{Gies} D.~R.,  {Bolton} C.~T.,  1986, \mn@doi [\apjs] {10.1086/191118}, \href
  {http://adsabs.harvard.edu/abs/1986ApJS...61..419G} {61, 419}

\bibitem[\protect\citeauthoryear{{H{\'e}nault-Brunet}
  et~al.,}{{H{\'e}nault-Brunet} et~al.}{2012}]{henaultbrunet2012}
{H{\'e}nault-Brunet} V.,  et~al., 2012, \mn@doi [\aap]
  {10.1051/0004-6361/201219471}, \href
  {http://adsabs.harvard.edu/abs/2012A%26A...546A..73H} {546, A73}

\bibitem[\protect\citeauthoryear{{Herbst}}{{Herbst}}{1976}]{herbst1976}
{Herbst} W.,  1976, \mn@doi [\apj] {10.1086/154681}, \href
  {http://adsabs.harvard.edu/abs/1976ApJ...208..923H} {208, 923}

\bibitem[\protect\citeauthoryear{{Hills}}{{Hills}}{1980}]{hills1980}
{Hills} J.~G.,  1980, \mn@doi [\apj] {10.1086/157703}, \href
  {http://adsabs.harvard.edu/abs/1980ApJ...235..986H} {235, 986}

\bibitem[\protect\citeauthoryear{{H{\"o}gbom}}{{H{\"o}gbom}}{1974}]{hoegbom1974}
{H{\"o}gbom} J.~A.,  1974, \aaps, \href
  {http://adsabs.harvard.edu/abs/1974A%26AS...15..417H} {15, 417}

\bibitem[\protect\citeauthoryear{{Huang} \& {Gies}}{{Huang} \&
  {Gies}}{2006}]{huanggies2006}
{Huang} W.,  {Gies} D.~R.,  2006, \mn@doi [\apj] {10.1086/505782}, \href
  {http://adsabs.harvard.edu/abs/2006ApJ...648..580H} {648, 580}

\bibitem[\protect\citeauthoryear{{Huchtmeier} \& {Day}}{{Huchtmeier} \&
  {Day}}{1975}]{huchtmeierday1975}
{Huchtmeier} W.~K.,  {Day} G.~A.,  1975, \aap, \href
  {http://adsabs.harvard.edu/abs/1975A%26A....41..153H} {41, 153}

\bibitem[\protect\citeauthoryear{{Humphreys}}{{Humphreys}}{1973}]{humphreys1973}
{Humphreys} R.~M.,  1973, \aaps, \href
  {http://adsabs.harvard.edu/abs/1973A%26AS....9...85H} {9, 85}

\bibitem[\protect\citeauthoryear{{Hur}, {Sung}  \& {Bessell}}{{Hur}
  et~al.}{2012}]{hur2012}
{Hur} H.,  {Sung} H.,   {Bessell} M.~S.,  2012, \mn@doi [\aj]
  {10.1088/0004-6256/143/2/41}, \href
  {http://adsabs.harvard.edu/abs/2012AJ....143...41H} {143, 41}

\bibitem[\protect\citeauthoryear{{Iglesias-Marzoa}, {L{\'o}pez-Morales}  \&
  {Jes{\'u}s Ar{\'e}valo Morales}}{{Iglesias-Marzoa}
  et~al.}{2015}]{iglesiasmarzoa2015}
{Iglesias-Marzoa} R.,  {L{\'o}pez-Morales} M.,   {Jes{\'u}s Ar{\'e}valo
  Morales} M.,  2015, \mn@doi [\pasp] {10.1086/682056}, \href
  {http://adsabs.harvard.edu/abs/2015PASP..127..567I} {127, 567}

\bibitem[\protect\citeauthoryear{{Kenyon} \& {Hartmann}}{{Kenyon} \&
  {Hartmann}}{1995}]{kenyonhartmann1995}
{Kenyon} S.~J.,  {Hartmann} L.,  1995, \mn@doi [\apjs] {10.1086/192235}, \href
  {http://adsabs.harvard.edu/abs/1995ApJS..101..117K} {101, 117}

\bibitem[\protect\citeauthoryear{{Kharchenko}}{{Kharchenko}}{2001}]{kharchenko2001}
{Kharchenko} N.~V.,  2001, Kinematika i Fizika Nebesnykh Tel, \href
  {http://adsabs.harvard.edu/abs/2001KFNT...17..409K} {17, 409}

\bibitem[\protect\citeauthoryear{{Kharchenko} \& {Roeser}}{{Kharchenko} \&
  {Roeser}}{2009}]{kharchenkoroeser2009}
{Kharchenko} N.~V.,  {Roeser} S.,  2009, VizieR Online Data Catalog, \href
  {http://adsabs.harvard.edu/abs/2009yCat.1280....0K} {1280}

\bibitem[\protect\citeauthoryear{{Kiminki} \& {Kobulnicky}}{{Kiminki} \&
  {Kobulnicky}}{2012}]{kiminkikobulnicky2012}
{Kiminki} D.~C.,  {Kobulnicky} H.~A.,  2012, \mn@doi [\apj]
  {10.1088/0004-637X/751/1/4}, \href
  {http://adsabs.harvard.edu/abs/2012ApJ...751....4K} {751, 4}

\bibitem[\protect\citeauthoryear{{Kiminki} et~al.,}{{Kiminki}
  et~al.}{2007}]{kiminki2007}
{Kiminki} D.~C.,  et~al., 2007, \mn@doi [\apj] {10.1086/513709}, \href
  {http://adsabs.harvard.edu/abs/2007ApJ...664.1102K} {664, 1102}

\bibitem[\protect\citeauthoryear{{Kiminki} et~al.,}{{Kiminki}
  et~al.}{2008}]{kiminki2008}
{Kiminki} D.~C.,  et~al., 2008, \mn@doi [\apj] {10.1086/588464}, \href
  {http://adsabs.harvard.edu/abs/2008ApJ...681..735K} {681, 735}

\bibitem[\protect\citeauthoryear{{Kiminki}, {Smith}, {Reiter}  \&
  {Bally}}{{Kiminki} et~al.}{2017}]{kiminki2017}
{Kiminki} M.~M.,  {Smith} N.,  {Reiter} M.,   {Bally} J.,  2017, \mn@doi
  [\mnras] {10.1093/mnras/stx607}, \href
  {http://adsabs.harvard.edu/abs/2017MNRAS.468.2469K} {468, 2469}

\bibitem[\protect\citeauthoryear{{Kobulnicky} et~al.,}{{Kobulnicky}
  et~al.}{2014}]{kobulnicky2014}
{Kobulnicky} H.~A.,  et~al., 2014, \mn@doi [\apjs]
  {10.1088/0067-0049/213/2/34}, \href
  {http://adsabs.harvard.edu/abs/2014ApJS..213...34K} {213, 34}

\bibitem[\protect\citeauthoryear{{Kramida}, {Ralchenko}, {Reader}  \& {NIST ASD
  Team}}{{Kramida} et~al.}{2016}]{kramida2016}
{Kramida} A.,  {Ralchenko} Y.,  {Reader} J.,   {NIST ASD Team} 2016, {NIST
  Atomic Spectra Database (version 5.4)}, \url {http://physics.nist.gov/asd}

\bibitem[\protect\citeauthoryear{{Kumar}, {Sharma}, {Manfroid}, {Gosset},
  {Rauw}, {Naz{\'e}}  \& {Kesh Yadav}}{{Kumar} et~al.}{2014}]{kumar2014}
{Kumar} B.,  {Sharma} S.,  {Manfroid} J.,  {Gosset} E.,  {Rauw} G.,  {Naz{\'e}}
  Y.,   {Kesh Yadav} R.,  2014, \mn@doi [\aap] {10.1051/0004-6361/201323027},
  \href {http://adsabs.harvard.edu/abs/2014A%26A...567A.109K} {567, A109}

\bibitem[\protect\citeauthoryear{{Lada} \& {Lada}}{{Lada} \&
  {Lada}}{1991}]{ladalada1991}
{Lada} C.~J.,  {Lada} E.~A.,  1991, in {Janes} K.,  ed.,  Astronomical Society
  of the Pacific Conference Series Vol. 13, The Formation and Evolution of Star
  Clusters. pp 3--22

\bibitem[\protect\citeauthoryear{{Lada} \& {Lada}}{{Lada} \&
  {Lada}}{2003}]{ladalada2003}
{Lada} C.~J.,  {Lada} E.~A.,  2003, \mn@doi [\araa]
  {10.1146/annurev.astro.41.011802.094844}, \href
  {http://adsabs.harvard.edu/abs/2003ARA%26A..41...57L} {41, 57}

\bibitem[\protect\citeauthoryear{{Levato}, {Garc{\'{\i}}a}, {Loust{\'o}},
  {Morrell}  \& {Saizar}}{{Levato} et~al.}{1986}]{levato1986}
{Levato} H.,  {Garc{\'{\i}}a} B.,  {Loust{\'o}} C.,  {Morrell} N.,   {Saizar}
  P.,  1986, \rmxaa, \href {http://adsabs.harvard.edu/abs/1986RMxAA..13....3L}
  {13, 3}

\bibitem[\protect\citeauthoryear{{Levato}, {Malaroda}, {Garcia}, {Morrell}  \&
  {Solivella}}{{Levato} et~al.}{1990}]{levato1990}
{Levato} H.,  {Malaroda} S.,  {Garcia} B.,  {Morrell} N.,   {Solivella} G.,
  1990, \mn@doi [\apjs] {10.1086/191419}, \href
  {http://adsabs.harvard.edu/abs/1990ApJS...72..323L} {72, 323}

\bibitem[\protect\citeauthoryear{{Levato}, {Malaroda}, {Morrell}, {Garcia}  \&
  {Hernandez}}{{Levato} et~al.}{1991a}]{levato1991a}
{Levato} H.,  {Malaroda} S.,  {Morrell} N.,  {Garcia} B.,   {Hernandez} C.,
  1991a, \mn@doi [\apjs] {10.1086/191551}, \href
  {http://adsabs.harvard.edu/abs/1991ApJS...75..869L} {75, 869}

\bibitem[\protect\citeauthoryear{{Levato}, {Malaroda}, {Garcia}, {Morrell},
  {Solivella}  \& {Grosso}}{{Levato} et~al.}{1991b}]{levato1991b}
{Levato} H.,  {Malaroda} S.,  {Garcia} B.,  {Morrell} N.,  {Solivella} G.,
  {Grosso} M.,  1991b, \mn@doi [\apss] {10.1007/BF00643023}, \href
  {http://adsabs.harvard.edu/abs/1991Ap%26SS.183..147L} {183, 147}

\bibitem[\protect\citeauthoryear{{Lindegren} et~al.,}{{Lindegren}
  et~al.}{2016}]{lindegren2016}
{Lindegren} L.,  et~al., 2016, \mn@doi [\aap] {10.1051/0004-6361/201628714},
  \href {http://adsabs.harvard.edu/abs/2016A%26A...595A...4L} {595, A4}

\bibitem[\protect\citeauthoryear{{Ma{\'{\i}}z Apell{\'a}niz}
  et~al.,}{{Ma{\'{\i}}z Apell{\'a}niz} et~al.}{2013}]{maizapellaniz2013}
{Ma{\'{\i}}z Apell{\'a}niz} J.,  et~al., 2013, in Massive Stars: From alpha to
  Omega. p.~198 (\mn@eprint {arXiv} {1306.6417})

\bibitem[\protect\citeauthoryear{{Ma{\'{\i}}z Apell{\'a}niz}
  et~al.,}{{Ma{\'{\i}}z Apell{\'a}niz} et~al.}{2016}]{maizapellaniz2016}
{Ma{\'{\i}}z Apell{\'a}niz} J.,  et~al., 2016, \mn@doi [\apjs]
  {10.3847/0067-0049/224/1/4}, \href
  {http://cdsads.u-strasbg.fr/abs/2016ApJS..224....4M} {224, 4}

\bibitem[\protect\citeauthoryear{{Markwardt}}{{Markwardt}}{2009}]{markwardt2009}
{Markwardt} C.~B.,  2009, in {Bohlender} D.~A.,  {Durand} D.,   {Dowler} P.,
  eds,  Astronomical Society of the Pacific Conference Series Vol. 411,
  Astronomical Data Analysis Software and Systems XVIII. p.~251 (\mn@eprint
  {arXiv} {0902.2850})

\bibitem[\protect\citeauthoryear{{Martins}, {Schaerer}  \& {Hillier}}{{Martins}
  et~al.}{2005}]{martins2005}
{Martins} F.,  {Schaerer} D.,   {Hillier} D.~J.,  2005, \mn@doi [\aap]
  {10.1051/0004-6361:20042386}, \href
  {http://adsabs.harvard.edu/abs/2005A%26A...436.1049M} {436, 1049}

\bibitem[\protect\citeauthoryear{{Martins}, {Marcolino}, {Hillier}, {Donati}
  \& {Bouret}}{{Martins} et~al.}{2015}]{martins2015}
{Martins} F.,  {Marcolino} W.,  {Hillier} D.~J.,  {Donati} J.-F.,   {Bouret}
  J.-C.,  2015, \mn@doi [\aap] {10.1051/0004-6361/201423882}, \href
  {http://adsabs.harvard.edu/abs/2015A%26A...574A.142M} {574, A142}

\bibitem[\protect\citeauthoryear{{Mason}, {Hartkopf}, {Gies}, {Henry}  \&
  {Helsel}}{{Mason} et~al.}{2009}]{mason2009}
{Mason} B.~D.,  {Hartkopf} W.~I.,  {Gies} D.~R.,  {Henry} T.~J.,   {Helsel}
  J.~W.,  2009, \mn@doi [\aj] {10.1088/0004-6256/137/2/3358}, \href
  {http://adsabs.harvard.edu/abs/2009AJ....137.3358M} {137, 3358}

\bibitem[\protect\citeauthoryear{{Massey}}{{Massey}}{1980}]{massey1980}
{Massey} P.,  1980, \mn@doi [\apj] {10.1086/157770}, \href
  {http://adsabs.harvard.edu/abs/1980ApJ...236..526M} {236, 526}

\bibitem[\protect\citeauthoryear{{Massey} \& {Conti}}{{Massey} \&
  {Conti}}{1981}]{masseyconti1981}
{Massey} P.,  {Conti} P.~S.,  1981, \mn@doi [\apj] {10.1086/158695}, \href
  {http://adsabs.harvard.edu/abs/1981ApJ...244..173M} {244, 173}

\bibitem[\protect\citeauthoryear{{Massey} \& {Johnson}}{{Massey} \&
  {Johnson}}{1993}]{masseyjohnson1993}
{Massey} P.,  {Johnson} J.,  1993, \mn@doi [\aj] {10.1086/116487}, \href
  {http://adsabs.harvard.edu/abs/1993AJ....105..980M} {105, 980}

\bibitem[\protect\citeauthoryear{{Massey} \& {Thompson}}{{Massey} \&
  {Thompson}}{1991}]{masseythompson1991}
{Massey} P.,  {Thompson} A.~B.,  1991, \mn@doi [\aj] {10.1086/115774}, \href
  {http://adsabs.harvard.edu/abs/1991AJ....101.1408M} {101, 1408}

\bibitem[\protect\citeauthoryear{{Massey}, {DeGioia-Eastwood}  \&
  {Waterhouse}}{{Massey} et~al.}{2001}]{massey2001}
{Massey} P.,  {DeGioia-Eastwood} K.,   {Waterhouse} E.,  2001, \mn@doi [\aj]
  {10.1086/318769}, \href {http://adsabs.harvard.edu/abs/2001AJ....121.1050M}
  {121, 1050}

\bibitem[\protect\citeauthoryear{{Mayer}, {Lorenz}, {Drechsel}  \&
  {Abseim}}{{Mayer} et~al.}{2001}]{mayer2001}
{Mayer} P.,  {Lorenz} R.,  {Drechsel} H.,   {Abseim} A.,  2001, \mn@doi [\aap]
  {10.1051/0004-6361:20000228}, \href
  {http://adsabs.harvard.edu/abs/2001A%26A...366..558M} {366, 558}

\bibitem[\protect\citeauthoryear{{McClure-Griffiths} \&
  {Dickey}}{{McClure-Griffiths} \& {Dickey}}{2007}]{mccluregriffithsdickey2007}
{McClure-Griffiths} N.~M.,  {Dickey} J.~M.,  2007, \mn@doi [\apj]
  {10.1086/522297}, \href {http://adsabs.harvard.edu/abs/2007ApJ...671..427M}
  {671, 427}

\bibitem[\protect\citeauthoryear{{Megeath}, {Cox}, {Bronfman}  \&
  {Roelfsema}}{{Megeath} et~al.}{1996}]{megeath1996}
{Megeath} S.~T.,  {Cox} P.,  {Bronfman} L.,   {Roelfsema} P.~R.,  1996, \aap,
  \href {http://adsabs.harvard.edu/abs/1996A%26A...305..296M} {305, 296}

\bibitem[\protect\citeauthoryear{{Moe} \& {Di Stefano}}{{Moe} \& {Di
  Stefano}}{2017}]{moedistefano2016}
{Moe} M.,  {Di Stefano} R.,  2017, \mn@doi [\apjs] {10.3847/1538-4365/aa6fb6},
  \href {http://adsabs.harvard.edu/abs/2017ApJS..230...15M} {230, 15}

\bibitem[\protect\citeauthoryear{{Moffat}}{{Moffat}}{1978}]{moffat1978}
{Moffat} A.~F.~J.,  1978, \aap, \href
  {http://adsabs.harvard.edu/abs/1978A%26A....68...41M} {68, 41}

\bibitem[\protect\citeauthoryear{{Moffat} \& {Seggewiss}}{{Moffat} \&
  {Seggewiss}}{1978}]{moffatseggewiss1978}
{Moffat} A.~F.~J.,  {Seggewiss} W.,  1978, \aap, \href
  {http://adsabs.harvard.edu/abs/1978A%26A....70...69M} {70, 69}

\bibitem[\protect\citeauthoryear{{Moffat} \& {Seggewiss}}{{Moffat} \&
  {Seggewiss}}{1979}]{moffatseggewiss1979}
{Moffat} A.~F.~J.,  {Seggewiss} W.,  1979, \aap, \href
  {http://adsabs.harvard.edu/abs/1979A%26A....77..128M} {77, 128}

\bibitem[\protect\citeauthoryear{{Mohr-Smith} et~al.,}{{Mohr-Smith}
  et~al.}{2017}]{mohrsmith2017}
{Mohr-Smith} M.,  et~al., 2017, \mn@doi [\mnras] {10.1093/mnras/stw2751}, \href
  {http://adsabs.harvard.edu/abs/2017MNRAS.465.1807M} {465, 1807}

\bibitem[\protect\citeauthoryear{{Morrell}, {Garcia}  \& {Levato}}{{Morrell}
  et~al.}{1988}]{morrell1988}
{Morrell} N.,  {Garcia} B.,   {Levato} H.,  1988, \mn@doi [\pasp]
  {10.1086/132344}, \href {http://adsabs.harvard.edu/abs/1988PASP..100.1431M}
  {100, 1431}

\bibitem[\protect\citeauthoryear{{Morrell} et~al.,}{{Morrell}
  et~al.}{2001}]{morrell2001}
{Morrell} N.~I.,  et~al., 2001, \mn@doi [\mnras]
  {10.1046/j.1365-8711.2001.04500.x}, \href
  {http://adsabs.harvard.edu/abs/2001MNRAS.326...85M} {326, 85}

\bibitem[\protect\citeauthoryear{{Morrison} \& {Conti}}{{Morrison} \&
  {Conti}}{1980}]{morrisonconti1980}
{Morrison} N.~D.,  {Conti} P.~S.,  1980, \mn@doi [\apj] {10.1086/158101}, \href
  {http://adsabs.harvard.edu/abs/1980ApJ...239..212M} {239, 212}

\bibitem[\protect\citeauthoryear{{Munoz}, {Moffat}, {Hill}, {Shenar},
  {Richardson}, {Pablo}, {St-Louis}  \& {Ramiaramanantsoa}}{{Munoz}
  et~al.}{2017}]{munoz2017}
{Munoz} M.,  {Moffat} A.~F.~J.,  {Hill} G.~M.,  {Shenar} T.,  {Richardson}
  N.~D.,  {Pablo} H.,  {St-Louis} N.,   {Ramiaramanantsoa} T.,  2017, \mn@doi
  [\mnras] {10.1093/mnras/stw2283}, \href
  {http://adsabs.harvard.edu/abs/2017MNRAS.467.3105M} {467, 3105}

\bibitem[\protect\citeauthoryear{{Naz{\'e}}, {Antokhin}, {Sana}, {Gosset}  \&
  {Rauw}}{{Naz{\'e}} et~al.}{2005}]{naze2005}
{Naz{\'e}} Y.,  {Antokhin} I.~I.,  {Sana} H.,  {Gosset} E.,   {Rauw} G.,  2005,
  \mn@doi [\mnras] {10.1111/j.1365-2966.2005.08945.x}, \href
  {http://adsabs.harvard.edu/abs/2005MNRAS.359..688N} {359, 688}

\bibitem[\protect\citeauthoryear{{Niemela} \& {Moffat}}{{Niemela} \&
  {Moffat}}{1982}]{niemelamoffat1982}
{Niemela} V.~S.,  {Moffat} A.~F.~J.,  1982, \mn@doi [\apj] {10.1086/160161},
  \href {http://adsabs.harvard.edu/abs/1982ApJ...259..213N} {259, 213}

\bibitem[\protect\citeauthoryear{{Niemela}, {Massey}  \& {Conti}}{{Niemela}
  et~al.}{1980}]{niemela1980}
{Niemela} V.~S.,  {Massey} P.,   {Conti} P.~S.,  1980, \mn@doi [\apj]
  {10.1086/158419}, \href {http://adsabs.harvard.edu/abs/1980ApJ...241.1050N}
  {241, 1050}

\bibitem[\protect\citeauthoryear{{Niemela}, {Morrell}, {Fern{\'a}ndez
  Laj{\'u}s}, {Barb{\'a}}, {Albacete Colombo}  \& {Orellana}}{{Niemela}
  et~al.}{2006}]{niemela2006}
{Niemela} V.~S.,  {Morrell} N.~I.,  {Fern{\'a}ndez Laj{\'u}s} E.,  {Barb{\'a}}
  R.,  {Albacete Colombo} J.~F.,   {Orellana} M.,  2006, \mn@doi [\mnras]
  {10.1111/j.1365-2966.2006.10046.x}, \href
  {http://adsabs.harvard.edu/abs/2006MNRAS.367.1450N} {367, 1450}

\bibitem[\protect\citeauthoryear{{Oort} \& {Spitzer}}{{Oort} \&
  {Spitzer}}{1955}]{oortspitzer1955}
{Oort} J.~H.,  {Spitzer} Jr. L.,  1955, \mn@doi [\apj] {10.1086/145958}, \href
  {http://adsabs.harvard.edu/abs/1955ApJ...121....6O} {121, 6}

\bibitem[\protect\citeauthoryear{{Otero}}{{Otero}}{2006}]{otero2006}
{Otero} S.~A.,  2006, Open European Journal on Variable Stars, \href
  {http://adsabs.harvard.edu/abs/2006OEJV...45....1O} {45, 1}

\bibitem[\protect\citeauthoryear{{Penny}, {Gies}, {Hartkopf}, {Mason}  \&
  {Turner}}{{Penny} et~al.}{1993}]{penny1993}
{Penny} L.~R.,  {Gies} D.~R.,  {Hartkopf} W.~I.,  {Mason} B.~D.,   {Turner}
  N.~H.,  1993, \mn@doi [\pasp] {10.1086/133200}, \href
  {http://adsabs.harvard.edu/abs/1993PASP..105..588P} {105, 588}

\bibitem[\protect\citeauthoryear{{Pojmanski}}{{Pojmanski}}{1997}]{pojmanski1997}
{Pojmanski} G.,  1997, \actaa, \href
  {http://adsabs.harvard.edu/abs/1997AcA....47..467P} {47, 467}

\bibitem[\protect\citeauthoryear{{Povich} et~al.,}{{Povich}
  et~al.}{2011a}]{povich2011b}
{Povich} M.~S.,  et~al., 2011a, \mn@doi [\apjs] {10.1088/0067-0049/194/1/6},
  \href {http://adsabs.harvard.edu/abs/2011ApJS..194....6P} {194, 6}

\bibitem[\protect\citeauthoryear{{Povich} et~al.,}{{Povich}
  et~al.}{2011b}]{povich2011}
{Povich} M.~S.,  et~al., 2011b, \mn@doi [\apjs] {10.1088/0067-0049/194/1/14},
  \href {http://adsabs.harvard.edu/abs/2011ApJS..194...14P} {194, 14}

\bibitem[\protect\citeauthoryear{{Preibisch} \& {Mamajek}}{{Preibisch} \&
  {Mamajek}}{2008}]{preibischmamajek2008}
{Preibisch} T.,  {Mamajek} E.,  2008, in {Reipurth} B.,  ed., , Handbook of
  Star Forming Regions, Volume II.
ASP, San Francisco, CA, p.~235

\bibitem[\protect\citeauthoryear{{Preibisch}, {Schuller}, {Ohlendorf},
  {Pekruhl}, {Menten}  \& {Zinnecker}}{{Preibisch}
  et~al.}{2011}]{preibisch2011d}
{Preibisch} T.,  {Schuller} F.,  {Ohlendorf} H.,  {Pekruhl} S.,  {Menten}
  K.~M.,   {Zinnecker} H.,  2011, \mn@doi [\aap] {10.1051/0004-6361/201015425},
  \href {http://adsabs.harvard.edu/abs/2011A%26A...525A..92P} {525, A92}

\bibitem[\protect\citeauthoryear{{Pr{\v s}a} \& {Zwitter}}{{Pr{\v s}a} \&
  {Zwitter}}{2005}]{prsazwitter2005}
{Pr{\v s}a} A.,  {Zwitter} T.,  2005, \mn@doi [\apj] {10.1086/430591}, \href
  {http://adsabs.harvard.edu/abs/2005ApJ...628..426P} {628, 426}

\bibitem[\protect\citeauthoryear{{Rathborne}, {Brooks}, {Burton}, {Cohen}  \&
  {Bontemps}}{{Rathborne} et~al.}{2004}]{rathborne2004}
{Rathborne} J.~M.,  {Brooks} K.~J.,  {Burton} M.~G.,  {Cohen} M.,   {Bontemps}
  S.,  2004, \mn@doi [\aap] {10.1051/0004-6361:20031631}, \href
  {http://adsabs.harvard.edu/abs/2004A%26A...418..563R} {418, 563}

\bibitem[\protect\citeauthoryear{{Rauw}, {Vreux}, {Gosset}, {Hutsemekers},
  {Magain}  \& {Rochowicz}}{{Rauw} et~al.}{1996}]{rauw1996}
{Rauw} G.,  {Vreux} J.-M.,  {Gosset} E.,  {Hutsemekers} D.,  {Magain} P.,
  {Rochowicz} K.,  1996, \aap, \href
  {http://cdsads.u-strasbg.fr/abs/1996A%26A...306..771R} {306, 771}

\bibitem[\protect\citeauthoryear{{Rauw}, {Sana}, {Gosset}, {Vreux}, {Jehin}  \&
  {Parmentier}}{{Rauw} et~al.}{2000}]{rauw2000}
{Rauw} G.,  {Sana} H.,  {Gosset} E.,  {Vreux} J.-M.,  {Jehin} E.,
  {Parmentier} G.,  2000, \aap, \href
  {http://adsabs.harvard.edu/abs/2000A%26A...360.1003R} {360, 1003}

\bibitem[\protect\citeauthoryear{{Rauw}, {Sana}, {Antokhin}, {Morrell},
  {Niemela}, {Albacete Colombo}, {Gosset}  \& {Vreux}}{{Rauw}
  et~al.}{2001}]{rauw2001}
{Rauw} G.,  {Sana} H.,  {Antokhin} I.~I.,  {Morrell} N.~I.,  {Niemela} V.~S.,
  {Albacete Colombo} J.~F.,  {Gosset} E.,   {Vreux} J.-M.,  2001, \mn@doi
  [\mnras] {10.1046/j.1365-8711.2001.04681.x}, \href
  {http://adsabs.harvard.edu/abs/2001MNRAS.326.1149R} {326, 1149}

\bibitem[\protect\citeauthoryear{{Rauw}, {Naz{\'e}}, {Fern{\'a}ndez Laj{\'u}s},
  {Lanotte}, {Solivella}, {Sana}  \& {Gosset}}{{Rauw} et~al.}{2009}]{rauw2009}
{Rauw} G.,  {Naz{\'e}} Y.,  {Fern{\'a}ndez Laj{\'u}s} E.,  {Lanotte} A.~A.,
  {Solivella} G.~R.,  {Sana} H.,   {Gosset} E.,  2009, \mn@doi [\mnras]
  {10.1111/j.1365-2966.2009.15226.x}, \href
  {http://adsabs.harvard.edu/abs/2009MNRAS.398.1582R} {398, 1582}

\bibitem[\protect\citeauthoryear{{Rebolledo} et~al.,}{{Rebolledo}
  et~al.}{2016}]{rebolledo2016}
{Rebolledo} D.,  et~al., 2016, \mn@doi [\mnras] {10.1093/mnras/stv2776}, \href
  {http://adsabs.harvard.edu/abs/2016MNRAS.456.2406R} {456, 2406}

\bibitem[\protect\citeauthoryear{{Ritchie}, {Clark}, {Negueruela}  \&
  {Crowther}}{{Ritchie} et~al.}{2009}]{ritchie2009}
{Ritchie} B.~W.,  {Clark} J.~S.,  {Negueruela} I.,   {Crowther} P.~A.,  2009,
  \mn@doi [\aap] {10.1051/0004-6361/200912686}, \href
  {http://adsabs.harvard.edu/abs/2009A%26A...507.1585R} {507, 1585}

\bibitem[\protect\citeauthoryear{{Roberts}, {Lehar}  \& {Dreher}}{{Roberts}
  et~al.}{1987}]{roberts1987}
{Roberts} D.~H.,  {Lehar} J.,   {Dreher} J.~W.,  1987, \mn@doi [\aj]
  {10.1086/114383}, \href {http://adsabs.harvard.edu/abs/1987AJ.....93..968R}
  {93, 968}

\bibitem[\protect\citeauthoryear{{Rochau}, {Brandner}, {Stolte}, {Gennaro},
  {Gouliermis}, {Da Rio}, {Dzyurkevich}  \& {Henning}}{{Rochau}
  et~al.}{2010}]{rochau2010}
{Rochau} B.,  {Brandner} W.,  {Stolte} A.,  {Gennaro} M.,  {Gouliermis} D.,
  {Da Rio} N.,  {Dzyurkevich} N.,   {Henning} T.,  2010, \mn@doi [\apjl]
  {10.1088/2041-8205/716/1/L90}, \href
  {http://adsabs.harvard.edu/abs/2010ApJ...716L..90R} {716, L90}

\bibitem[\protect\citeauthoryear{{Rochau} et~al.,}{{Rochau}
  et~al.}{2011}]{rochau2011}
{Rochau} B.,  et~al., 2011, \mn@doi [\mnras]
  {10.1111/j.1365-2966.2011.19561.x}, \href
  {http://adsabs.harvard.edu/abs/2011MNRAS.418..949R} {418, 949}

\bibitem[\protect\citeauthoryear{{Sab{\'{\i}}n-Sanjuli{\'a}n}
  et~al.,}{{Sab{\'{\i}}n-Sanjuli{\'a}n} et~al.}{2014}]{sabinsanjulian2014}
{Sab{\'{\i}}n-Sanjuli{\'a}n} C.,  et~al., 2014, \mn@doi [\aap]
  {10.1051/0004-6361/201322798}, \href
  {http://adsabs.harvard.edu/abs/2014A%26A...564A..39S} {564, A39}

\bibitem[\protect\citeauthoryear{{Sana} et~al.,}{{Sana}
  et~al.}{2012}]{sana2012}
{Sana} H.,  et~al., 2012, \mn@doi [Science] {10.1126/science.1223344}, \href
  {http://adsabs.harvard.edu/abs/2012Sci...337..444S} {337, 444}

\bibitem[\protect\citeauthoryear{{Sana} et~al.,}{{Sana}
  et~al.}{2013}]{sana2013}
{Sana} H.,  et~al., 2013, \mn@doi [\aap] {10.1051/0004-6361/201219621}, \href
  {http://adsabs.harvard.edu/abs/2013A%26A...550A.107S} {550, A107}

\bibitem[\protect\citeauthoryear{{Sana} et~al.,}{{Sana}
  et~al.}{2014}]{sana2014}
{Sana} H.,  et~al., 2014, \mn@doi [\apjs] {10.1088/0067-0049/215/1/15}, \href
  {http://adsabs.harvard.edu/abs/2014ApJS..215...15S} {215, 15}

\bibitem[\protect\citeauthoryear{{Schnurr}, {Moffat}, {St-Louis}, {Morrell}  \&
  {Guerrero}}{{Schnurr} et~al.}{2008}]{schnurr2008a}
{Schnurr} O.,  {Moffat} A.~F.~J.,  {St-Louis} N.,  {Morrell} N.~I.,
  {Guerrero} M.~A.,  2008, \mn@doi [\mnras] {10.1111/j.1365-2966.2008.13584.x},
  \href {http://adsabs.harvard.edu/abs/2008MNRAS.389..806S} {389, 806}

\bibitem[\protect\citeauthoryear{{Schweickhardt}, {Schmutz}, {Stahl},
  {Szeifert}  \& {Wolf}}{{Schweickhardt} et~al.}{1999}]{schweickhardt1999}
{Schweickhardt} J.,  {Schmutz} W.,  {Stahl} O.,  {Szeifert} T.,   {Wolf} B.,
  1999, \aap, \href {http://cdsads.u-strasbg.fr/abs/1999A%26A...347..127S}
  {347, 127}

\bibitem[\protect\citeauthoryear{{Sexton}, {Povich}, {Smith}, {Babler}, {Meade}
   \& {Rudolph}}{{Sexton} et~al.}{2015}]{sexton2015}
{Sexton} R.~O.,  {Povich} M.~S.,  {Smith} N.,  {Babler} B.~L.,  {Meade} M.~R.,
   {Rudolph} A.~L.,  2015, \mn@doi [\mnras] {10.1093/mnras/stu2143}, \href
  {http://adsabs.harvard.edu/abs/2015MNRAS.446.1047S} {446, 1047}

\bibitem[\protect\citeauthoryear{{Smith}}{{Smith}}{1987}]{smith1987}
{Smith} R.~G.,  1987, \mn@doi [\mnras] {10.1093/mnras/227.4.943}, \href
  {http://adsabs.harvard.edu/abs/1987MNRAS.227..943S} {227, 943}

\bibitem[\protect\citeauthoryear{{Smith}}{{Smith}}{2004}]{smith2004}
{Smith} N.,  2004, \mn@doi [\mnras] {10.1111/j.1365-2966.2004.07943.x}, \href
  {http://adsabs.harvard.edu/abs/2004MNRAS.351L..15S} {351, L15}

\bibitem[\protect\citeauthoryear{{Smith}}{{Smith}}{2006}]{smith2006a}
{Smith} N.,  2006, \mn@doi [\mnras] {10.1111/j.1365-2966.2006.10007.x}, \href
  {http://adsabs.harvard.edu/abs/2006MNRAS.367..763S} {367, 763}

\bibitem[\protect\citeauthoryear{{Smith} \& {Brooks}}{{Smith} \&
  {Brooks}}{2007}]{smithbrooks2007}
{Smith} N.,  {Brooks} K.~J.,  2007, \mn@doi [\mnras]
  {10.1111/j.1365-2966.2007.12021.x}, \href
  {http://adsabs.harvard.edu/abs/2007MNRAS.379.1279S} {379, 1279}

\bibitem[\protect\citeauthoryear{{Smith} \& {Brooks}}{{Smith} \&
  {Brooks}}{2008}]{smithbrooks2008}
{Smith} N.,  {Brooks} K.~J.,  2008, in {Reipurth} B.,  ed., , Handbook of Star
  Forming Regions, Volume II.
ASP, San Francisco, CA, p.~138

\bibitem[\protect\citeauthoryear{{Smith} \& {Conti}}{{Smith} \&
  {Conti}}{2008}]{smithconti2008}
{Smith} N.,  {Conti} P.~S.,  2008, \mn@doi [\apj] {10.1086/586885}, \href
  {http://adsabs.harvard.edu/abs/2008ApJ...679.1467S} {679, 1467}

\bibitem[\protect\citeauthoryear{{Smith} \& {Stassun}}{{Smith} \&
  {Stassun}}{2017}]{smithstassun2017}
{Smith} N.,  {Stassun} K.~G.,  2017, \mn@doi [\aj] {10.3847/1538-3881/aa5d0c},
  \href {http://adsabs.harvard.edu/abs/2017AJ....153..125S} {153, 125}

\bibitem[\protect\citeauthoryear{{Smith}, {Egan}, {Carey}, {Price}, {Morse}  \&
  {Price}}{{Smith} et~al.}{2000}]{smith2000}
{Smith} N.,  {Egan} M.~P.,  {Carey} S.,  {Price} S.~D.,  {Morse} J.~A.,
  {Price} P.~A.,  2000, \mn@doi [\apjl] {10.1086/312578}, \href
  {http://adsabs.harvard.edu/abs/2000ApJ...532L.145S} {532, L145}

\bibitem[\protect\citeauthoryear{{Smith}, {Bally}  \& {Brooks}}{{Smith}
  et~al.}{2004}]{smith2004b}
{Smith} N.,  {Bally} J.,   {Brooks} K.~J.,  2004, \mn@doi [\aj]
  {10.1086/383291}, \href {http://adsabs.harvard.edu/abs/2004AJ....127.2793S}
  {127, 2793}

\bibitem[\protect\citeauthoryear{{Smith}, {Stassun}  \& {Bally}}{{Smith}
  et~al.}{2005}]{smith2005c}
{Smith} N.,  {Stassun} K.~G.,   {Bally} J.,  2005, \mn@doi [\aj]
  {10.1086/427249}, \href {http://adsabs.harvard.edu/abs/2005AJ....129..888S}
  {129, 888}

\bibitem[\protect\citeauthoryear{{Smith}, {Bally}  \& {Walborn}}{{Smith}
  et~al.}{2010a}]{smith2010a}
{Smith} N.,  {Bally} J.,   {Walborn} N.~R.,  2010a, \mn@doi [\mnras]
  {10.1111/j.1365-2966.2010.16520.x}, \href
  {http://adsabs.harvard.edu/abs/2010MNRAS.405.1153S} {405, 1153}

\bibitem[\protect\citeauthoryear{{Smith} et~al.,}{{Smith}
  et~al.}{2010b}]{smith2010b}
{Smith} N.,  et~al., 2010b, \mn@doi [\mnras]
  {10.1111/j.1365-2966.2010.16792.x}, \href
  {http://adsabs.harvard.edu/abs/2010MNRAS.406..952S} {406, 952}

\bibitem[\protect\citeauthoryear{{Solivella} \& {Niemela}}{{Solivella} \&
  {Niemela}}{1999}]{solivellaniemela1999}
{Solivella} G.~R.,  {Niemela} V.~S.,  1999, in {Morrell} N.~I.,  {Niemela}
  V.~S.,   {Barb{\'a}} R.~H.,  eds,  Revista Mexicana de Astronomia y
  Astrofisica Conference Series Vol. 8, Workshop on Hot Stars in Open Clusters
  of the Galaxy and the Magellenic Clouds. pp 145--147

\bibitem[\protect\citeauthoryear{{Sota}, {Ma{\'{\i}}z Apell{\'a}niz},
  {Morrell}, {Barb{\'a}}, {Walborn}, {Gamen}, {Arias}  \& {Alfaro}}{{Sota}
  et~al.}{2014}]{sota2014}
{Sota} A.,  {Ma{\'{\i}}z Apell{\'a}niz} J.,  {Morrell} N.~I.,  {Barb{\'a}}
  R.~H.,  {Walborn} N.~R.,  {Gamen} R.~C.,  {Arias} J.~I.,   {Alfaro} E.~J.,
  2014, \mn@doi [\apjs] {10.1088/0067-0049/211/1/10}, \href
  {http://adsabs.harvard.edu/abs/2014ApJS..211...10S} {211, 10}

\bibitem[\protect\citeauthoryear{{Soubiran}, {Jasniewicz}, {Chemin}, {Crifo},
  {Udry}, {Hestroffer}  \& {Katz}}{{Soubiran} et~al.}{2013}]{soubiran2013}
{Soubiran} C.,  {Jasniewicz} G.,  {Chemin} L.,  {Crifo} F.,  {Udry} S.,
  {Hestroffer} D.,   {Katz} D.,  2013, \mn@doi [\aap]
  {10.1051/0004-6361/201220927}, \href
  {http://adsabs.harvard.edu/abs/2013A%26A...552A..64S} {552, A64}

\bibitem[\protect\citeauthoryear{{Spitzer}}{{Spitzer}}{1978}]{spitzer1978}
{Spitzer} L.,  1978, {Physical processes in the interstellar medium}.
Wiley-Interscience, New York, NY, \mn@doi{10.1002/9783527617722}

\bibitem[\protect\citeauthoryear{{Steenbrugge}, {de Bruijne}, {Hoogerwerf}  \&
  {de Zeeuw}}{{Steenbrugge} et~al.}{2003}]{steenbrugge2003}
{Steenbrugge} K.~C.,  {de Bruijne} J.~H.~J.,  {Hoogerwerf} R.,   {de Zeeuw}
  P.~T.,  2003, \mn@doi [\aap] {10.1051/0004-6361:20030277}, \href
  {http://adsabs.harvard.edu/abs/2003A%26A...402..587S} {402, 587}

\bibitem[\protect\citeauthoryear{{Tapia}, {Roth}, {Marraco}  \& {Ruiz}}{{Tapia}
  et~al.}{1988}]{tapia1988}
{Tapia} M.,  {Roth} M.,  {Marraco} H.,   {Ruiz} M.~T.,  1988, \mn@doi [\mnras]
  {10.1093/mnras/232.3.661}, \href
  {http://cdsads.u-strasbg.fr/abs/1988MNRAS.232..661T} {232, 661}

\bibitem[\protect\citeauthoryear{{Tapia}, {Roth}, {V{\'a}zquez}  \&
  {Feinstein}}{{Tapia} et~al.}{2003}]{tapia2003}
{Tapia} M.,  {Roth} M.,  {V{\'a}zquez} R.~A.,   {Feinstein} A.,  2003, \mn@doi
  [\mnras] {10.1046/j.1365-8711.2003.06186.x}, \href
  {http://adsabs.harvard.edu/abs/2003MNRAS.339...44T} {339, 44}

\bibitem[\protect\citeauthoryear{{Thackeray}, {Tritton}  \&
  {Walker}}{{Thackeray} et~al.}{1973}]{thackeray1973}
{Thackeray} A.~D.,  {Tritton} S.~B.,   {Walker} E.~N.,  1973, \memras, \href
  {http://adsabs.harvard.edu/abs/1973MmRAS..77..199T} {77, 199}

\bibitem[\protect\citeauthoryear{{Th{\'e}} \& {Vleeming}}{{Th{\'e}} \&
  {Vleeming}}{1971}]{thevleeming1971}
{Th{\'e}} P.~S.,  {Vleeming} G.,  1971, \aap, \href
  {http://adsabs.harvard.edu/abs/1971A%26A....14..120T} {14, 120}

\bibitem[\protect\citeauthoryear{{The}, {Bakker}  \& {Antalova}}{{The}
  et~al.}{1980a}]{the1980d}
{The} P.~S.,  {Bakker} R.,   {Antalova} A.,  1980a, \aaps, \href
  {http://adsabs.harvard.edu/abs/1980A%26AS...41...93T} {41, 93}

\bibitem[\protect\citeauthoryear{{The}, {Bakker}  \& {Tjin A Djie}}{{The}
  et~al.}{1980b}]{the1980b}
{The} P.~S.,  {Bakker} R.,   {Tjin A Djie} H.~R.~E.,  1980b, \aap, \href
  {http://adsabs.harvard.edu/abs/1980A%26A....89..209T} {89, 209}

\bibitem[\protect\citeauthoryear{{Tokovinin}, {Fischer}, {Bonati}, {Giguere},
  {Moore}, {Schwab}, {Spronck}  \& {Szymkowiak}}{{Tokovinin}
  et~al.}{2013}]{tokovinin2013}
{Tokovinin} A.,  {Fischer} D.~A.,  {Bonati} M.,  {Giguere} M.~J.,  {Moore} P.,
  {Schwab} C.,  {Spronck} J.~F.~P.,   {Szymkowiak} A.,  2013, \mn@doi [\pasp]
  {10.1086/674012}, \href {http://adsabs.harvard.edu/abs/2013PASP..125.1336T}
  {125, 1336}

\bibitem[\protect\citeauthoryear{{Tovmassian}, {Hovhannessian}  \&
  {Epremian}}{{Tovmassian} et~al.}{1994}]{tovmassian1994}
{Tovmassian} H.~M.,  {Hovhannessian} R.~K.,   {Epremian} R.~A.,  1994, \mn@doi
  [\apss] {10.1007/BF00658208}, \href
  {http://adsabs.harvard.edu/abs/1994Ap%26SS.213..175T} {213, 175}

\bibitem[\protect\citeauthoryear{{Turner} \& {Moffat}}{{Turner} \&
  {Moffat}}{1980}]{turnermoffat1980}
{Turner} D.~G.,  {Moffat} A.~F.~J.,  1980, \mn@doi [\mnras]
  {10.1093/mnras/192.2.283}, \href
  {http://adsabs.harvard.edu/abs/1980MNRAS.192..283T} {192, 283}

\bibitem[\protect\citeauthoryear{{Tutukov}}{{Tutukov}}{1978}]{tutukov1978}
{Tutukov} A.~V.,  1978, \aap, \href
  {http://adsabs.harvard.edu/abs/1978A%26A....70...57T} {70, 57}

\bibitem[\protect\citeauthoryear{{Vall{\'e}e}}{{Vall{\'e}e}}{2014}]{vallee2014}
{Vall{\'e}e} J.~P.,  2014, \mn@doi [\aj] {10.1088/0004-6256/148/1/5}, \href
  {http://adsabs.harvard.edu/abs/2014AJ....148....5V} {148, 5}

\bibitem[\protect\citeauthoryear{{Vazquez}, {Baume}, {Feinstein}  \&
  {Prado}}{{Vazquez} et~al.}{1996}]{vazquez1996}
{Vazquez} R.~A.,  {Baume} G.,  {Feinstein} A.,   {Prado} P.,  1996, \aaps,
  \href {http://adsabs.harvard.edu/abs/1996A%26AS..116...75V} {116, 75}

\bibitem[\protect\citeauthoryear{{Walborn}}{{Walborn}}{1973}]{walborn1973}
{Walborn} N.~R.,  1973, \mn@doi [\apj] {10.1086/151891}, \href
  {http://adsabs.harvard.edu/abs/1973ApJ...179..517W} {179, 517}

\bibitem[\protect\citeauthoryear{{Walborn}}{{Walborn}}{1982a}]{walborn1982b}
{Walborn} N.~R.,  1982a, \mn@doi [\aj] {10.1086/113216}, \href
  {http://adsabs.harvard.edu/abs/1982AJ.....87.1300W} {87, 1300}

\bibitem[\protect\citeauthoryear{{Walborn}}{{Walborn}}{1982b}]{walborn1982a}
{Walborn} N.~R.,  1982b, \mn@doi [\apjl] {10.1086/183747}, \href
  {http://adsabs.harvard.edu/abs/1982ApJ...254L..15W} {254, L15}

\bibitem[\protect\citeauthoryear{{Walborn}}{{Walborn}}{1995}]{walborn1995}
{Walborn} N.~R.,  1995, in {Niemela} V.,  {Morrell} N.,   {Feinstein} A.,  eds,
   Revista Mexicana de Astronomia y Astrofisica Conference Series Vol. 2, The
  Eta Carinae Region: A Laboratory of Stellar Evolution. p.~51

\bibitem[\protect\citeauthoryear{{Walborn}}{{Walborn}}{2009}]{walborn2009}
{Walborn} N.~R.,  2009, in {Livio} M.,  {Villaver} E.,  eds,  Space Telescope
  Science Institute Symposium Series Vol. 20, Massive Stars: From Pop III and
  GRBs to the Milky Way. Cambridge University Press, Cambridge, UK, pp
  167--177, \mn@doi{10.1017/CBO9780511770593.012}

\bibitem[\protect\citeauthoryear{{Walborn} \& {Hesser}}{{Walborn} \&
  {Hesser}}{1975}]{walbornhesser1975}
{Walborn} N.~R.,  {Hesser} J.~E.,  1975, \mn@doi [\apj] {10.1086/153720}, \href
  {http://adsabs.harvard.edu/abs/1975ApJ...199..535W} {199, 535}

\bibitem[\protect\citeauthoryear{{Walborn} \& {Liller}}{{Walborn} \&
  {Liller}}{1977}]{walbornliller1977}
{Walborn} N.~R.,  {Liller} M.~H.,  1977, \mn@doi [\apj] {10.1086/154917}, \href
  {http://adsabs.harvard.edu/abs/1977ApJ...211..181W} {211, 181}

\bibitem[\protect\citeauthoryear{{Walborn} et~al.,}{{Walborn}
  et~al.}{2002a}]{walborn2002a}
{Walborn} N.~R.,  et~al., 2002a, \mn@doi [\aj] {10.1086/339831}, \href
  {http://adsabs.harvard.edu/abs/2002AJ....123.2754W} {123, 2754}

\bibitem[\protect\citeauthoryear{{Walborn}, {Danks}, {Vieira}  \&
  {Landsman}}{{Walborn} et~al.}{2002b}]{walborn2002b}
{Walborn} N.~R.,  {Danks} A.~C.,  {Vieira} G.,   {Landsman} W.~B.,  2002b,
  \mn@doi [\apjs] {10.1086/339373}, \href
  {http://adsabs.harvard.edu/abs/2002ApJS..140..407W} {140, 407}

\bibitem[\protect\citeauthoryear{{Walborn}, {Smith}, {Howarth}, {Vieira Kober},
  {Gull}  \& {Morse}}{{Walborn} et~al.}{2007}]{walborn2007}
{Walborn} N.~R.,  {Smith} N.,  {Howarth} I.~D.,  {Vieira Kober} G.,  {Gull}
  T.~R.,   {Morse} J.~A.,  2007, \mn@doi [\pasp] {10.1086/511756}, \href
  {http://adsabs.harvard.edu/abs/2007PASP..119..156W} {119, 156}

\bibitem[\protect\citeauthoryear{{Walborn} et~al.,}{{Walborn}
  et~al.}{2014}]{walborn2014}
{Walborn} N.~R.,  et~al., 2014, \mn@doi [\aap] {10.1051/0004-6361/201323082},
  \href {http://adsabs.harvard.edu/abs/2014A%26A...564A..40W} {564, A40}

\bibitem[\protect\citeauthoryear{{Walsh}}{{Walsh}}{1984}]{walsh1984}
{Walsh} J.~R.,  1984, \aap, \href
  {http://adsabs.harvard.edu/abs/1984A%26A...138..380W} {138, 380}

\bibitem[\protect\citeauthoryear{{Wang} et~al.,}{{Wang}
  et~al.}{2011}]{wang2011}
{Wang} J.,  et~al., 2011, \mn@doi [\apjs] {10.1088/0067-0049/194/1/11}, \href
  {http://adsabs.harvard.edu/abs/2011ApJS..194...11W} {194, 11}

\bibitem[\protect\citeauthoryear{{Whiteoak}}{{Whiteoak}}{1994}]{whiteoak1994}
{Whiteoak} J.~B.~Z.,  1994, \mn@doi [\apj] {10.1086/174313}, \href
  {http://adsabs.harvard.edu/abs/1994ApJ...429..225W} {429, 225}

\bibitem[\protect\citeauthoryear{{Williams}, {Gies}, {Hillwig}, {McSwain}  \&
  {Huang}}{{Williams} et~al.}{2011}]{williams2011}
{Williams} S.~J.,  {Gies} D.~R.,  {Hillwig} T.~C.,  {McSwain} M.~V.,   {Huang}
  W.,  2011, \mn@doi [\aj] {10.1088/0004-6256/142/5/146}, \href
  {http://adsabs.harvard.edu/abs/2011AJ....142..146W} {142, 146}

\bibitem[\protect\citeauthoryear{{Wright}, {Drew}  \& {Mohr-Smith}}{{Wright}
  et~al.}{2015}]{wright2015}
{Wright} N.~J.,  {Drew} J.~E.,   {Mohr-Smith} M.,  2015, \mn@doi [\mnras]
  {10.1093/mnras/stv323}, \href
  {http://adsabs.harvard.edu/abs/2015MNRAS.449..741W} {449, 741}

\bibitem[\protect\citeauthoryear{{Wright}, {Bouy}, {Drew}, {Sarro}, {Bertin},
  {Cuillandre}  \& {Barrado}}{{Wright} et~al.}{2016}]{wright2016}
{Wright} N.~J.,  {Bouy} H.,  {Drew} J.~E.,  {Sarro} L.~M.,  {Bertin} E.,
  {Cuillandre} J.-C.,   {Barrado} D.,  2016, \mn@doi [\mnras]
  {10.1093/mnras/stw1148}, \href
  {http://adsabs.harvard.edu/abs/2016MNRAS.460.2593W} {460, 2593}

\bibitem[\protect\citeauthoryear{{Wu}, {Zhou}, {Ma}  \& {Du}}{{Wu}
  et~al.}{2009}]{wu2009}
{Wu} Z.-Y.,  {Zhou} X.,  {Ma} J.,   {Du} C.-H.,  2009, \mn@doi [\mnras]
  {10.1111/j.1365-2966.2009.15416.x}, \href
  {http://adsabs.harvard.edu/abs/2009MNRAS.399.2146W} {399, 2146}

\bibitem[\protect\citeauthoryear{{Yonekura}, {Asayama}, {Kimura}, {Ogawa},
  {Kanai}, {Yamaguchi}, {Barnes}  \& {Fukui}}{{Yonekura}
  et~al.}{2005}]{yonekura2005}
{Yonekura} Y.,  {Asayama} S.,  {Kimura} K.,  {Ogawa} H.,  {Kanai} Y.,
  {Yamaguchi} N.,  {Barnes} P.~J.,   {Fukui} Y.,  2005, \mn@doi [\apj]
  {10.1086/496869}, \href {http://adsabs.harvard.edu/abs/2005ApJ...634..476Y}
  {634, 476}

\bibitem[\protect\citeauthoryear{{de Bruijne}}{{de
  Bruijne}}{1999}]{debruijne1999}
{de Bruijne} J.~H.~J.,  1999, \mn@doi [\mnras]
  {10.1046/j.1365-8711.1999.02953.x}, \href
  {http://adsabs.harvard.edu/abs/1999MNRAS.310..585D} {310, 585}

\bibitem[\protect\citeauthoryear{{de Graauw}, {Lidholm}, {Fitton}, {Beckman},
  {Israel}, {Nieuwenhuijzen}  \& {Vermue}}{{de Graauw}
  et~al.}{1981}]{degraauw1981}
{de Graauw} T.,  {Lidholm} S.,  {Fitton} B.,  {Beckman} J.,  {Israel} F.~P.,
  {Nieuwenhuijzen} H.,   {Vermue} J.,  1981, \aap, \href
  {http://adsabs.harvard.edu/abs/1981A%26A...102..257D} {102, 257}

\makeatother
\end{thebibliography}

%%%%%%%%%%%%%%%%%%%%%%%%%%%%%%%%%%%%%%%%%%%%%%%%%%%%%%%%%%%%%%%%%%%%%%%
\appendix

\section{Periodograms}
\label{sec:ostars-appendix}

\btxt{Here we present frequency power spectra for the binary systems fit in
  Section \ref{sec:ostars-binaries}.  We use the discrete Fourier
  transform (DFT) and one-dimensional {\tt CLEAN}
  algorithm\footnote{The two-dimensional {\tt CLEAN} algorithm was
    first developed by \citet{hoegbom1974}.} of \citet{roberts1987} as
  implemented in IDL by A. W. Fullerton \citep[see][]{fullerton1997}.
  The DFT is computed up to a maximum frequency of
  $f_{\textrm{max}}=1/ (2\Delta t_{\textrm{min}})$, where $\Delta
  t_{\textrm{min}}$ is the minimum time interval between observations.
  {\tt CLEAN} deconvolves the window function from the DFT, removing
  aliases introduced by uneven time sampling.  We run {\tt CLEAN} for
  100 iterations with a gain of 0.5.}

\btxt{The resulting periodograms are used first to check for
  periodicity and second to corroborate the periods found by the
  orbital fitting package {\tt rvfit} \citep{iglesiasmarzoa2015} for
  HD 92607, HD 93576, and HDE 305536.  {\tt rvfit} does not require an
  intial estimate of the period, and we leave the period as a free
  parameter when fitting these three systems.  For the eclipsing
  binary HDE 303312, we fix the period in {\tt rvfit} at the
  photometric period as described in Section
  \ref{subsec:ostars-303312}.}

%----------------------------------------------------------------------
\begin{figure}
  \centering
  \includegraphics[width=\columnwidth,trim=6mm 2mm 6mm 6mm,clip]{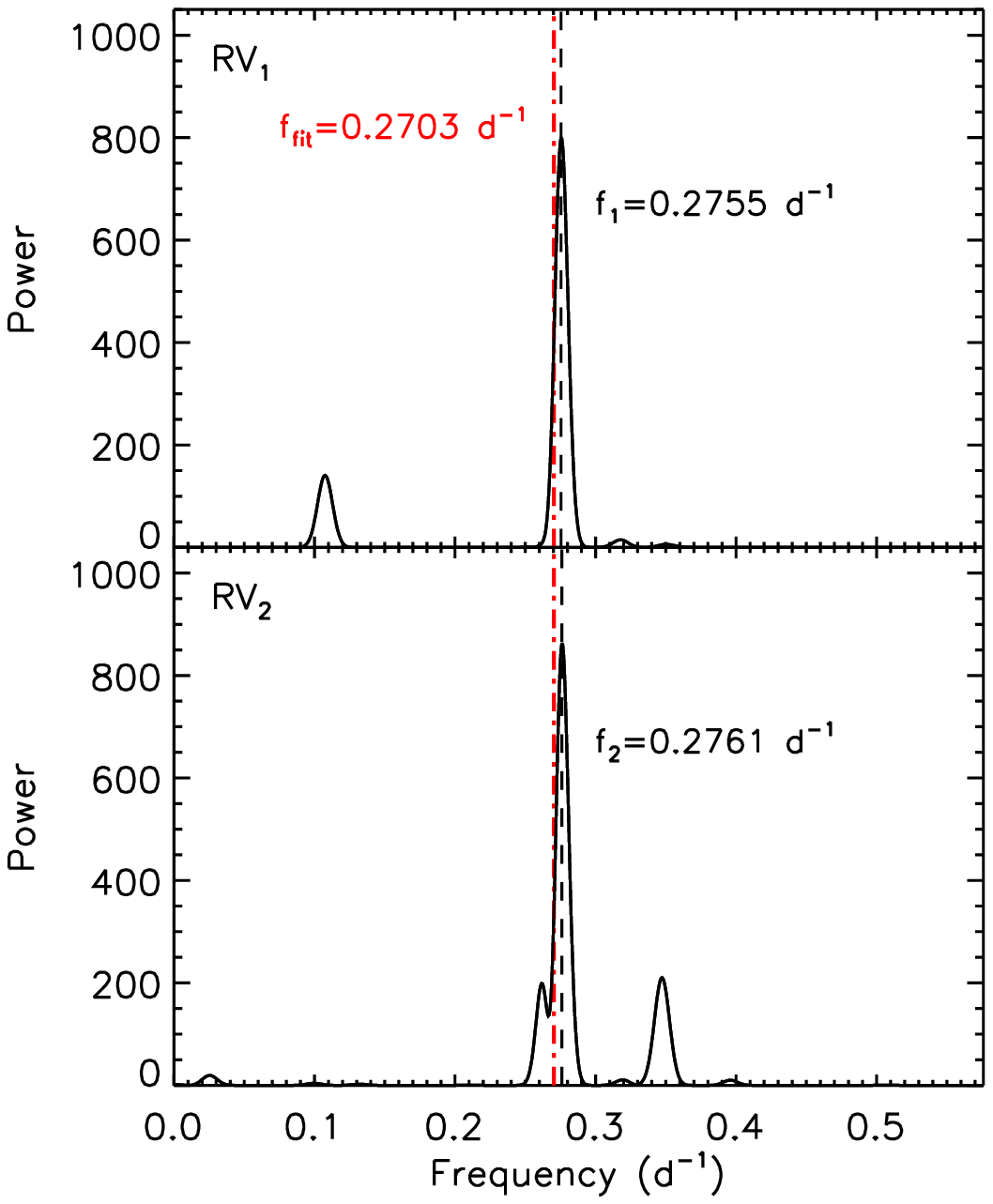}
  \caption{\btxt{{\tt CLEAN}ed power spectra of the primary (top) and
      secondary (bottom) RV components of the SB2 HD 92607.  The black
      dashed lines mark the frequencies $f_1$ and $f_2$ corresponding
      to the strongest peak in each power spectrum.  The red dot-dash
      line in both panels indicates the frequency $f_{\textrm{fit}}$
      corresponding to the final best-fitting orbital period from an
      unconstrained simultaneous fit to both components with {\tt
        rvfit}.} }
  \label{fig:ostars-app92607}  
\end{figure}  
%----------------------------------------------------------------------

\btxt{The periodograms for the primary and secondary components of HD
  92607 (Figure \ref{fig:ostars-app92607}) both have clear peaks at
  frequencies of $\approx0.276$~d~$^{-1}$, corresponding to a period
  of $\sim3.6$~d.  The full, unconstrained fit with {\tt rvfit} (see
  Section \ref{subsec:ostars-92607}) finds a period of 3.6993~d.
  Visual inspection of phase-folded RV curves rules out most common
  possible alias periods.  We cannot definitively exclude the $1+f$
  alias, which is above our maximum searchable frequency and
  corresponds to a period of $\sim0.79$~d; however, a sub-day period
  would be extremely unusual for a main-sequence O+O binary
  \citep{kiminkikobulnicky2012,sana2012,moedistefano2016,almeida2017}.

  The components of HD 92607 are of very similar spectral type
  \citep[O8.5 V + O9 V;][]{sexton2015}.  If we swap the values of
  RV$_{1}$ and RV$_{2}$ at several epochs, the periodograms of both
  components show clear peaks at 0.538~d$^{-1}$, or a period of
  1.859~d.  With these component assignments, {\tt rvfit} finds a
  plausible solution for a similar period of 1.84957~d.  This fit
  requires a substantial eccentricity ($e=0.35$) and an approximate
  inclination of $45\degr$.}

%----------------------------------------------------------------------
\begin{figure}
  \centering
  \includegraphics[width=\columnwidth,trim=6mm 2mm 6mm 6mm,clip]{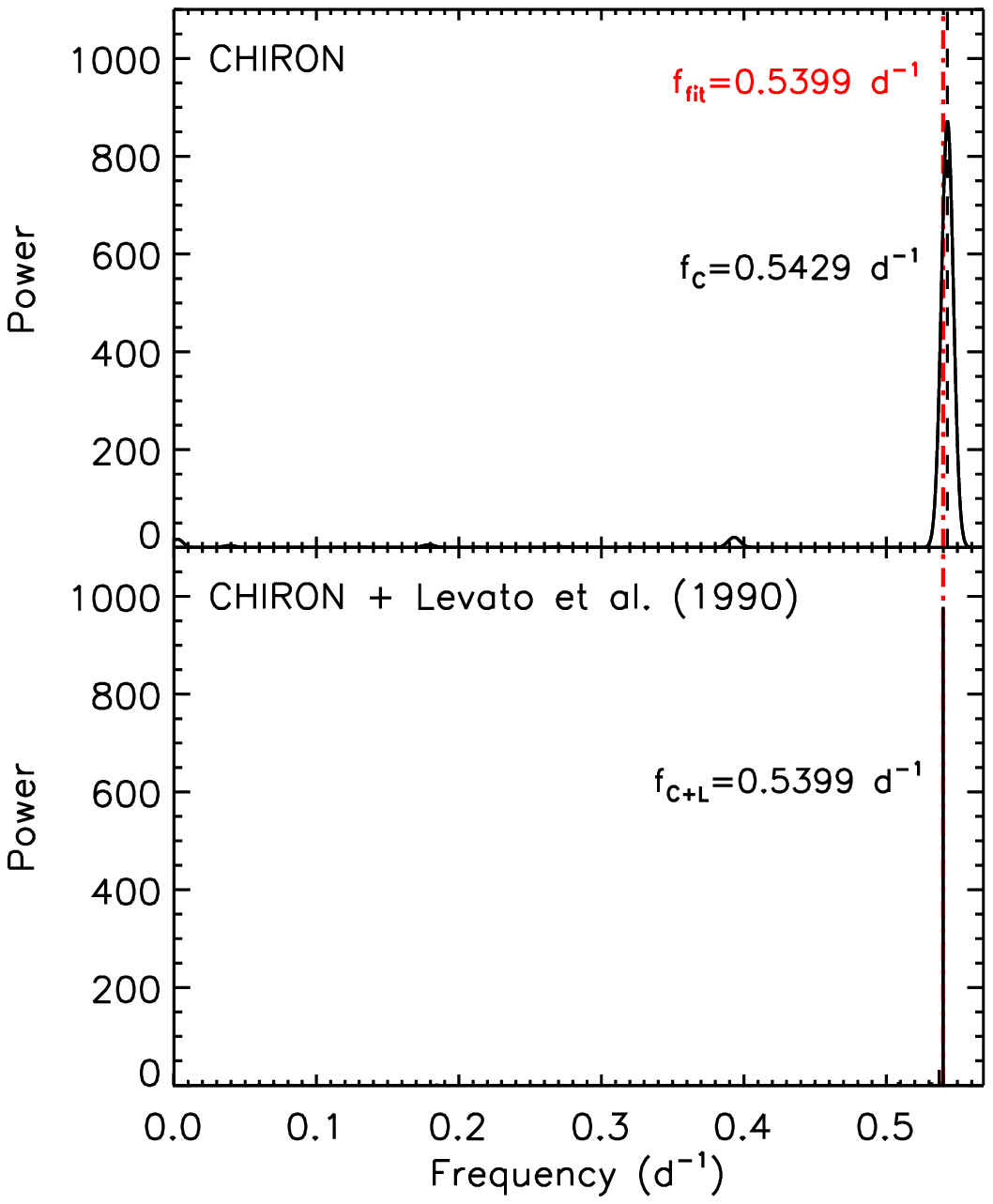}
  \caption{\btxt{Top: {\tt CLEAN}ed power spectrum of our CHIRON RV
      data for the SB1 HD 93576.  The black dashed line marks the
      frequency $f_{\textrm{C}}$ corresponding to the strongest peak.
      Bottom: as above, including RV data from \citet{levato1990}.
      The frequency $f_{\textrm{C+L}}$ corresponds to the strongest
      peak.  The red dot-dash line in both panels indicates the
      frequency $f_{\textrm{fit}}$ corresponding to the final
      best-fitting orbital period from an unconstrained fit to the
      combined data set with {\tt rvfit}.}}
  \label{fig:ostars-app93576}
\end{figure}
%----------------------------------------------------------------------

\btxt{The periodogram derived from our CHIRON data for HD 93576 (upper
  panel in Figure \ref{fig:ostars-app93576}) shows a single strong peak
  at a frequency of 0.543~d$^{-1}$, or a period of 1.84~d.  A similar
  peak appears when this system's data from \citet{levato1990} are
  included (lower panel of Figure \ref{fig:ostars-app93576}).  The
  full, unconstrained fit with {\tt rvfit} to the combined data set
  (see Section \ref{subsec:ostars-93576}) finds a period of
  1.852102~d.  Before {\tt CLEAN}ing, the DFT for HD 93576 also shows
  a peak at the $1-f$ alias of 0.460~d$^{-1}$, or a period of 2.174~d.
  We can force a plausible solution at this period in {\tt rvfit} if
  we constrain the range of allowed periods and fit only to our CHIRON
  data; the combined data set including \citet{levato1990} RVs is
  incompatible with this period.  As this alias peak is removed by the
  {\tt CLEAN} deconvolution, it appears to be an artifact of our
  observational time sampling.
}

%----------------------------------------------------------------------
\begin{figure}
  \centering
  \includegraphics[width=\columnwidth,trim=6mm 2mm 6mm 6mm,clip]{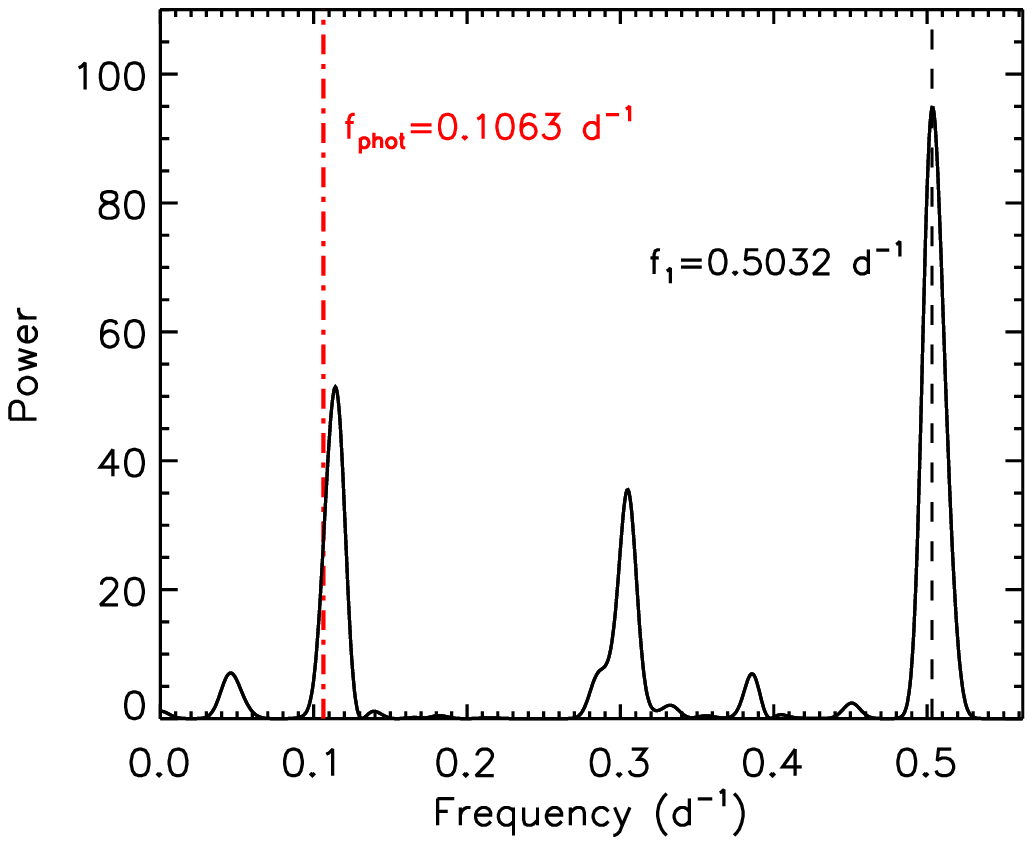}
  \caption{\btxt{{\tt CLEAN}ed power spectrum of our CHIRON RV data for the
      eclipsing SB1 HDE 303312.  The black dashed line marks the
      frequency $f_{1}$ corresponding to the strongest peak.  The red
      dot-dash line indicates the frequency $f_{\textrm{phot}}$ of the
      ASAS $V$-band light curve.}}
  \label{fig:ostars-app303312}  
\end{figure}
%----------------------------------------------------------------------

\btxt{The RV periodogram for the eclipsing binary HDE 303312 (Figure
  \ref{fig:ostars-app303312}) has several peaks.  The strongest peak
  is at a frequency of 0.503~d$^{-1}$ or a period of 1.99~d.  This
  period---which, as a near-integer multiple of days, is likely a
  sampling artifact---is ruled out by visual inspection of the
  phase-folded RV curve and by constrained tests in {\tt rvfit}. The
  second-highest peak in the periodogram is at a frequency of
  0.114~d$^{-1}$, or a period of $\sim8.8$~d.  Unconstrained fits in
  {\tt rvfit} converge on a period between 9.41 and 9.65~d, close to
  the photometric period of 9.4111~d (which corresponds to a frequency
  of 0.10626~d$^{-1}$).  As described in Section
  \ref{subsec:ostars-303312}, we fix the period at the photometric
  period for subsequent fitting.  }

%----------------------------------------------------------------------
\begin{figure}
  \centering
  \includegraphics[width=\columnwidth,trim=6mm 2mm 6mm 6mm,clip]{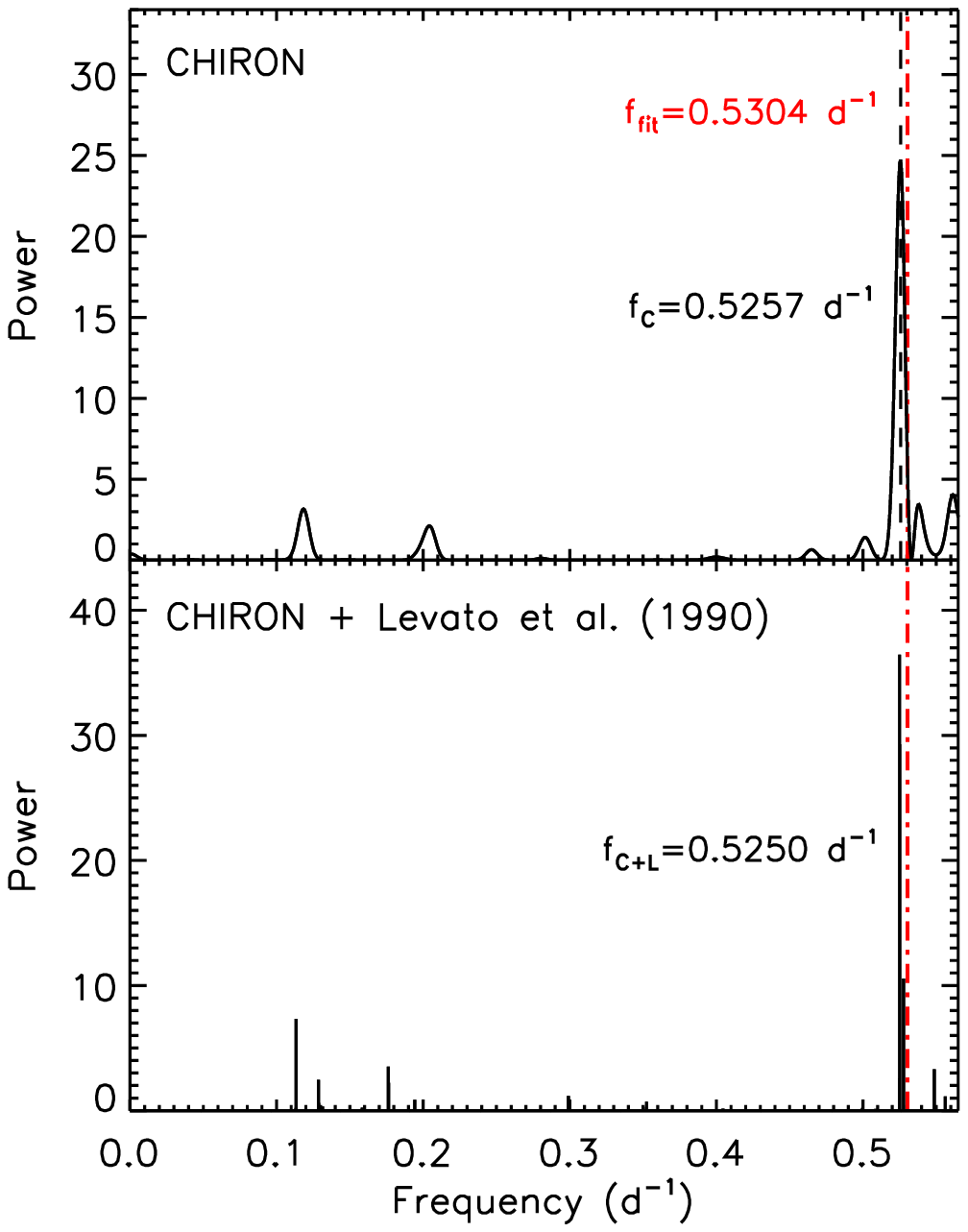}
  \caption{\btxt{Top: {\tt CLEAN}ed power spectrum of our CHIRON RV
      data for the SB1 HDE 305536.  The black dashed line marks the
      frequency $f_{\textrm{C}}$ corresponding to the strongest peak.
      Bottom: as above, including RV data from \citet{levato1990}.
      The frequency $f_{\textrm{C+L}}$ corresponds to the strongest
      peak.  The red dot-dash line in both panels indicates the
      frequency $f_{\textrm{fit}}$ corresponding to the final
      best-fitting orbital period from an unconstrained fit to our
      CHIRON data with {\tt rvfit}.}}
  \label{fig:ostars-app305536}
\end{figure}
%----------------------------------------------------------------------

\btxt{The periodogram derived from our CHIRON data for HDE 305536 (top
  panel of Figure \ref{fig:ostars-app305536}) has its highest peak at
  a frequency of 0.526~d$^{-1}$ or a period of 1.90~d.
  A similar peak frequency is found when this system's data from
  \citet{levato1990} are included (bottom panel of Figure
  \ref{fig:ostars-app305536}).  Visual inspection of phase-folded RV
  curves and constrained tests in {\tt rvfit} rule out common possible
  aliases.  The full, unconstrained fit to our CHIRON data with {\tt
    rvfit} finds an orbital period of 1.88535~d, corresponding to a
  frequency of 0.53041~d$^{-1}$.  As described in Section
  \ref{subsec:ostars-305536}, {\tt rvfit} is unable to converge on a
  solution that includes the \citet{levato1990} data.  }

%%%%%%%%%%%%%%%%%%%%%%%%%%%%%%%%%%%%%%%%%%%%%%%%%%%%%%%%%%%%%%%%%%%%%%%

\bsp
\label{lastpage}
\end{document}